\documentclass[12pt]{article}

\pdfoutput=1
\usepackage[dvipsnames]{xcolor}
\usepackage{amsfonts}
\usepackage{amsmath}
\usepackage{epsfig}
\usepackage{latexsym}
\usepackage{graphicx}
\usepackage{color}
\usepackage{amssymb}
\usepackage{epstopdf,mathtools,braket}
\numberwithin{equation}{section}
\usepackage{tikz}
\usetikzlibrary{intersections, calc}
\usetikzlibrary {arrows.meta,bending,positioning}
\usepackage[hang,small,bf]{caption}
\usepackage[subrefformat=parens]{subcaption}
\allowdisplaybreaks
\usepackage{afterpage}

\usepackage{url}

\usepackage{bm}
\usepackage{cancel}

\usepackage{hyperref}
\hypersetup{colorlinks,linkcolor=blue,citecolor=BrickRed,urlcolor=cyan}

\def\be{\begin{equation}}
\def\ee{\end{equation}}
\def\bea{\begin{eqnarray}}
\def\eea{\end{eqnarray}}
\def\bequ{\begin{equation}}
\def\eequ{\end{equation}}

\def\del{\partial}

\def\d{\text{d}}
\def\i{\text{i}}
\renewcommand{\thefootnote}{\fnsymbol{footnote}}

\newcommand{\D}{\text{d}}

\begin{document}
\hfuzz=100pt
\title{New Field Theories with Foliation Structure
and Subdimensional Particles from Godbillon-Vey Invariant}

\author{
\large 
Hiromi Ebisu$^{1}$\footnote{hiromi.ebisu(at)yukawa.kyoto-u.ac.jp},\quad
Masazumi Honda$^{2}$\footnote{masazumi.honda(at)riken.jp}, \\
Taiichi Nakanishi$^{1,2}$\footnote{taiichi.nakanishi(at)yukawa.kyoto-u.ac.jp}  \ \  {\rm and}\ \ Soichiro Shimamori$^{3}$\footnote{s(underbar)shimamori(at)het.phys.sci.osaka-u.ac.jp}
\vspace{1em} \\
\\
$^{1}${\small\it Center for Gravitational Physics and Quantum Information,} \\
{\small\it Yukawa Institute for Theoretical Physics, Kyoto University,}\\
{\small\it  Sakyo-ku, Kyoto 606-8502, Japan}
\\
$^{2}${\small\it Interdisciplinary Theoretical and Mathematical Sciences Program (iTHEMS), }\\
{\small\it RIKEN, Wako 351-0198, Japan}
\\
$^{3}${\small\it Department of Physics, Osaka University, Machikaneyama-Cho 1-1, Toyonaka 560-0043, Japan}
}

\date{\small{August 2024}}

\maketitle
\thispagestyle{empty}
\centerline{}

\begin{abstract}
    Recently, subdimensional particles including fractons have attracted much attention from various areas. Notable features of such matter phases are mobility constraints and subextensive ground state degeneracies (GSDs). 
    In this paper, we propose a BF-like theory motivated by the Godbillon-Vey invariant, which is a mathematical invariant of the foliated manifold. 
    Our theory hosts subsystem higher form symmetries which manifestly ensure the mobility constraint
    and subextensive GSD through the spontaneous symmetry breaking. We also discuss some lattice spin models which realize the same low energy behaviours as the BF-like theory. Furthermore, we explore dynamical matter theories 
    which are coupled to
    the BF-like theory. 
\end{abstract}

\vfill
\noindent

{\small YITP-24-78, RIKEN-iTHEMS-Report-24, OU-HET-1237}

\renewcommand{\thefootnote}{\arabic{footnote}}
\setcounter{footnote}{0}

\newpage
\pagenumbering{arabic}
\setcounter{page}{1}
\tableofcontents

\newpage

\section{Introduction}\label{sec:introduction}
Originally proposed in the context of quantum information science \cite{Chamon:2004lew,Haah:2011drr}, fracton phases have attracted plethora of attentions in view of various research areas. 
The striking feature of these phases is that they admit fractionalized quasiparticle excitations with mobility constraints, giving rise to subextensive ground state degeneracy (GSD). 
According to their mobility constraints, there are several types of such particles. 
Some of the examples are \textit{planons} which can move through a two-dimensional surface, \textit{lineons} which can move through a one-dimensional line, and \textit{fractons} which are completely immobile.
Throughout this work, we term such particles 
subject to mobility constraints as \textit{subdimensional particles}.
The discovery of subdimensional particles has deepened
our understanding of 
phases of 
matter.
For instance, the relationship between the ground state Hilbert space the foliation structure of the space manifold \cite{Shirley:2017suz,Shirley:2018nhn,Shirley:2018vtc,Shirley:2019hzt} has been introduced in the context of the X-cube model \cite{Vijay:2016phm}, which is a prototypical example of the fracton phases. 
Also, 
studies of the fracton phases have introduced a new type of symmetries, \textit{spatially modulated symmetries}, such as the dipole symmetry associated with conservation of dipole moments \cite{BEEKMAN20171,griffin2015scalar,Pretko:2016kxt,Pretko:2016lgv,Pretko:2018jbi,Seiberg:2019vrp}, 
and the subsystem symmetry whose symmetry defects are topological only in certain directions \cite{PhysRevB.66.054526,plqt_ising_2004,Seiberg:2020bhn,Seiberg:2020wsg,Seiberg:2020cxy,Gorantla:2020xap,Katsura:2022xkg,Yamaguchi:2021qrx,Honda:2022shd}.

So far, there have been many attempts to describe subdimensional particles in the framework of quantum field theory.
One of them is the \textit{foliated BF theory}, which can be constructed via stacking BF theories in layers \cite{Slagle:2018swq,Slagle:2020ugk,Hsin:2021mjn,Hirono:2022dci,Cao:2023rrb,2023foliated,2024multipole,Ohmori:2022rzz,Shimamura:2024kwf,Spieler:2023wkz,Hsin:2024eyg,anomaly2024}. 
Although this theory has succeeded in capturing various properties of subdimensional particles in a field theoretic manner, there are some subtleties in the following sense.
First, as pointed out in \cite{2023foliated}, at least some kinds of foliated BF theories are not defined as rigorous U(1) gauge theories on a smooth manifold. 
In particular, there are some issues in global structure of the gauge group coming from regularization of theories with dipole symmetry.
In continuum, dipole charge is not quantized as there is no minimal unit of length, implying the corresponding gauge group cannot take compact topology. 
Second, the foliated BF theories that have been studied so far are not genuinely characterized by the foliation structure of the manifold. 
Indeed, one has to include the redundancy of the choice of normal 1-form field as a gauge redundancy in the field theory, as the foliation structure should be independent of the particular choice. 
In this regard, we should think that these theories of the field theory are characterized by the normal 1-form field, rather than the foliation structure.

The main aim of this work is to present a field theoretical description of subdimensional particles with proper foliation structures which resolves the subtleties mentioned above. 
In our approach, we characterize the phases of subdimensional particles by the genuine foliation structure starting from the \emph{Godbillon-Vey invariant} which is the mathematical invariant of the manifold equipped with the foliation structure \cite{1572261549792201088}~\footnote{See e.g.~\cite{tamura1992topology} for a primer to this subject.}.  

\begin{figure}[t]
    \begin{center}
        \begin{tikzpicture}    
            \foreach \y in {0, 1.5, 3} {
            
            \draw[ultra thick] (3,0+\y) sin (4,0.3+\y) cos (5,\y) sin (6,-0.3+\y) cos (7,\y) sin (8,0.3+\y);
        }
        \draw[very thick,->, red!80] (5, 0) -- ++(1/5, 2/3)  node [above right] {$\omega$};
    \end{tikzpicture}
        \caption{Schematic picture of the foliation structure of the manifold. The foliated manifold is piled up by an infinite number of leaves which are depicted as wavy lines. The leaves are characterized by the normal 1-form field $\omega$.}
        \label{fig:foliation}
    \end{center}
\end{figure}
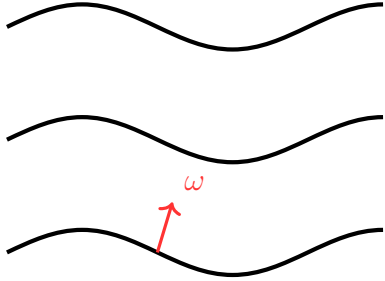

To be more specific, let us outline how to build up our theory by taking a clue from the Chern-Simons theory. 
In the Chern-Simons case, we get a topological field theory by interpreting the characteristic class\footnote{
Strictly speaking, more appropriate terminology would be the secondary characteristic class but we simply call it the characteristic class throughout this paper.
} of topological structure as a field theory.
In the same manner, we make use of the characteristic class of the foliation structure called Godbillon-Vey class to obtain the field theory for the matter phases. The foliation structure of the manifold is specified by a 1-form field $\omega$ normal to the leaves of the foliation as illustrated in Fig.\,\ref{fig:foliation}.

We emphasize that a normal 1-form field $\omega$ and another
1-form field $\omega'$ define the same foliation structure
when $\omega' = c\,\omega$ with some positive function $c$.
Thus, a normal 1-form field $\omega$ and another normal 1-form field $c\,\omega$ mean the same foliation structure:
\begin{align}
    \omega \sim c\,\omega \quad (c: \mathrm{positive\ function})\ , 
\end{align}
which can be regarded as a gauge redundancy.
Once given a normal 1-form field $\omega$ which describes the foliation structure,
we can find a 1-form field $\eta$ satisfying\footnote{The existence of $\eta$ is guaranteed by the consistency condition
called \textit{integrability condition}.}
\begin{align} \label{eq:defofeta_intro}
    \d \omega = \omega \wedge \eta\ .
\end{align}
By using $\eta$, we can construct a characteristic class
of the foliation structure, called Godbillon-Vey class as
\begin{align}
    \mathrm{(Godbillon\text{-}Vey\ class)} := [\eta \wedge (\d\eta)^\alpha] \qquad (\alpha: \mathrm{integer})\ , 
\end{align}
where the square bracket means an element of the de Rham cohomology. 
One may naively expect that this Godbillon-Vey class can be used to construct a field theory describing phases of matter of subdimensional particles. 
However, as we will see later, it turns out that we cannot reproduce the subextensive GSD just by promoting this class to a field theory. 
In this paper, we overcome this problem by making some modifications and obtain a suitable field theory for foliation matter phases, particularly focusing on $\alpha=1$ case.

This work is organized as follows.
In section~\ref{sec:GV_field_theory_for_single_foliation},
we introduce the mathematical definition of the Godbillon-Vey class of a codimension one single foliation, and naively promote it to a field theory.
Afterwards, we modify the theory to describe the desired foliation matter phases which show the subextensive GSD to its system size.
In section~\ref{sec:lattice_spin_models}, we investigate lattice spin models whose low energy behavior can be described by the Godbillon-Vey field theory.
In section~\ref{sec:coupling_matter_theory}, we construct matter field theories which can be coupled to the Godbillon-Vey field theory, and discuss the subdimensional particle
behavior.
Section~\ref{sec:conclusion} is devoted for conclusion and future directions.
In appendices, we provide some detailed calculations of the canonical structure in Godbillon-Vey field theory, and mathematical properties of foliation structures.
In appendix~\ref{sec:canonical_quantization}, we calculate the canonical commutation relations of non-trivial loop operators in Godbillon-Vey field theory via the quantization of constrained system.
In appendix~\ref{sec:precise_formulation}, we present more rigorous formulation of the Godbillon-Vey field theories in the same split as a work by Dijkgraaf and Witten~\cite{Dijkgraaf:1989pz}.
In appendix~\ref{sec:higher_codimension}, we propose possible generalizations of the Godbillon-Vey field theory to higher codimension foliations. Remarkably, the Godbillon-Vey field theory with a higher codimension foliation possesses the non-abelian gauge group structures rather than the abelian one, which is developed in the main text.
In appendix~\ref{sec:g-structure}, we mention generalized mathematical structures beyond foliation structures. We argue the spacetime symmetries, which is referred to as \textit{G-structure}, when the spacetime manifold is equipped with these mathematical structures.

\section{Godbillon-Vey Field Theory for Codimension One Single Foliation}
\label{sec:GV_field_theory_for_single_foliation}
In this section, we construct a field theory motivated by a characteristic class of the foliation structure.
Before delving into our theories which require several detailed analyzes, we present summary of the results in this section. 

As explained in section~\ref{sec:introduction}, we start from the Godbillon-Vey invariant defined by
\begin{align}
    \label{eq:GV_invariant}
    \text{GV} = \int_{M^3}
    \Bigl[ \eta \wedge \d\eta - \lambda \wedge (\d\omega - \omega \wedge \eta) \Bigr]\ ,
\end{align}
which is a mathematical invariant associated with the foliation structure of the $(2+1)$-dimensional spacetime manifold $M^3$. Here, $\lambda$ is the Lagrange multiplier for the relation~\eqref{eq:defofeta_intro}. 
As we will see, even if we interpret this number as a physical action, this Godbillon-Vey invariant itself does not reproduce subextensive GSD, and is not suited for a field theory describing subdimensional particle phases. 
However, motivated by this invariant, we can construct a field theory which shows the subextensive GSD in $2+1$ dimensions:\footnote{In this action, $b$, $c$, $\lambda$, and $\phi$ are dynamical variables, while $\omega$ is a background field.}
\begin{align}
    \label{eq:action_U(1)_GVBF}
     S[b,c, \lambda, \phi] = \frac{k}{2\pi} \int_{M^3} 
     \Bigl[ b\wedge \d c - \lambda \wedge (\d\phi \wedge \omega - \omega \wedge b) \Bigr]\ .
\end{align}
Here, $b$ and $c$ are U(1) gauge fields, $\lambda$ is a 1-form field, and $\phi$ is a compact scalar. Also, $k$ is the level quantized to be integer, which is an analogue of the Chern-Simons level.
Note that $\lambda$ is no longer a Lagrange multiplier as it involves the canonical momentum constructed from the dynamical field $\phi$.
The GSD of this theory is given by
\begin{align}
    \text{GSD} = k^{\beta_1^\omega L_\omega} (=\infty)\ ,
\end{align}
where $\beta_1^\omega$ is the first Betti number of the \textit{leaf manifold}, which is stretched in transverse directions to the normal 1-form $\omega$, and $L_\omega$ is a lattice size along $\omega$-direction, although which diverges in the continuum theory.
This subextensive GSD can be understood as spontaneous symmetry breaking of 1-form subsystem symmetries, whose symmetry operators are topological only in a leaf of the foliation.

We can moreover generalize the action \eqref{eq:action_U(1)_GVBF} in the following way: 
\begin{align}
    \label{eq:action_U(1)_pqGVBF}
    S = \frac{k}{2\pi} \int_{M^{d+1}} 
    \Bigl[ b^{(d-p)} \wedge \d c^{(p)} - \lambda^{(p)} \wedge (\d\zeta^{(d-p-1)} \wedge \omega - \omega \wedge b^{(d-p)}) \Bigr]\ ,
\end{align}
where $M^{d+1}$ is a ($d+1$)-dimensional spacetime manifold whose codimension one foliation structure is characterized by the background normal $1$-form field $\omega$. In this action, $b^{(d-p)}$ and $c^{(p)}$ are ($d-p$)-form and $p$-form U(1) gauge fields respectively. Also, $\zeta^{(d-p-1)}$ is a ($d-p-1$)-form field, and $k$ is again an integer parameter.
In this theory, we will see that the GSD is given by
\begin{align}
    \text{GSD} = k^{\beta_p^\omega L_\omega} (=\infty)\ ,
\end{align}
where $\beta_p^\omega$ is a $p$-th Betti number of the leaf, and $L_\omega$ is lattice size along $\omega$-direction, although it is infinity in continuum field theory case.
This GSD is again understood as spontaneous symmetry breaking of $(d-p)$-form subsystem symmetries.

We dedicate the rest of this section to present the detailed explanation on the above outline. (See also appendix~\ref{sec:canonical_quantization}.)

\subsection{Codimension One Foliation and Godbillon-Vey Class}
A codimension one foliation structure can be characterized by
a normal 1-form $\omega$, which denotes the direction normal to stacked leaf manifolds at every point.
The normal 1-form $\omega$ must satisfy the following
integrability condition:
\begin{align}
    \label{eq:omega_integrability}
    \omega \wedge \d\omega = 0\ ,
\end{align}
which implies that this normal form can be written \textit{locally} as
\begin{align}
    \label{eq:omega_local}
    \omega \overset{\text{locally}}{=} g\, \d f\ .
\end{align}
Here, \textit{locally} means that this equality is only valid for
a certain local coordinate system. Also,~$f$ and $g$ are local functions
defined on this neighborhood system.
Given this expression, a leaf manifold $L^{d}_{\alpha}$ is defined by the patchworks of locally defined constant $d$-dimensional hypersurfaces $f=\alpha$:
\begin{align}
    L^{d}_\alpha = \bigcup_{\substack{\mathrm{local\ coordinate}\\ \mathrm{systems}}} \{ x | f(x) = \alpha \}\ . 
\end{align}
Then, the foliation structure on a spacetime manifold $M^{d+1}$ can be defined by the stacking of these leaves on the manifold:
\begin{align}
    M^{d+1} = \bigsqcup_{\alpha} L^{d}_\alpha\ .
\end{align}
In the local expression \eqref{eq:omega_local} of $\omega$, we find that the local function $g$ is irrelevant to the 
definition of leaf manifolds.
This corresponds to the fact that the magnitude of the normal 1-form field does not have any sense. When we refer to the foliation structure, we only focus on the properties of leaf manifolds and how they are 
stacked, forming layers
and not on the explicit form of normal 1-form $\omega$.
Hence, even if we multiply $\omega$
by a positive function $c$, we obtain the same foliation structure.
In other words, the folioation structure always has the following redundancy:
\begin{align}
    \label{eq:omega_redundancy}
    \omega \sim c\, \omega \quad (c: \mathrm{positive\ function})\ .
\end{align}
This redundancy gives rise to the gauge symmetry when 
we construct a theory coupling to the foliation structure, which is discussed
in the later sections.

From now on, we seek for a de Rham cohomology class which can be made from the foliation structure of the manifold. From the integrability condition \eqref{eq:omega_integrability},
there exists a 1-form $\eta$, which satisfies the following relation: 
\begin{align}
    \label{eq:eta_definition}
    \d\omega = \omega \wedge \eta\ .
\end{align} 
This is not invariant under the change of the normal 1-form field defining the same foliation structure
because if we pick $\omega':=c\,\omega$ in \eqref{eq:omega_redundancy} as a normal 1-form field, then $\d \omega$ is changed as
\begin{align}
\begin{aligned}
    \d\omega \ \rightarrow\ \d \omega' 
    = \d c \wedge \omega + c\,\d\omega  
    =\omega' \wedge (\eta - \d (\log c))\ . 
\end{aligned}
\end{align}
This implies that we must introduce an 1-form field $\eta' := \eta - \d(\log c)$ to satisfy the relation~\eqref{eq:eta_definition}:\footnote{The function $\d(\log c)$ is well-defined function since $c$ is a globally positive function.}
\begin{align}
    \d \omega' = \omega' \wedge \eta' \ .
\end{align}
Note also that the $1$-form field $\eta$ is not uniquely determined. 
Indeed, if we shift $\eta$ by $\omega$ multiplied with some scalar function $u$ as
\begin{align}
    \label{eq:eta_redundancy}
    \eta \rightarrow \tilde{\eta} = \eta + u\,\omega \ ,
\end{align}
then we have the same relation for $\tilde{\eta}$:
\begin{align}
    \d \omega = \omega \wedge \tilde{\eta}\ .
\end{align}

In terms of the 1-form $\eta$ defined as \eqref{eq:eta_definition},
we can define the Godbillon-Vey class in terms of a de Rham cohomology class:
\begin{align}
    \label{eq:Godbillon-Vey_definition}
        \mathrm{(Godbillon\text{-}Vey\ class)} := [\eta \wedge (\d\eta)^\alpha] \qquad (\alpha  :  \mathrm{integer})\ .
\end{align}
We can easily check that the de Rham cohomology class is invariant 
under the change of the normal 1-form \eqref{eq:omega_redundancy} and the redundancy of $\eta$ \eqref{eq:eta_redundancy}.
For simplicity, we restrict to $\alpha = 1$ while the generalization to generic $\alpha$ is straightforward.
We can prove the invariance under the redundancy of the normal 1-form $\omega$ \eqref{eq:omega_redundancy} as follows: 
\begin{align}
    \begin{aligned}
    \eta \wedge \d\eta \ \rightarrow\ 
    \eta' \wedge \d\eta' &= \left( \eta - \d(\log c) \right) \wedge \d \left( \eta-\d(\log c) \right) \\
    &= \eta \wedge \d\eta - \d\left( (\log c) \d\eta \right)\ .
    \end{aligned}
\end{align}
Since the difference between $\eta\wedge \d \eta$ and $\eta' \wedge \d \eta'$ is just an exact form, we can regard these two classes as the same cohomology class:
\begin{align}\label{eq:derham_invariance1}
    [\eta\wedge \d \eta]=[\eta' \wedge \d \eta']\ .
\end{align}
One can verify the invariance of \eqref{eq:derham_invariance1} under the redundancy of $\eta$ as follows. 
We first note the following identity:
\begin{align}
     \omega \wedge \d\eta =0 \ .
\label{eq:omega_deta}
\end{align}
This can be derived by differentiating both hand sides of \eqref{eq:eta_definition}:
\begin{align}
    \d (\d\omega ) = \d(\omega \wedge \eta) = 0\quad \Longrightarrow \quad \d\omega \wedge \eta - \omega \wedge \d\eta = -\omega \wedge \d\eta = 0\ ,
\end{align}
where in the middle step, we have used the identity $\d \omega \wedge \eta = \omega \wedge \eta \wedge \eta = 0$.
Exploiting this identity \eqref{eq:omega_deta} jointly with the integrability condition \eqref{eq:omega_integrability}, the Godbillon-Vey class transforms under the change \eqref{eq:eta_redundancy} of $\eta$ as 
\begin{align}
    \begin{aligned}
    \eta \wedge d\eta \rightarrow \tilde{\eta} \wedge \d\tilde{\eta} &= (\eta + u\omega) \wedge \d(\eta + u\omega) \\
    &= \eta \wedge \d\eta + u\omega \wedge \d\eta + \eta \wedge \d(u\omega) + u\omega \wedge \d(u\omega) \\
    &= \eta \wedge \d\eta - \d(u \eta \wedge \omega)\ , 
    \end{aligned}
\end{align}
which again implies that $\eta\wedge \d \eta$ and $\tilde{\eta} \wedge \d \tilde{\eta}$ belong to the same de Rham cohomology class:
\begin{align}\label{eq:derham_invariance2}
    [\eta\wedge \d \eta]=[\tilde{\eta} \wedge \d \tilde{\eta}]\ .
\end{align}
By using the Godbillon-Vey class defined in \eqref{eq:Godbillon-Vey_definition}, we can construct a mathematical index on the odd-dimensional spacetime manifold with the foliation structure called \textit{Godbillon-Vey invariant}:
\begin{align}
    \text{GV}[M^{2\alpha+1}, \langle\omega\rangle] = \int_{M^{2\alpha+1}} \eta \wedge (\d\eta)^\alpha\ .
\end{align}
Here, we denote the dependence on the foliation structure as $\langle \omega \rangle$ to emphasize that the quantity depends on its foliation structure but not on the specific choice of the normal 1-form~$\omega$.
While the form of this invariant somewhat resembles 
the Chern-Simons invariant on a manifold with principal bundle structure,
this quantity is a geometrical invariant under transformations which respect the foliation structure as seen from \eqref{eq:derham_invariance1} and \eqref{eq:derham_invariance2} in the above discussions.%
\footnote{Let $f:M\rightarrow N$ be a map which respects the foliation structure, which means that 
the pullback $f^{*}: T^{*}M\rightarrow T^{*}N$ maps the normal 1-form $\omega$ of the foliation on $M$ to the normal 1-form $f_{*}\omega$ of the foliation on $N$. 
The pullback commutes with the external derivative and the wedge product.
This leads to
\begin{align}
    \d(f^{*} \omega) = f^{*}(\d \omega) = f^{*} (\omega \wedge \eta) = f^{*}\omega \wedge f^{*} \eta,
\end{align}
which means that $f^{*}\eta$ is a suitable 1-form field in \eqref{eq:eta_definition} for the normal 1-form $f^{*}\omega$.
The uniqueness following from \eqref{eq:derham_invariance1} and \eqref{eq:derham_invariance2} shows that the number
 \begin{align}
     \int_{N^{2\alpha+1}} f^{*}\eta \wedge \left(\d (f^{*}\eta)\right)^\alpha = \int_{M^{2\alpha+1}} \eta \wedge (\d\eta)^{\alpha},
 \end{align}
coincides with the Godbillon-Vey invariant associated with $N^{2\alpha+1}$ and $\langle f_{*}\omega \rangle$.
}

\subsection{First Attempt to Construct Field Theory Motivated by Godbillon-Vey Class}
In the last subsection, we have seen 
that the Godbillon-Vey class has two kinds of redundancies stemming from the foliation structure. 
Toward constructing a field theory which respects the foliation structure, we regard these redundancies as the gauge symmetries. 
The most natural candidate possessing these gauge redundancies can be obtained by promoting the 1-form field~$\eta$ defined in \eqref{eq:eta_definition} to a physical degree of freedom, and the Godbillon-Vey invariant to an physical action:
\begin{align}
    \label{eq:GV_action}
    S[\eta, \lambda ;M^3,\langle\omega\rangle] = C\,\int_{M^3} 
    \Bigl[ \eta \wedge \d\eta - \lambda \wedge (\d\omega - \omega \wedge \eta) \Bigr]\ ,
\end{align}
where $C$ is some constant.
As expected, this theory has the following two types of gauge symmetries:
\begin{align}
    \label{eq:GV_gauge_shift}
    &\eta \rightarrow \eta + u \omega\ , \quad \lambda \rightarrow \lambda + 2 u\, \eta - 2\, \d u\ , \\
    \label{eq:GV_gauge_exact}
    &\eta \rightarrow \eta - \d(\log c)\ , \quad \lambda \rightarrow c^{-1}\,\lambda\ , \quad \omega \rightarrow c\,\omega \ , 
\end{align}
where $u$ and $v$ are any spacetime functions and $c$ is 
a spacetime \emph{positive} function.%
\footnote{
The spacetime function $u$ transforms under \eqref{eq:GV_gauge_exact} as $u\rightarrow c^{-1}u$ to keep the transformation law of $\lambda$.
}
The first and second gauge redundancies correspond to the ones of the normal 1-form \eqref{eq:omega_redundancy} and $\eta$ \eqref{eq:eta_redundancy}, respectively. In addition, this theory also possesses the following symmetry:
\begin{align}
        \label{eq:GV_gauge_lambda}
    &\lambda \rightarrow \lambda + v\,\omega\ .
\end{align}

In spite of this apparent success, we fail to build up a proper matter phase which shows the subextensive GSD since the above theory is reduced to
an invertible phase, where everything is determined by the classical configuration of fields.
To see this, we first integrate out the Lagrange multiplier $\lambda$, which gives rise to the constraint \eqref{eq:eta_definition}.
Given this constraint, the configuration of $\eta$ is completely determined up to the gauge transformations \eqref{eq:GV_gauge_shift} and \eqref{eq:GV_gauge_exact}.
In this sense, the model \eqref{eq:GV_action} itself cannot be used as an effective field theory with the subextensive GSD. 
In the next subsections, we will see that by modifying the above theory, and obtain a more intriguing theory which is not merely an invertible phase. 
Nevertheless, there are important lessons to be learned from the above model \eqref{eq:GV_action}, which convince ourselves that the Godbillon-Vey action is a good starting point.   

First, as opposed to the Chern-Simons theory, the coefficient $C$ appearing in front of the integral is not quantized. 
In the case of the Chern-Simon theory,
the level quantization comes from the U(1) gauge
invariance of the action.
In our case, whereas, the gauge group must be $\mathbf{R}_{+}$ (set of positive real numbers) rather than U(1) which is clearly seen from \eqref{eq:GV_gauge_exact}.
This non-compactness of the gauge group admits any real number
in front of the integral, hence there is no level quantization, and nor 
charge quantization which is valid for ordinary~U(1) gauge
theory case.

Secondly, the model \eqref{eq:GV_action} possesses the 1-form subsystem symmetry \cite{Gaiotto:2014kfa}, meaning that there exists a codimension two defect that is topological in certain directions, but not in other ones. To see this, it is convenient first to write down the equations of motion with respect to $\eta$:\footnote{If we consider the equation of motion for the Lagrange multiplier $\lambda$, the constraint \eqref{eq:eta_definition} is obtained. As said above, the second equation fixes the configuration of the field $\eta$. }
\begin{align}
    2\,\d\eta - \lambda \wedge \omega = 0\ , \quad \text{for }\   ^\forall\,\lambda \ , 
\end{align}
which is rewritten as
\begin{align}
    \label{eq:eta_1-form_symm}
     \d\eta\bigr|_{\omega^\text{T}} = 0\ .
\end{align}
Here, the symbol $(\cdots )\bigr|_{\omega^\text{T}}$ stands for the restriction of the form to the leaf manifold which is spread in the transverse directions to the foliation normal field $\omega$.
This equation of motion can be regarded as the flat connection condition in leaf directions. This implies the existence of the $\mathbf{R}_{+}$ 1-form subsystem symmetry and the corresponding symmetry defect is given by
\begin{align}
    \label{eq:eta_topological_op}
    \exp\left(\i\alpha \int_\mathcal{C} \eta \right)\ , \quad \alpha\in\mathbf{R}\ , 
\end{align}
where the integral region $\mathcal{C}$ is a closed loop embedded in a certain leaf. Indeed, thanks to the flat connection condition, this defect is invariant under the continuous
deformation of $\mathcal{C}$ inside the leaf.
This property of the above symmetry defect is a reminiscent of the global symmetries of the subdimensional particle theories, where the mobility is restricted to the certain submanifold.

We can see the relationship between the above symmetry defect and the subdimensional particles in an alternative perspective.
By coupling the gauge field $\eta$ with some classical matter current $J$, 
the whole action takes the following form:
\begin{align}
    S = S_{\text{gauge}} + \int \eta \wedge \star J \ ,
    \label{eq:coupling_current}
\end{align}
where $S_{\text{gauge}}$ is some field theories which have the gauge redundncies \eqref{eq:GV_gauge_shift} and \eqref{eq:GV_gauge_exact} such as the Godbillon-Vey action.
To keep the gauge invariance, we must impose several conditions on
the matter current $J$.
Firstly, the action must be invariant under the gauge transformation~\eqref{eq:GV_gauge_exact}.
Hence, the matter current must satisfy
\begin{align}
    \d\star J = 0\  ,
    \label{eq:continuous_equation}
\end{align}
which is nothing but the current conservation law.
Secondly, the gauge invariance under~\eqref{eq:GV_gauge_shift} imposes the following 
stringent condition on the matter current:
\begin{align}
    \label{eq:current_omega_ortho}
    \omega \wedge \star J = 0\ ,
\end{align}
which is rewritten as
\begin{align}
    \label{eq:current_continuous_eq}
    \omega_\mu J^\mu = 0 \ . 
\end{align}
This indicates that the matter current must be orthogonal
to the foliation field $\omega$, namely the matter current only flows in the leaf directions.
This feature is typically seen in the subdimensional particle theories, where our matter current $J$ serves as the lineon which can move in one space direction
in (2+1) dimensions.

\subsection{Modified Godbillon-Vey Field Theory with Subextensive GSD}
As discussed in the last
subsection, the Godbillon-Vey action \eqref{eq:GV_action} itself is not suited for the foliation matter phase, which shows the subextensive GSD since it is reduced to just an invertible phase. 
To obtain the subextensive GSD, we should overcome the following problems of the action \eqref{eq:GV_action}:
\begin{itemize}
    \item \textbf{Problem 1}: non-compactness of gauge group.\\
    As seen before, the Godbillon-Vey action \eqref{eq:GV_action} has $\mathbf{R}_{+}$ gauge symmetry \eqref{eq:GV_gauge_exact}. Due to this non-compactness of the gauge group, subsystem symmetry algebras are infinitely generated by loop operators \eqref{eq:eta_topological_op}, which is the immediate cause of non-subextensive GSD. To achieve the subextensive GSD, therefore, we should have a compact gauge group rather than noncompact one. We should also comment that if the gauge group is compact, the level which is termed by $C$ in the above action \eqref{eq:GV_action} is quantized. 
    \item \textbf{Problem 2}: invertible phase. \\
    As we discussed, the action \eqref{eq:GV_action} describes an invertible phase. 
    The Godbillon-Vey functional is
    an invariant defined on the \emph{spacetime manifold} rather than principal bundles. 
    Hence, once we fix the foliated spacetime manifold, namely solving the constraint \eqref{eq:eta_definition} with respect to $\eta$, then there are no seeds giving rise to the non-trivial GSD.%
    \footnote{There is a
    crucial difference between topological orders such as Chern-Simons and Godbillon-Vey action \eqref{eq:GV_action}.
    The Chern-Simons functional is defined on the \emph{principal bundle structure}. Therefore, for the Chern-Simons case, even if we fix the spacetime manifold, the structures of the principal bundles can be non-trivial, which leads to the GSD of the topological order.}  
    To avoid this problem, we need other dynamical gauge fields which are not completely fixed only from the constraints coming from the foliation structure. 
\end{itemize}

Our main goal in this section is to obtain the theory \eqref{eq:action_U(1)_GVBF} by modifying the Godbillon-Vey action \eqref{eq:GV_action} such that the above two problems are resolved based on the following schematic program:
\begin{align}
\label{eq:strategy}
    \begin{matrix}
    \mathbf{R}_{+}\ \eqref{eq:GV_action}\ \overset{\text{Step }1}{\xlongrightarrow[]{\hspace{3em}}} \ \mathbf{C} \backslash \{0\}\ \overset{\text{Step }2}{\xlongrightarrow[]{\hspace{3em}}}\ \text{U(1)} \times \mathbf{R}_{+} \ \overset{\text{Step 3}}{\xlongrightarrow[]{\hspace{3em}}} \ 
    & \hspace{-10mm}\text{U}(1)\ \eqref{eq:GV_action_step3} \\
    & \hspace{-10mm}{\text{\scriptsize Step 4}}{\bigg\downarrow} \\
    & \hspace{-10mm}\text{U(1) BF-like}\ \eqref{eq:action_U(1)_GVBF} \ .
     \end{matrix}
\end{align}
Our strategy roughly consists of two parts. From Step 1 to Step 3, we first modify the non-compact gauge group $\mathbf{R}_{+}$ to the compact one U(1) to resolve \textbf{Problem 1}. Although we can arrive at the U(1) Chern-Simons-like gauge theory whose action is given in \eqref{eq:GV_action_step3} after these three steps, this theory is again reduced to the invertible phase. 
To resolve \textbf{Problem 2}, we generalize 
the obtained U(1) Chern-Simons-like action to BF-like one \eqref{eq:action_U(1)_GVBF} keeping the foliation structure. 
Below, we provide the detailed explanation for each step.

\subsubsection*{Step 1. Extension of the Gauge Group}
As repeatedly stressed so far, the level quantization condition is not guaranteed in the Godbillon-Vey action \eqref{eq:GV_action} due to the non-compactness of the gauge group.  In Step 1, we analytically continue the real gauge group $\mathbf{R}_{+}$ to the complex one $\mathbf{C}\backslash\{0\}$ with a puncture at the origin. 
Notice that this extended complex gauge group $\mathbf{C}\backslash\{0\}$ is homotopically equivalent to the compact group U(1). This fact plays a crucial role in resolving \textbf{Problem 1} as we will see later.
Although the geometrical meaning is elusive, we consider a (nonzero) complex-valued normal 1-form $\omega_{\mathbf{C}}$, which is an analytic
continuation of the real valued one $\omega$ and satisfies the \textit{extended} integrability condition
\begin{align}
    \omega_{\mathbf{C}} \wedge \d \omega_{\mathbf{C}} = 0 \ .
\end{align}
Correspondingly, the one-form gauge field $\eta$ should also be lifted up to the complex gauge field $\eta_{\mathbf{C}}$ satisfying
\begin{align}
    \d \omega_{\mathbf{C}} = \omega_{\mathbf{C}} \wedge \eta_{\mathbf{C}} \ .
\end{align}
Under these preparations, we arrive at
\begin{align}\label{eq:GV_action_step1}
    S[\eta_{\mathbf{C}}, \lambda_{\mathbf{C}}; M^3, \langle\omega_{\mathbf{C}}\rangle] = \frac{k}{4\pi} \mathrm{Re}\left[\int_{M^3} \eta_{\mathbf{C}} \wedge \d\eta_{\mathbf{C}} - \lambda_{\mathbf{C}} \wedge (\d\omega_{\mathbf{C}} - \omega_{\mathbf{C}} \wedge \eta_{\mathbf{C}} ) \right].
\end{align}
This action has three kinds of gauge redundancies which are complex analogue of \eqref{eq:GV_gauge_shift}, \eqref{eq:GV_gauge_exact} and \eqref{eq:GV_gauge_lambda}:
\begin{align}
    \label{eq:GV_gauge_shift_cpx}
    &\eta_{\mathbf{C}} \rightarrow \eta_{\mathbf{C}} + u\, \omega_{\mathbf{C}}\ , \quad \lambda_{\mathbf{C}} \rightarrow \lambda_{\mathbf{C}} + 2 u\, \eta_{\mathbf{C}} - 2\, \d u\ , \\
    \label{eq:GV_gauge_exact_cpx}
    &\eta_{\mathbf{C}} \rightarrow \eta_{\mathbf{C}} - \d(\log c)\ , \quad \lambda_{\mathbf{C}} \rightarrow c^{-1}\,\lambda_{\mathbf{C}}\ , \quad \omega_{\mathbf{C}} \rightarrow c\,\omega_{\mathbf{C}} \ , \\
    \label{eq:GV_gauge_lambda_cpx}
    &\lambda_{\mathbf{C}} \rightarrow \lambda_{\mathbf{C}} + v\,\omega_{\mathbf{C}}\ .
\end{align}
where $u$ and $v$ are any complex-valued functions and $c$ is 
a nonzero complex valued function.

\subsubsection*{Step 2. Polar Decomposition}
\begin{figure}[t]
    \centering
            \begin{tikzpicture}[scale=0.6]
        \coordinate (A) at (0 , 0) {};
        \coordinate (B) at (8 , 0) {};
        \fill[red!15] ($(A)+(0, -4)$)--($(A)+(8, -4)$)--($(A)+(8, 4)$)--($(A)+(0, 4)$);
        \fill[red!15] ($(A)!0.5!(B)+(-4, 4)$)--($(A)!0.5!(B)$)--($(A)!0.5!(B)+(-4, -4)$);
         \filldraw[fill=white] ($(A)!0.5!(B)$) circle (0.15) node[below right] {$O$};
         \node at ($(A)!0.5!(B)+(0, -5)$) {$\mathbf{C}\backslash\{0\}$}; 
         
         \node at (10,0) {{\Large $\cong$}};
         
         \coordinate (C) at (12 , 0) {};
        \coordinate (D) at (20 , 0) {};
        \foreach \x in {0,1,2, 3, 4, 5, 6, 7, 8, 9, 10, 11}
                {
                    \draw[thick, blue!80] ($(C)!0.5!(D)$) -- ({16+4*cos(180*\x/6)}, {4*sin(180*\x/6)});
                }
        \node[thick, blue!80, right=0.1cm] at ($(C)!0.5!(D)+(4,0)$) {$\mathbf{R}_{+}$};
        \node[thick, OliveGreen] at ($(C)!0.5!(D)+({2.9*cos(195)},{2.9*sin(195)})$) {U(1)};
        \node at ($(C)!0.5!(D)+(0, -5)$) {$\text{U(1)}\times\mathbf{R}_{+}$};
         \filldraw[fill=white] ($(C)!0.5!(D)$) circle (0.15);
         \fill[white] ($(C)!0.5!(D) +(0.5, -0.5)$) circle (10pt) node[black] {$O$};
         \draw[OliveGreen,thick, arrows = {-Stealth[scale=1.1]}] ($(C)!0.5!(D) +({2*cos(15)}, {2*sin(15)})$) arc(15:345:2);
         \end{tikzpicture}
    \caption{Pictorial understanding of the polar decomposition \eqref{eq:polar_decomposition}. The left panel dictates the region of $\mathbf{C}\backslash\{0\}$ where the origin is cut from the complex plane $\mathbf{C}$. On the other hand, the right panel denotes that the positive real line $\mathbf{R}_{+}$ (colored in blue) is piled up along with the circle U(1) (in green). The polar decomposition \eqref{eq:polar_decomposition} asserts that these two perspectives must be equivalent.}
    \label{fig:picture_of_polar_decomposition}
\end{figure}
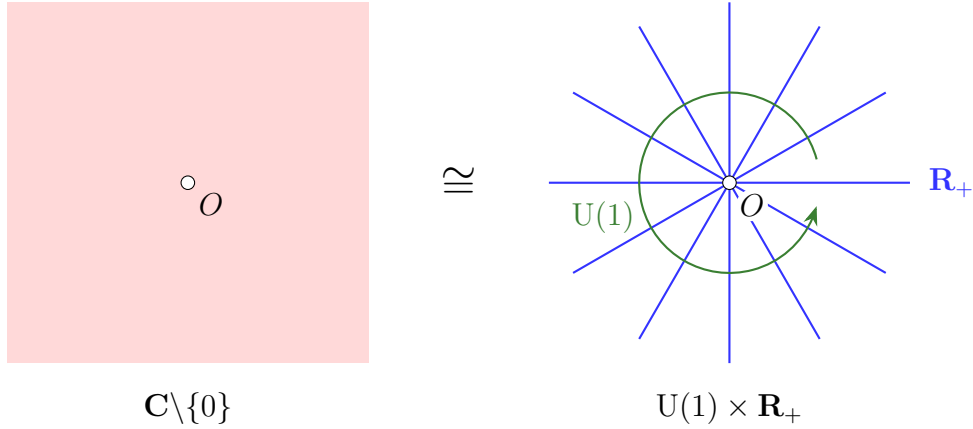
In Step 1, we have extended the gauge group to $\mathbf{C}\backslash\{0\}$ and constructed the action \eqref{eq:GV_action_step1}. However, we have no way to interpret the complex-valued normal 1-form in a geometrical way. To resolve this problem, we should first notice that the group $\mathbf{C}\backslash\{0\}$ has the polar decomposition: (see Fig.\,\ref{fig:picture_of_polar_decomposition})
\begin{align}\label{eq:polar_decomposition}
    \mathbf{C}\backslash\{0\}\cong \text{U}(1)\times \mathbf{R}_{+} \ . 
\end{align}
Keeping this decomposition in mind, it is convenient to restrict ourselves to the following complex-valued field $\omega_{\mathbf{C}}$:
\begin{align}
    \omega_{\mathbf{C}} = e^{\i\phi}\, \omega\ , 
\end{align}
where $\phi$ is the dynamical compact boson field with the periodicity $2\pi$, and $\omega$ is the real-valued 1-form background field, which is identified as the original normal 1-form of the foliation structure.
Accordingly, we decompose the other complex-valued fields into real-valued ones as
\begin{align}
    \eta_{\mathbf{C}} := a+\i b\ , \quad \lambda_{\mathbf{C}} := e^{\i (\delta - \phi)} \lambda\  , 
\end{align}
where $\delta$ is just a real constant.
By plugging these decomposed forms into \eqref{eq:GV_action_step1}, the action becomes
\begin{align}
    \label{eq:GV_action_U(1)}
    \begin{aligned}
    &S[a, b, \lambda, \phi; M^3, \langle\omega\rangle] \\
    &\ = \frac{k}{4\pi}\int_{M^3} \left(a\wedge \d a - b\wedge \d b - (\cos\delta)\ \lambda \wedge (\d\omega - \omega \wedge a) + (\sin\delta)\ \lambda \wedge(\d\phi \wedge \omega - \omega \wedge b)\right).
    \end{aligned}
\end{align}
Now, the 1-form $\lambda$ is no longer an Lagrange multiplier since $\lambda$ has its conjugate momentum made of the dynamical field $\phi$.
The above action possesses the following four kinds of gauge redundancies corresponding to \eqref{eq:GV_gauge_shift_cpx}-\eqref{eq:GV_gauge_lambda_cpx}: 
\begin{align}
 \label{eq:GV_gauge}
\begin{split}
    &a \rightarrow a + (\cos \delta) u\,\omega, \ b \rightarrow b + (\sin \delta) u\,\omega\ , \ \lambda \rightarrow \lambda + 2 u a -2(\tan\delta) u\,b -2(\tan\delta)u\d\phi -2\d u \ ,  \\
    &a \rightarrow a - \d (\log \rho), \quad \lambda \rightarrow \rho^{-1}\, \lambda\ , \quad \omega \rightarrow \rho\, \omega\ , \\
    &b \rightarrow b - \d \theta\ , \quad \phi \rightarrow \phi + \theta\ , \\
    &\lambda \rightarrow \lambda + v\, \omega\ ,
\end{split}
\end{align}
where $u$, $v$ are real spacetime functions. Also, $\rho\in\mathbf{R}_{+}$ and $\theta\in[0, 2\pi)$ are parameters of the $\mathbf{R}_{+}$ and U(1) gauge transformations, respectively. 
Putting the second and third transformations together, the original $\mathbf{C}\backslash\{0\}$ gauge transformation \eqref{eq:GV_gauge_exact_cpx} is restored. 
Note that to maintain the gauge invariance under the U(1) gauge transformation given by the third line of \eqref{eq:GV_gauge},
the level $k$ is restricted to be an integer. 
The equations of motion of this theory are
\begin{align}
\begin{split}
    &2\d a + (\cos \delta) \lambda \wedge \omega = 0\ , \\\
    &2\d b + (\sin \delta) \lambda \wedge \omega = 0\ , \\
    &(\cos \delta) (\d\omega - \omega \wedge a) - (\sin \delta) (\d\phi \wedge \omega - \omega \wedge b) = 0\ , \\
    &(\sin \delta) \d(\lambda \wedge \omega ) = 0\ .
\end{split}
\end{align}
Similar to \eqref{eq:eta_1-form_symm},
the first two equations represent the conservation laws of
subsystem 1-form symmetries:
\begin{align}
    \d a \bigr|_{\omega^\text{T}} = 0\ , \qquad \d b \bigr|_{\omega^\text{T}} = 0\ .
\end{align}
Due to these conservation laws, we have the following gauge invariant topological operators
\begin{align}\label{eq:symop_decompsed}
    \exp \left(\i\beta \int_{\mathcal{C}} a\right), \quad \exp \left(\i\gamma \int_{\mathcal{C'} }b \right)\ , 
\end{align}
where $\mathcal{C}$ and $\mathcal{C}'$ are closed loops embedded in a leaf of the foliation. Indeed, these operators are invariant under smooth deformations
of the loops as long as embedded in the same leaf, just like \eqref{eq:eta_topological_op}.
Moreover, the coefficient $\beta$ can take any real values since $a$ is the $\mathbf{R}_{+}$ gauge field and there is no non-trivial winding structure. On the one hand, $b$ is the U(1) gauge field and hence the coefficient $\gamma$ is quantized to be integer.

We should remark that if we set $\delta=0$ or $\pi$ in the action \eqref{eq:GV_action_U(1)}, then the action is reduced to
\begin{align}
    S = \frac{k}{4\pi} \int_{M^3} 
    \Bigl[ a\wedge \d a \pm \lambda \wedge (\d \omega - \omega \wedge a) \Bigr]
    - \frac{k}{4\pi} \int_{M^3}  b\wedge \d b \ .
\end{align}
While the first term represents the original Godbillon-Vey action \eqref{eq:GV_action}, the last one containing 
$b$ denotes the decoupled U$(1)_{k}$ Chern-Simons theory. 
As we said before, the Godbillon-Vey action \eqref{eq:GV_action} describes just an invertible phase,
which does not exhibit the
subextensive GSD for $\delta=0$ or $\pi$,
hence we do not
consider these cases any further. 

Things get more interesting when
$\delta=\pi/2$ or $3\pi/2$. In such cases, 
the action \eqref{eq:GV_action_U(1)} becomes
\begin{align}
\label{eq:GV_decoupled}
    S = \frac{k}{4\pi} \int_{M^3} a\wedge \d a 
    - \frac{k}{4\pi} \int_{M^3} 
    \Bigl[ b\wedge \d b \pm \lambda \wedge (\d\phi \wedge \omega - \omega \wedge b)  \Bigr] \ .
\end{align}
Unlike the previous cases, the theory is composed of the decoupled $\mathbf{R}_{+}$ Chern-Simons theory and some U(1) gauge theory coupled with the foliation structure. 
In this reduced action, the first transformation of \eqref{eq:GV_gauge}
is ill-defined 
since the factor $\tan\delta$ in the transformation of $\lambda$ becomes singular. 
Instead, however, we have 
the following gauge redundancy: 
\begin{align}
\label{eq:GV_gauge_emergent}
    b \rightarrow b + u\,\omega, \quad \lambda \rightarrow \lambda + 2u\,\xi -2u\,b  -2u\d\phi - 2\d u\ ,
\end{align}
where $\xi$ is a 1-form background gauge field defined by%
\footnote{The existence of the background field $\xi$ is guaranteed by the integrability condition \eqref{eq:omega_integrability}.}
\begin{align}
\label{eq:xi_definition}
    \d\omega = \omega \wedge \xi\ .
\end{align}

\subsubsection*{Step 3. Extraction of U(1) Gauge Theory}
Let us focus on $\delta=\pi/2$ corresponding to take the negative signature in~\eqref{eq:GV_decoupled}.%
\footnote{Note that the case of $\delta=3\pi/2$ is equivalent to the one of $\delta=\pi/2$ up to the redefinition of $\lambda$.} 
In Step 3, we extract
the U(1) gauge theory part from this action. 
This can be easily done since the $\mathbf{R}_{+}$ gauge theory is completely decoupled from the U(1) theory. 
By dropping off the decoupled $\mathbf{R}_{+}$ Chern-Simons action, we can obtain the following U(1) gauge theory: 
\begin{align}\label{eq:GV_action_step3}
    S = \frac{k}{4\pi} \int_{M^3} 
    \Bigl[ b\wedge \d b - \lambda \wedge (\d\phi \wedge \omega - \omega \wedge b) \Bigr] \ .
\end{align}
This theory has the following gauge symmetries:
\begin{align}
\begin{split}
    &b \rightarrow b + u\,\omega\ , \quad \lambda \rightarrow \lambda + 2u\, \xi -2u\,b -2u\d\phi - 2\d u\ , \\
    &b \rightarrow b - \d\theta\ , \quad \phi \rightarrow \phi + \theta\ , \\
    &\lambda \rightarrow \rho^{-1}\lambda\ , \quad \omega \rightarrow \rho\,\omega\ , \\
    &\lambda \rightarrow \lambda + v\,\omega\ ,
\end{split}
\end{align}
where the $\xi$ is the background field defined in \eqref{eq:GV_gauge_emergent}-\eqref{eq:xi_definition}. 
The equations of motion are
\begin{align}
    2\d b + \lambda \wedge \omega = 0\ , \quad
    \d\phi \wedge \omega - \omega \wedge b = 0\ , \quad
    \d(\lambda \wedge \omega) = 0\ .
\end{align}
Following the 
similar discussion around~\eqref{eq:symop_decompsed}, we can construct a gauge invariant subsystem symmetry operator:
\begin{align}
    \exp\left(\i \gamma \int_{\mathcal{C}} b\right)\ , \quad \gamma \in \mathbf{Z}\ , 
\end{align}
where $\mathcal{C}$ is a closed loop embedded in a leaf of the foliation.

Although above compact gauge theory \eqref{eq:GV_action_step3} resolves \textbf{Problem 1}, 
we again face an issue that 
the theory is reduced to an invertible phase.
To see how, 
in the coordinate patch where the normal 1-form $\omega$ is written as $\omega = \omega_1 \d x^1$,
the Lagrangian can be written as
\begin{align}\label{eq:GV_action_coord}
\begin{aligned}
    \mathcal{L} = \frac{k}{4\pi}& 
    \Bigl[ b_0(\partial_1 b_2 - \partial_2 b_1) + b_1 (\partial_2 b_0 - \partial_0 b_2) + b_2(\partial_0 b_1 - \partial_1 b_0) \\
    &-\lambda_0(-\partial_2 \phi \omega_1 - \omega_1 b_2) - \lambda_2 (\partial_0 \phi \omega_1 + \omega_1 b_0) \Bigr] \ .
\end{aligned}
\end{align}
Here, $b_0$ and $\lambda_0$ work as Lagrange multipliers since the Lagrangian do not involve the time derivatives of them. By integrating out them, we obtain the following constraints:
\begin{align}
    2(\partial_1 b_2 - \partial_2 b_1) - \lambda_2 \omega_1 = 0\ , \quad
    \omega_1 b_2 + \partial_2 \phi \omega_1 = 0\ .
\end{align}
By substituting these constraints into \eqref{eq:GV_action_coord}, the Lagrangian becomes trivial: 
\begin{align}
\begin{aligned}
    \mathcal{L} = \frac{k}{4\pi} (-b_1 \partial_0 b_2 + b_2 \partial_0 b_1 - \lambda_2 \partial_0 \phi \omega_1) 
    \sim \frac{k}{2\pi} \partial_0 \phi \partial_1 \partial_2 \phi 
    \sim 0\ .
\end{aligned}
\end{align}
Here, $\sim$ represents the two quantities are equal up to total derivative terms. Thus, this theory describes an invertible phase.

\subsubsection*{Step 4. Generalization to U(1) BF-like Theory}
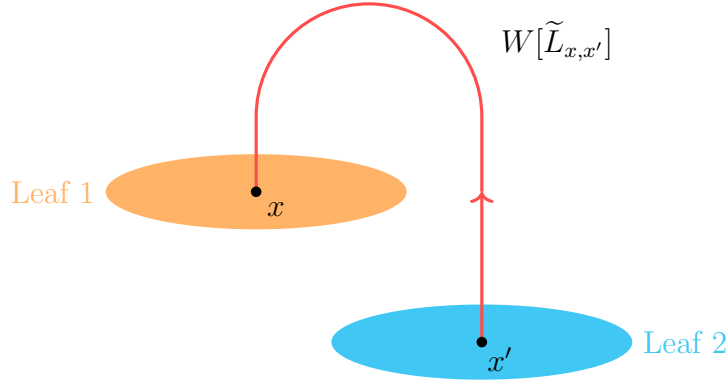
\begin{figure}[t]
    \begin{center}
        \begin{tikzpicture}    
        \fill[orange!60] (0,5) circle(2 and 0.5) node[left=2cm] {Leaf 1};
        \fill[cyan!60] (3,3) circle(2 and 0.5) node[right=2cm] {Leaf 2};
        \draw[very thick, red!70] (0, 5) -- (0, 6) arc(180:0:1.5) --(3, 5);
        \draw[very thick, red!70, <-](3, 5)--(3, 3);
        \fill (0, 5) circle(2pt) node[below right] {$x$};
        \fill (3, 3) circle(2pt) node[below right=-0.1cm] {$x'$};
        \node at (4, 7) {$W[\widetilde{L}_{x, x'}]$}; 
    \end{tikzpicture}
        \caption{Non-trivial cofiguration of the loop operator $W[\widetilde{L}_{x, x'}]$. In this picture, the two endpoints $x$ and $x'$ of the loop operator reside in different leaves whose local patches are depicted by colored ellipses. }
        \label{fig:non_trivial_config_line_W}
    \end{center}
\end{figure}

\begin{figure}[t]
    \begin{center}
        \begin{tikzpicture}  
        \shadedraw[bottom color=green!42, top color=green!5, line width=1pt] (0.5, 6.5)arc(90:0:0.5) -- (1, 3)--(5, 3)--(5, 6) arc(0:90:0.5);
        \shadedraw[bottom color=green!42, top color=green!5, line width=1pt] (0, 5)-- (0, 6) arc(180:0:0.5)--(1, 5)--cycle;
        \draw[very thick, OliveGreen] (0, 5)--(1, 5);
        \draw[very thick, dotted, OliveGreen] (1, 5) --(4, 5);
        \draw[very thick, OliveGreen] (5, 3)--(1, 3);
        \draw[very thick, red!70] (0, 5) -- (0, 6) arc(180:0:0.5) --(1, 4.5);
        \draw[very thick, red!70, <-](1, 4.5)--(1, 3);
        \node at (-1, 6) {$W[\widetilde{L}_{x, x'}]$};

        \draw[very thick, dotted, red!70] (4, 5) -- (4, 6) arc(180:90:0.5);
        \draw[very thick, red!70, ->] (4.5, 6.5) arc(90:0:0.5) -- (5, 4.5);
        \draw[very thick, red!70](5, 4.5)--(5, 3);
        \node at (6, 6) {$W[\widetilde{L}_{y', y}]$};
        \fill (0, 5) circle(2pt) node[below left=0.05cm] {$x$};
        \fill (1, 3) circle(2pt) node[below left=0.025cm] {$x'$};
        \fill (4, 5) circle(2pt) node[below right=0.1cm] {$y$};
        \fill (5, 3) circle(2pt) node[below right=0.025cm] {$y'$};
        \fill[opacity=0.3, orange] (2,5) circle(2.5 and 0.5);
        \node[black] at (2.5, 6) {$S_{x, x', y', y}$} ;
        \fill[opacity=0.3, cyan] (3,3) circle(2.5 and 0.5);
    \end{tikzpicture}
        \caption{Picture of the surface $S_{x, x', y', y}$ (colored in green) which connects two lines $\widetilde{L}_{x, x'}$ and $\widetilde{L}_{y, y'}$. The two red lines mean the topological operators $W[\widetilde{L}_{x, x'}]$ and $W[\widetilde{L}_{y', y}]$. In this figure, while two points $x$ and $y$ belong to the same leaf, the other two points $x'$ and $y'$ do to the other same leaf.}
        \label{fig:topological_condition}
    \end{center}
\end{figure}
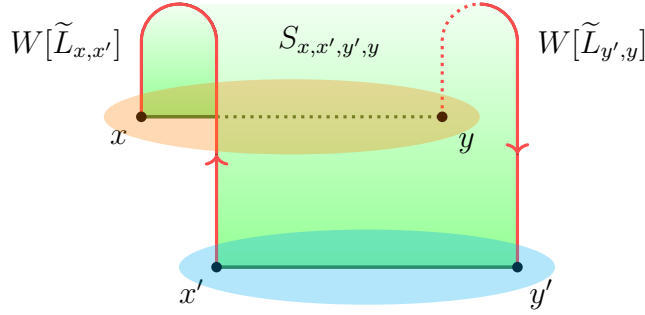

The theory obtained after Step 3 is again just an invertible phase. To make our foliation field theory non-trivial, we add slight modifications to the invertible theory~\eqref{eq:GV_action_step3} by generalizing the Chern-Simons-like term to the BF-like one:
\begin{align}
     S = \frac{k}{2\pi} \int_{M^3} 
     \Bigl[ b\wedge \d c - \lambda \wedge (\d\phi \wedge \omega - \omega \wedge b) \Bigr] \ ,
     \quad k\in \mathbf{Z}\ , 
     \label{eq:GV_action_step4}
\end{align}
where $c$ is a newly introduced U(1) gauge field. 
The gauge symmetries in this theory are given by
\begin{align}
    \label{eq:GVBF_gauge_1}
    &c \rightarrow c + u\,\omega\ , \quad \lambda \rightarrow \lambda + u\xi - \d u\ , \\
    &b \rightarrow b-\d\theta\ , \quad \phi \rightarrow \phi + \theta\ , \\
    &c \rightarrow c + \d\tilde{\theta}\ , \label{eq:GVBF_gauge_3} \\
    &\lambda \rightarrow \rho^{-1}\lambda\ , \quad \omega \rightarrow \rho\, \omega\ , \\
    &\lambda \rightarrow \lambda + v\,\omega\ .  \label{eq:GVBF_gauge_5}
\end{align}
The equations of motion read
\begin{align}
    &\d c + \lambda \wedge \omega = 0\ , \\
    &\d b = 0\ , \label{eq:GVBF_eom2_step1} \\
    &\d\phi \wedge \omega - \omega \wedge b = 0\ , \label{eq:GVBF_eom3_step4} \\
    &\d(\lambda \wedge \omega) = 0\ . 
\end{align}
Also, we have the following gauge invariant operators:
\begin{align}
\label{eq:symmetry_defect_subsystem}
    U_{\beta}[\mathcal{C}]:=\exp \left(\i\beta \int_\mathcal{C} c\right)\ ,
\end{align}
and%
\footnote{
The coefficient $\gamma'$ 
needs to be integer to keep the gauge invariance: $\phi \rightarrow \phi + 2\pi f(x)$ with $f(x) \in \mathbf{Z}$.
}
\begin{align}
    \label{eq:symmetry_defect_line}
    V_{\gamma}[\widetilde{\mathcal{C}}]:=\exp\left(\i\gamma \int_{\widetilde{\mathcal{C}}} b\right)\ , \quad W_{\gamma'}[\widetilde{L}_{x, x'}]:=\exp\left(\i\gamma' \phi(x)\right) \exp\left(\i\gamma' \int_{\widetilde{L}_{x,x'}} b\right) \exp\left(-\i\gamma' \phi(x')\right)\ , 
\end{align}
where $\mathcal{C}$ is an embedded loop in a leaf. 
Also, $\widetilde{\mathcal{C}}$ is an loop, and $\widetilde{L}_{x,x'}$ is a line connected from a point $x'$ to $x$, both of which are not necessarily restricted on a single leaf. 
The last line operator $W[\widetilde{L}_{x, x'}]$ becomes trivial when the line $\widetilde{L}_{x, x'}$ resides on a leaf due to the equation of motion \eqref{eq:GVBF_eom3_step4}. 
See Fig.\,\ref{fig:non_trivial_config_line_W} for a schematic picture of a non-trivial configuration of the line operator $W[\widetilde{L}_{x, x'}]$. 
Also, $W[\widetilde{L}_{x, x'}]$ is invariant under the smooth deformation such that each endpoint moves within
leaf. By using \eqref{eq:GVBF_eom2_step1}-\eqref{eq:GVBF_eom3_step4}, this can be verified as
\begin{align}
    W_{\gamma'}[\widetilde{L}_{x, x'}]\,W_{\gamma'}[\widetilde{L}_{y, y'}]^{-1} = W_{\gamma'}[\widetilde{L}_{x, x'}]\,W_{\gamma'}[\widetilde{L}_{y', y}] = \exp\left(\i \gamma' \int_{S_{x, x', y', y}} \d b\right)=1\ ,  
\end{align}
where $x$ and $y$ are located in the same leaf, and $x'$ and $y'$ are in the other same leaf. Also, $S_{x, x', y', y}$ is the surface connecting two lines $\widetilde{L}_{x, x'}$ and $\widetilde{L}_{y, y'}$ as depicted in Fig.\,\ref{fig:topological_condition}. 


Note that although the gauge field $c$ is not directly coupled with the normal 1-form $\omega$, the field $c$ possesses the desirable shift transformation \eqref{eq:GVBF_gauge_1}, which was originally a redundancy~\eqref{eq:GV_gauge_shift} of the Godbillon-Vey invariant. In this sense, the information of the foliation structure is inherited by $c$ rather than $b$. 

In what follows, we explain that this theory describes non-trivial ground state properties in the sense that the theory is not merely an invertible phase. In a coordinate patch where the normal 1-form $\omega$ is written as $\omega = \omega_1 \d x^1$, the Lagrangian of \eqref{eq:GV_action_step4} is written as
\begin{align}
\label{eq:Lagrangian_w1}
\begin{aligned}
    \mathcal{L} &= \frac{k}{2\pi} 
    \Bigl[ b_0(\partial_1 c_2 - \partial_2 c_1) + b_1(\partial_2 c_0 - \partial_0 c_2) + b_2 (\partial_0 c_1 - \partial_1 c_0) \\
    &\quad\quad\quad - \lambda_0 (-\partial_2 \phi \omega_1 - \omega_1 b_2) - \lambda_2 (\partial_0 \phi \omega_1 + \omega_1 b_0) \Bigr] \ .
\end{aligned}
\end{align}
Constraints from Lagrange multipliers $b_0$, $c_0$ and $\lambda_0$ are
\begin{align}
    &\partial_1 c_2 - \partial_2 c_1 - \lambda_2 \omega_1 = 0\ , \label{eq:constraints_BFlike_1} \\
    &\partial_1 b_2 - \partial_2 b_1 = 0\ , \label{eq:constraints_BFlike_2} \\
    &\partial_2\phi \omega_1 + \omega_1 b_2 = 0\ . \label{eq:constraints_BFlike_3} 
\end{align}
Integrating out these auxiliary fields and using the constraints \eqref{eq:constraints_BFlike_1} and \eqref{eq:constraints_BFlike_3}, the resulting theory is given by
\begin{align}\label{eq:lagrangian_step4}
    \mathcal{L} &= \frac{k}{2\pi} (-b_1 \partial_0 c_2 + b_2 \partial_0 c_1 - \lambda_2 \partial_0 \phi \omega_1)\sim \frac{k}{2\pi} (c_2 (\partial_0 b_1 + \partial_0 \partial_1 \phi)) \ , 
\end{align}
where $\sim$ represents the two quantities are equal up to total derivative terms.
The other constraint \eqref{eq:constraints_BFlike_2} together with \eqref{eq:constraints_BFlike_3} implies 
\begin{align}
    \partial_2(\partial_1  \phi +  b_1) = 0\ \Longrightarrow \ \partial_1 \phi + b_1 = f(x^0,x^1)\ , 
\end{align}
where $f(x^0,x^1)$ is some function which depends only on the coordinates $x^0$ and $x^1$.
By substituting this expression into \eqref{eq:lagrangian_step4}, we finally arrive at
\begin{align}
    \mathcal{L} \sim \frac{k}{2\pi}
  c_2 \partial_0 f(x^0,x^1) \ .
\end{align}
As opposed to \eqref{eq:GV_action_step3}, this Lagrangian admits a nontrivial canonical structure where $c_{2}$ and $f$ serve as a role of the canonical quantities. 
From this final expression, we can expect that the BF-like theory \eqref{eq:GV_action_step4} can produce the subextensive GSD, which will be investigated in the next subsection and appendix~\ref{sec:canonical_quantization}.

\subsubsection*{Generalization to higher dimensions}
We can generalize the $(2+1)$-dimensional U(1) BF-like theory \eqref{eq:GV_action_step4} to the higher dimensional one:
\begin{align}
\label{eq:generalized_BF_theory}
    S = \frac{k}{2\pi} \int_{M^{d+1}} 
    \Bigl[ b^{(d-p)}\wedge \d c^{(p)} - \lambda^{(p)} \wedge (\d\zeta^{(d-p-1)} \wedge \omega - \omega \wedge b^{(d-p)}) \Bigr] \ ,
    \quad k\in\mathbf{Z}\ .
\end{align}
Here, $b^{(d-p)}$ and $c^{(p)}$ are U(1) higher form gauge fields, $\lambda^{(p)}$ is a $p$-form field, and $\zeta^{(d-p-1)}$ is a $(d-p-1)$-form U(1) gauge field. This theory also possesses similar gauge redundancies to the three-dimensional case (see from \eqref{eq:GVBF_gauge_1} to \eqref{eq:GVBF_gauge_5}.) in addition to 
\begin{align}
    \label{eq:zeta_gauge}
    \zeta^{(d-p-1)}\ \rightarrow\ \zeta^{(d-p-1)} + \d \chi^{(d-p-2)}\ , 
\end{align}
where $\chi^{(d-p-2)}$ is a $(d-p-2)$-form U(1) gauge field. Likewise in the three dimensions, we find that the following gauge invariant operators:
\begin{align}
    \exp\left(\i\gamma\int_{\mathcal{U}^p} c^{(p)}\right),
\end{align}
and\footnote{
The parameter $\gamma'$ needs to be integer just like the case of \eqref{eq:symmetry_defect_line} because of the gauge symmetry \eqref{eq:zeta_gauge} of $\zeta^{(d-p-1)}$.
}
\begin{align}
    \exp\left(\i\gamma \int_{\widetilde{\mathcal{U}}^q} b\right), \quad \exp\left(\i\gamma'\int_{\widetilde{\mathcal{V}}_{1}^{q-1}} \zeta^{(q-1)}\right) \exp\left(\i\gamma'\int_{\mathcal{\widetilde{U}}^{q}} b^{(q)}\right) \exp\left(-\i\gamma'\int_{\widetilde{\mathcal{V}}_{2}^{q-1}} \zeta^{(q-1)} \right)\ ,
\end{align}
where $\mathcal{U}^{(p)}$ is an embedded closed $p$-dimensional manifold in a leaf. Also, $\widetilde{\mathcal{U}}^{(p)}$ is a closed $p$-dimensional manifold, and $\widetilde{\mathcal{U}}^{(q)}$ is a $q$-dimensional manifold connected from its orientation reversed boundary $-\widetilde{\mathcal{V}}_{2}^{q-1}$ to the other one $\widetilde{\mathcal{V}}_{1}^{q-1}$, both of which are not necessarily restricted on a single leaf. These gauge invariant operators correspond to the $(d-p)$-form subsystem symmetry.%
\footnote{
Note that we denote the spacetime dimensions as $d+1$. Therefore, the ($d-p$)-form symmetry is associated to the $p$-dimensional symmetry operator.
}

\subsection{Subextensive Ground State Degeneracy}\label{subsec:subGSD}
Here, we argue the GSD of the U(1) BF-like theory \eqref{eq:GV_action_step4} based on the canonical quantization.
Since our theory is a constrained system which contains various constraints including gauge redundancies, we must be careful of the canonical quantization. 
We relegate the rigorous discussions on the quantization to appendix~\ref{sec:canonical_quantization} and just digest our result of the GSD in the main text.
When we take the coordinate patch where the normal 1-form $\omega$ is written as $\omega = \omega_1 \d x^1$, the most important equal time commutation relation is
\begin{align}\label{eq:com_relation}
   \Bigl[ (b_1 + \partial_1 \phi)(x^1,x^2)\ ,\ c_2(y^1,y^2) \Bigr]    
   = \frac{2\pi\i}{k}\delta(x^1-y^1)\delta(x^2-y^2)\ .
\end{align}

As the Hamiltonian of the theory become zero, the GSD is calculated by the algebra of the gauge invariant operators.
As discussed in appendix \ref{sec:canonical_quantization}, $c_2$ and $b_1 + \partial_1 \phi$ are canonical momenta of each other.
This leads to the following commutation relation of gauge invariant operators:
\begin{align}
    U_{\beta}[\mathcal{C}] W_{\gamma'}[\widetilde{L}_{x, x'}]=e^{\frac{2\pi\beta\gamma'}{k}\,n(C, \widetilde{L}_{x, x'})}\,W_{\gamma'}[\widetilde{L}_{x, x'}]U_{\beta}[\mathcal{C}]\ ,
\end{align}
where $n(\mathcal{C}, \widetilde{L}_{x,x'})$ represents the intersection number between the loop $\mathcal{C}$ and the line $\widetilde{L}_{x,x'}$. 
As a result of the spontaneous symmetry breaking of the subsystem 1-form symmetry  generated by \eqref{eq:symmetry_defect_subsystem}, the ground state Hilbert space should be characterized by the representation of this algebra. After a similar discussion to the derivation of the GSD of the Chern-Simons theory,%
\footnote{See for example lecture notes \cite{Tong:2016kpv,Tong_gauge}.} we arrive at the following subextensive GSD 
\begin{align}
    \mathrm{GSD} = k^{\beta_1^\omega L_\omega} (=\infty)\ ,
\end{align}
where $\beta_1^\omega$ is the first Betti number of the leaf manifold, and $L_\omega$ is the lattice size along the normal 1-form $\omega$.

Likewise, we can also calculate the GSD of the U(1) BF-like theory in higher dimensions~\eqref{eq:generalized_BF_theory}:
\begin{align}
    \mathrm{GSD} = k^{\beta_p^\omega L_\omega} (=\infty)\ .
\end{align}
where $\beta_p^\omega$ is the $p$-th Betti number of the leaf manifold.

\section{Lattice Spin Models in $2+1$ dimensions}
\label{sec:lattice_spin_models}
So far, we have established the U(1) BF-like theory \eqref{eq:GV_action_step4} with the subextensive GSD based on the Godbillon-Vey invariants. 
In this section, we explore lattice models in $(2+1)$ dimensions which are expected as the UV completions of the BF-like theory and its variants.
In the first two subsections, we discuss lattice models which host a single foliation structure, and check that their low energy behaviors are indeed the same as the ones of the BF-like theory.
After discussing these models, we present other lattice models with two independent codimension one foliations. 

\subsection{Single Codimension One Foliation}
We start with the case where a lattice hosts single codimension one foliation.

\subsubsection{Stack of ($1+1$)-dimensional Ising Models}
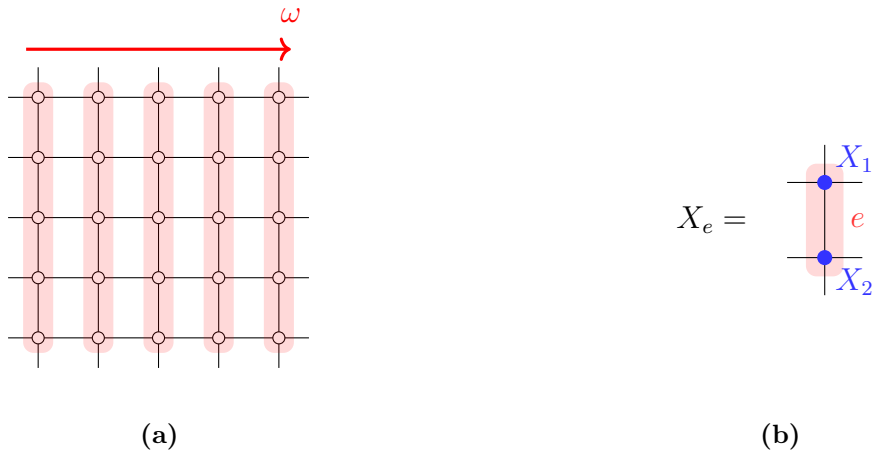
\begin{figure}[t]
    \begin{center}
        \begin{minipage}[b]{0.5\linewidth}
        \begin{center}
        \begin{tikzpicture}[scale=0.8]
            \draw[->,very thick, red](-2.2,2.8)--(2.2,2.8) node [above=0.2cm] {$\omega$};
            \draw[step=1cm](-2.5,-2.5) grid (2.5,2.5);
            \foreach \x in {-2,-1,0,1,2}
            {
                \foreach \y in {-2,-1,0,1,2}
                {
                    \filldraw[fill=white](\x,\y)circle(0.1);
                }
                \fill[opacity=0.15,red, rounded corners] (\x-0.25,-2.25)--(\x-0.25,2.25)--(\x+0.25,2.25)--(\x+0.25,-2.25)--cycle;
            }
        \end{tikzpicture}
        \end{center}
        \subcaption{}
        \label{fig:foliated_lattice_spins}
    \end{minipage}%
        \begin{minipage}[b]{0.5\linewidth}
        \begin{center}
                \begin{tikzpicture}
                    \fill[opacity=0.15,red,rounded corners](-0.25,-0.25)--(-0.25,1.25)--(0.25,1.25)--(0.25,-0.25)--cycle;
                    \node [] at (-1.5,0.5) {$X_e=$};
                    \draw[step=1cm] (-0.5,-0.5) grid (0.5, 1.5);
                    \fill[fill=blue!80](0,0)circle(0.1);
                    \fill[fill=blue!80](0,1)circle(0.1);
                    \node [below right,blue!80] at (0,0) {$X_2$};
                    \node [above right,blue!80] at (0,1) {$X_1$};
                    \node [right, red!80] at (0.2,0.5) {$e$};
                \end{tikzpicture}
                \vspace{1.5\baselineskip}
        \end{center}
        \subcaption{}
        \label{fig:X_e^omega^T}
        \end{minipage}
        \caption{\textbf{(a)} Lattice spins hosting a single foliation structure in the model \eqref{eq:2d_ising_stack_lattice}. The physical degrees of freedom are qubits denoted by circles ($\circ$). The normal 1-form $\omega$ points in a horizontal direction as depicted by the red arrow. The red rectangle regions which are expanded in a vertival direction mean the leaves corresponding to $\omega$. \textbf{(b)} The term $X_{e}$ constituting the Ising stacked Hamiltonian~\eqref{eq:2d_ising_stack_lattice}.
        This is expressed as the tensor product $X_{e} = X_1 X_2$, where $X_1$ and $X_2$ are the $X$ operators placed at the two endpoints of the edge $e$.}
    \end{center}
\end{figure}
The field theory \eqref{eq:GV_action_step4} can be considered as a low energy effective theory of a ($2+1$)-dimensional lattice model which consists of the stacking of ($1+1$)-dimensional Ising models.
This lattice spin model is defined on a two-dimensional spatial lattice which hosts a foliation structure (see Fig.\,\,\ref{fig:foliated_lattice_spins}).
The physical degrees of freedom are qubits; one qubit is placed on each vertex.
We assume that the spatial manifold and each leaf are all closed.
A typical example is a stacking of $S^1$ leaves in the two-dimensional torus $T^2$.
On such a lattice, the Hamiltonian of the stacked Ising models is given by the sum of local terms assigned to edges transverse to $\omega$:
\begin{align}
\label{eq:2d_ising_stack_lattice}
 H = - \sum_{e\,\in\, E_{\omega}} X_e\ ,
\end{align}
where
$E_{\omega}$ is the set of all the edges transverse to $\omega$. 
The local operator $X_e$ is the tensor product of $X$ operators (Pauli matrix $\sigma_X$) over the ends of the edge $e$, which is the same as for the ($1+1$)-dimensional Ising model living in leaves of the foliation (see Fig.\,\ref{fig:X_e^omega^T}).

The model~\eqref{eq:2d_ising_stack_lattice} is well-known  -- after all, the model describes stack of spontaneous symmetry breaking phases of the Ising model. 
Yet, for later convenience, we discuss the model~\eqref{eq:2d_ising_stack_lattice} in the language of quantum information. 
Indeed, the model~\eqref{eq:2d_ising_stack_lattice} is regarded as a stabilizer code model
since all of the local terms in the Hamiltonian mutually commute.%
\footnote{See e.g., \cite{Gottesman:1997zz} for a review on stabilizer codes.}
Also, the ground state of the Hamiltonian~\eqref{eq:2d_ising_stack_lattice} is a stabilized state with all the eigenvalues of $X_e$ being one.
It is immediate to
compute the GSD by counting the dimensions of full Hilbert space and the number of independent conditions, which ground states must obey.
The result is
\begin{align}
    \mathrm{GSD} = 2^{\beta_1^{\omega} L_{\omega}} \ ,
\end{align}
where $\beta_1^{\omega}$ is the first Betti number of a leaf
transverse to $\omega$, and $L_{\omega}$ is the lattice size along $\omega$-direction.
This GSD can be also obtained by taking the product of that of $(1+1)$-dimensional Ising model along $\omega$-direction.

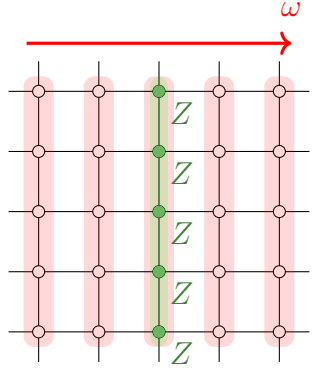
\begin{figure}[t]
    \begin{center}
        \begin{tikzpicture}[scale=0.8]
            \fill[opacity=0.15, red, rounded corners] (0-0.25,-2.25)--(0-0.25,2.25)--(0+0.25,2.25)--(0+0.25,-2.25)--cycle;
            \draw[->,very thick,red](-2.2,2.8)--(2.2,2.8) node [above=0.2cm] {$\omega$};
            \draw[step=1cm](-2.5,-2.5) grid (2.5,2.5);
            \foreach \y in {-2,-1,0,1,2}
            {
                \foreach \x in {-2,-1,0,1,2}
                {
                    \filldraw[fill=white](\x,\y)circle(0.1);
                }
                \fill[fill=OliveGreen!60](0,\y)circle(0.1);
                \node[below right, OliveGreen] at (0,\y) {$Z$};
            } 
            \fill[opacity=0.15, red, rounded corners] (-2-0.25,-2.25)--(-2-0.25,2.25)--(-2+0.25,2.25)--(-2+0.25,-2.25)--cycle;
            \fill[opacity=0.15, red, rounded corners] (-1-0.25,-2.25)--(-1-0.25,2.25)--(-1+0.25,2.25)--(-1+0.25,-2.25)--cycle;
            \fill[opacity=0.15, red, rounded corners] (1-0.25,-2.25)--(1-0.25,2.25)--(1+0.25,2.25)--(1+0.25,-2.25)--cycle;
            \fill[opacity=0.15, red, rounded corners] (2-0.25,-2.25)--(2-0.25,2.25)--(2+0.25,2.25)--(2+0.25,-2.25)--cycle;
            
            \fill[opacity=0.15, green, rounded corners] (0-0.15,-2.25)--(0-0.15,2.25)--(0+0.15,2.25)--(0+0.15,-2.25)--cycle;
        \end{tikzpicture}
        \caption{The logical $Z$ rigid loop operator in the Ising stack model \eqref{eq:2d_ising_stack_lattice} marked by green dots. It is embedded in a leaf of the foliation structure defined by the normal 1-form $\omega$. }
        \label{fig:ising_stacking_logical_operator}
    \end{center}
\end{figure}

The logical operators, which contribute to non-trivial GSD, are easy to identify.%
\footnote{To be more precise, the logical operators are operators that commute with the Hamiltonian which cannot be generated by the terms in the stabilizer model. }
They are embedding of ones of $(1+1)$-dimensional Ising models.
While local $X$ operators become logical operators because they commute with the Hamiltonian, the logical $Z$ operators are loop operators embedded in each leaf (Fig.\,\ref{fig:ising_stacking_logical_operator}).
In our lattice model \eqref{eq:2d_ising_stack_lattice}, the fractional excitations at the endpoints of $Z$-string operators extending transverse to $\omega$ can be interpreted as subdimensional particles.
Such energy excitations can move only along the $\omega$-transverse direction, indicating that they are lineon excitations.

\subsubsection{Twisted Plaquette Ising Model}
 \begin{figure}[t]
     \begin{center}
        \begin{minipage}[b]{0.5\linewidth}
        \begin{center}
           \begin{tikzpicture}[scale=0.8]
              \draw[->,very thick,red](-2.2,3.8)--(2.2,3.8) node [above=0.2cm] {$\omega$};
              \foreach \x in {-2,-1,0,1,2}
              {
                  \draw (\x,-2.3)--(\x, 3.3);
              }
             \foreach \y in {-1,0,1,2}
             {
                 \draw (-2.5,\y-0.5)--(2.5,\y+0.5);
             }
             \draw (-2.5,2.5)--(0.5,2.5+3/5);
             \draw (-0.5,-1.5-3/5)--(2.5,-1.5);
             \foreach \x in {-2,-1,0}
             {
                 \filldraw[fill=white](\x,3+\x*0.2)circle(0.1);
             }
             \foreach \x in {0,1,2}
             {
                 \filldraw[fill=white](\x,-2+\x*0.2)circle(0.1);
             }
             \foreach \x in {-2,-1,0,1,2}
             {
                 \foreach \y in {-1,0,1,2}
                 {
                     \filldraw[fill=white](\x,\y+\x*0.2)circle(0.1);
                 }
                 \fill[opacity=0.15, red, rounded corners] (\x-0.25,-2.25)--(\x-0.25,3.25)--(\x+0.25,3.25)--(\x+0.25,-2.25)--cycle;
             }
             \foreach \y in {-1.5,-0.5,1.5,2.5}
             {
                 \draw[very thick,dotted] (-3,\y)--(-2.7,\y);
                 \draw[very thick,dotted](3,\y)--(2.7,\y);
             }
             \draw[very thick,red,dotted] (-3,0.5)--(-2.7,0.5);
             \draw[very thick,red,dotted] (3,0.5)--(2.7,0.5);
        \end{tikzpicture}
        \end{center}
        \subcaption{}
        \label{fig:twisted_foliated_lattice_spins}
        \end{minipage}%
        \begin{minipage}[b]{0.5\linewidth}
        \begin{center}
            \begin{tikzpicture}[scale=0.9]
            \node [] at (-1.5,0.2) {$X_p=$};
            \draw (-0.5,-0.5)--(1.5,-0.1);
            \draw (-0.5,0.5)--(1.5,0.9);
            \draw (0,-0.9)--(0,1.1);
            \draw (1,-0.7)--(1,1.3);
            \fill[fill=blue!80](0,0.6)circle(0.1);
            \node [above left, color=blue!80] at (0,0.6) {$X_1$};
            \fill[fill=blue!80](0,-0.4)circle(0.1);
            \node [below left, color=blue!80] at (0,-0.4) {$X_2$};
            \fill[fill=blue!80](1,0.8)circle(0.1);
            \node [above right, color=blue!80] at (1,0.8) {$X_4$};
            \fill[fill=blue!80](1,-0.2)circle(0.1);
            \node [below right, color=blue!80] at (1,-0.2) {$X_3$};
            \fill [opacity=0.155, red, rounded corners] (-0.25,-0.4-0.25*0.2-0.25)--(-0.25,0.6-0.25*0.2+0.25)--(1.25,0.8+0.25*0.2+0.25)--(1.25,-0.2+0.25*0.2-0.25)--cycle;
            \fill [white, rounded corners] (0.25,-0.4+0.25*0.2+0.25)--(0.25,0.6+0.25*0.2-0.25)--(0.75,0.8-0.25*0.2-0.25)--(0.75,-0.2-0.25*0.2+0.25)--cycle;
            \node [red] at (0.5, 0.2) {$p$};
        \end{tikzpicture}
        \vspace{2.5\baselineskip}
        \subcaption{}
        \label{fig:twisted_plaquette_ising_local_operator}
        \end{center}
        \end{minipage}
        \caption{\textbf{(a)} Twisted lattice spins hosting a single foliation structure in the model \eqref{eq:twisted_Ising_plaquette}. While we employ the periodic boundary condition in the vertical direction, we take the twisted boundary condition in the horizontal one. Red dotted lines in this figure are identified with each other (This identification holds for other black dotted lines in the same way.). \textbf{(b)} The term $X_{p}$ given in the twisted plaquette Ising model \eqref{eq:twisted_Ising_plaquette} defined by $X_p = X_1 X_2 X_3 X_4$, where $X_i$'s are $X$ operators acting on the corner of the plaquette $p$.}
    \end{center}
\end{figure}
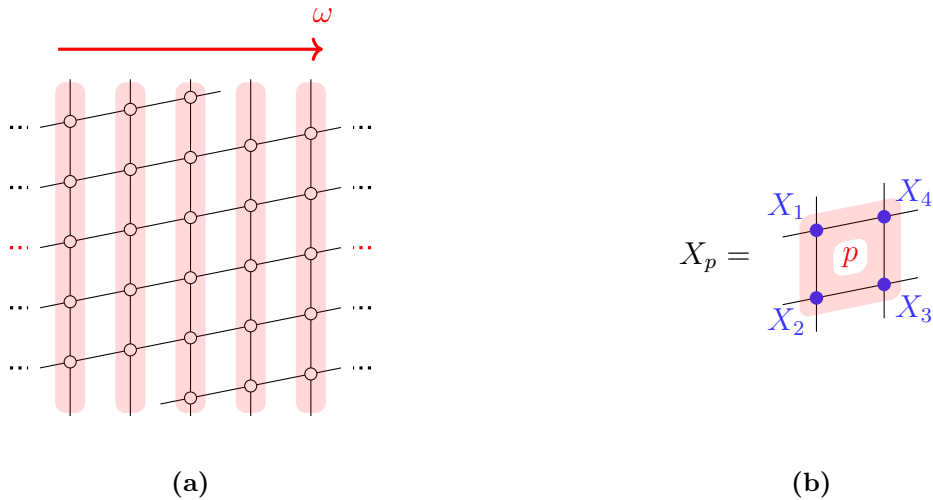
\begin{figure}[t]
    \begin{center}
        \begin{tabular}{cc}
        \begin{minipage}{0.4\linewidth}
        \centering
            \begin{tikzpicture}[scale=0.8]
                \draw[->,very thick,red](-2.2,3.8)--(2.2,3.8) node [above=0.2cm] {$\omega$};
                \foreach \x in {-2,-1,0,1,2}
                {
                    \draw (\x,-2.3)--(\x, 3.3);
                }
                \foreach \y in {-1,0,1,2}
                {
                    \draw (-2.5,\y-0.5)--(2.5,\y+0.5);
                }
                \draw (-2.5,2.5)--(0.5,2.5+3/5);
                \draw (-0.5,-1.5-3/5)--(2.5,-1.5);
                \foreach \x in {-2,-1,0}
                {
                    \filldraw[fill=white](\x,3+\x*0.2)circle(0.1);                    
                }
                \foreach \x in {0,1,2}
                {
                    \filldraw[fill=white](\x,-2+\x*0.2)circle(0.1);
                }
                \foreach \y in {-1,0,1,2}
                {
                    \foreach \x in {-2,-1,0,1,2}
                    {
                        \filldraw[fill=white](\x,\y+\x*0.2)circle(0.1);
                    }
                }
                \foreach \x in {-2,-1,0,1,2}
                {
                    \fill[opacity=0.15, red, rounded corners] (\x-0.25,-2.25)--(\x-0.25,3.25)--(\x+0.25,3.25)--(\x+0.25,-2.25)--cycle;
                }
                \foreach \y in {-1.5,-0.5,1.5,2.5}
                {
                    \draw[very thick,dashed] (-3,\y)--(-2.7,\y);
                    \draw[very thick,dashed](3,\y)--(2.7,\y);
                }
                \draw[very thick,red,dashed] (-3,0.5)--(-2.7,0.5);
                \draw[very thick,red,dashed] (3,0.5)--(2.7,0.5);
                \foreach \y in {-1,0,1,2}
                {
                \fill[OliveGreen!60](0, \y)circle(0.1) node [below right, OliveGreen] {$Z$};
                }
                \fill[OliveGreen!60](0,3)circle(0.1) node [below right, OliveGreen] {$Z$};
                \fill[OliveGreen!60](0,-2)circle(0.1) node [below right, OliveGreen] {$Z$};
                \fill[opacity=0.15, green, rounded corners] (0-0.15,-2.25)--(0-0.15,3.25)--(0+0.15,3.25)--(0+0.15,-2.25)--cycle;
            \end{tikzpicture}
            \subcaption{}
            \label{fig:twisted_plaquette_ising_logical_loop}
        \end{minipage}
        \hspace{1cm}
        &
        \begin{minipage}{0.4\linewidth}
        \centering
            \begin{tikzpicture}[scale=0.8]
                \draw[->,very thick,red](-2.2,3.8)--(2.2,3.8) node [above=0.2cm] {$\omega$};
                \foreach \x in {-2,-1,0,1,2}
                {
                    \draw (\x,-2.3)--(\x, 3.3);
                }
                \foreach \y in {,-1,0,1,2}
                {
                    \draw (-2.5,\y-0.5)--(2.5,\y+0.5);
                }
                \draw (-2.5,2.5)--(0.5,2.5+3/5);
                \draw (-0.5,-1.5-3/5)--(2.5,-1.5);
                \foreach \x in {-2,-1,0}
                {
                    \filldraw[fill=white](\x,3+\x*0.2)circle(0.1);
                    \fill[OliveGreen!60](\x,3+\x*0.2)circle(0.1)node[below right, OliveGreen] {$Z$};
                    
                }
                \foreach \x in {0,1,2}
                {
                    \filldraw[fill=white](\x,-2+\x*0.2)circle(0.1);
                    \fill[OliveGreen!60](\x,-2+\x*0.2)circle(0.1)node[below right, OliveGreen]{$Z$};
                }
                \foreach \y in {-1,0,1,2}
                {
                    \foreach \x in {-2,-1,0,1,2}
                    {
                        \filldraw[fill=white](\x,\y+\x*0.2)circle(0.1);
                        \fill[OliveGreen!60](\x,\y+\x*0.2)circle(0.1)node[below right, OliveGreen] {$Z$};
                    }
                    \filldraw[fill=white](0, \y)circle(0.1);
                    \fill[OliveGreen!60](0, \y)circle(0.1) node [below right, OliveGreen] {$Z$};
                }
                \foreach \y in {-1.5,-0.5,1.5,2.5}
                {
                    \draw[very thick,dashed] (-3,\y)--(-2.7,\y);
                    \draw[very thick,dashed](3,\y)--(2.7,\y);
                }
                \draw[very thick,red,dashed] (-3,0.5)--(-2.7,0.5);
                \draw[very thick,red,dashed] (3,0.5)--(2.7,0.5);
                \foreach \y in {-1,0,1,2}
                    {
                        \fill [opacity=0.15,green,rounded corners] (-2-0.25,\y-0.2*2.25-0.25)--(-2-0.25,\y-0.2*2.25+0.25)--(2+0.25,\y+0.2*2.25+0.25)--(2+0.25,\y+0.2*2.25-0.25)--cycle;
                    }
                \fill [opacity=0.15,green,rounded corners] (0-0.25,-2-0.2*0.25-0.25)--(0-0.25,-2-0.2*0.25+0.25)--(2+0.25,-2+0.2*2.25+0.25)--(2+0.25,-2+0.2*2.25-0.25)--cycle;
                \fill [opacity=0.15,green,rounded corners] (-2-0.25,3-0.2*2.25-0.25)--(-2-0.25,3-0.2*2.25+0.25)--(0+0.25,3+0.2*0.25+0.25)--(0+0.25,3+0.2*0.25-0.25)--cycle;
            \end{tikzpicture}
            \subcaption{}
            \label{fig:twisted_plaquette_ising_logical_filling}
        \end{minipage}
        \end{tabular}
        \caption{\textbf{(a)} The logical $Z$ rigid loop operator. In this figure, the $Z$ operators are depicted as green circles. This loop operator, which is pictured as green realm, is embedded in a leaf of the foliation structure defined by the normal 1-form $\omega$. \textbf{(b)} The logical $Z$ rigid loop operator extending along the horizontal direction. Due to the twisted boundary condition, the loop operator extends spirally to be enclosed.
        }
        \label{fig:logical_operators_in_twisted_plaquette_ising_model}
    \end{center}
\end{figure}
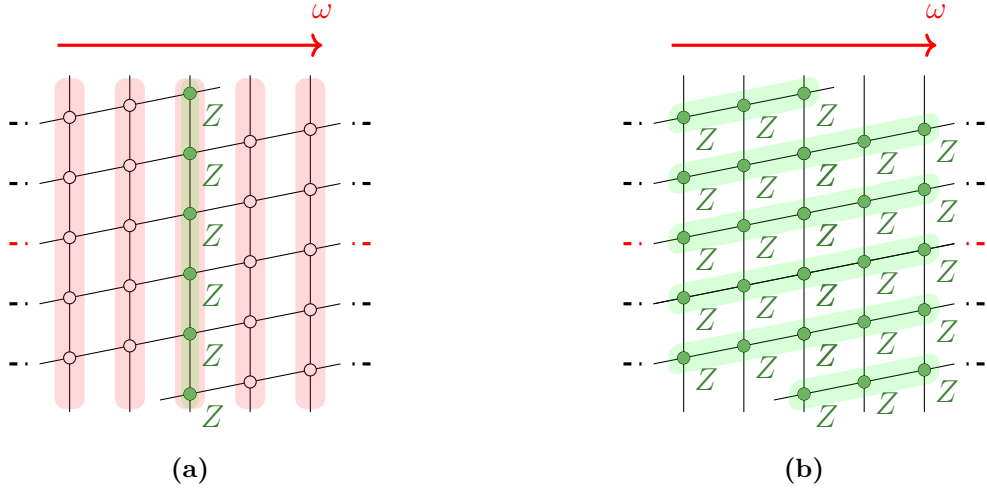
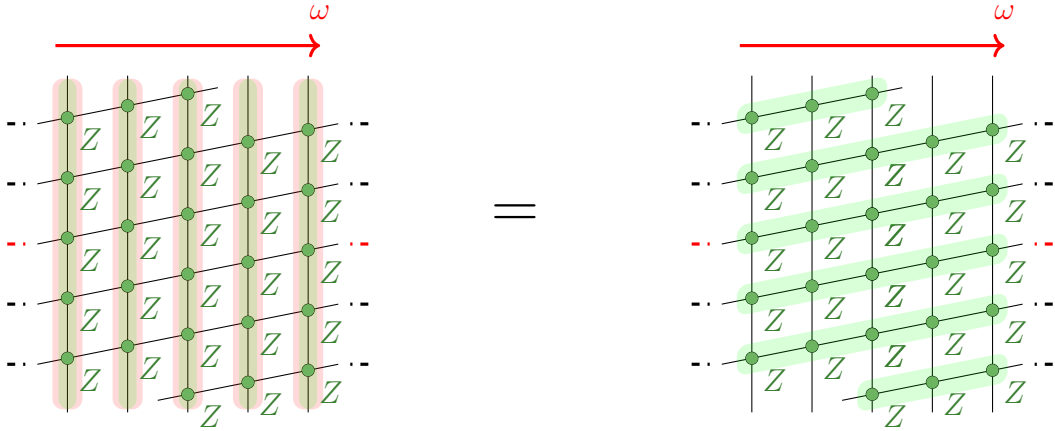
\begin{figure}[t]
    \begin{center}
        \begin{tabular}{ccc}
           \begin{minipage}{0.4\linewidth}
           \centering
            \begin{tikzpicture}[scale=0.8]
                \draw[->,very thick,red](-2.2,3.8)--(2.2,3.8) node [above=0.2cm] {$\omega$};
                \foreach \x in {-2,-1,0,1,2}
                {
                    \draw (\x,-2.3)--(\x, 3.3);
                }
                \foreach \y in {-1,0,1,2}
                {
                    \draw (-2.5,\y-0.5)--(2.5,\y+0.5);
                }
                \draw (-2.5,2.5)--(0.5,2.5+3/5);
                \draw (-0.5,-1.5-3/5)--(2.5,-1.5);
                \foreach \x in {-2,-1,0}
                {
                    \filldraw[fill=white](\x,3+\x*0.2)circle(0.1);
                    
                }
                \foreach \x in {0,1,2}
                {
                    \filldraw[fill=white](\x,-2+\x*0.2)circle(0.1);
                }
                \foreach \y in {-1,0,1,2}
                {
                    \foreach \x in {-2,-1,0,1,2}
                    {
                        \filldraw[fill=white](\x,\y+\x*0.2)circle(0.1);
                    }
                }
                \foreach \x in {-2,-1,0,1,2}
                {
                    \fill[opacity=0.15, red, rounded corners] (\x-0.25,-2.25)--(\x-0.25,3.25)--(\x+0.25,3.25)--(\x+0.25,-2.25)--cycle;
                }
                \foreach \y in {-1.5,-0.5,1.5,2.5}
                {
                    \draw[very thick,dashed] (-3,\y)--(-2.7,\y);
                    \draw[very thick,dashed](3,\y)--(2.7,\y);
                }
                \draw[very thick,red,dashed] (-3,0.5)--(-2.7,0.5);
                \draw[very thick,red,dashed] (3,0.5)--(2.7,0.5);
                \foreach \x in {-2,-1,0,1,2}{
                    \foreach \y in {-1,0,1,2}
                    {
                    \fill[OliveGreen!60](\x, \y+\x*0.2)circle(0.1) node [below right, OliveGreen] {$Z$};
                    }
                }
                \foreach \x in {-2,-1,0} {
                \fill[OliveGreen!60](\x,3+\x*0.2)circle(0.1) node [below right, OliveGreen] {$Z$};
                }
                \foreach \x in {0,1,2} {
                \fill[OliveGreen!60](\x,-2+\x*0.2)circle(0.1) node [below right, OliveGreen] {$Z$};
                }
                \foreach \x in {-2,-1,0,1,2}{
                    \fill[opacity=0.15, green, rounded corners] (\x-0.15,-2.25)--(\x-0.15,3.25)--(\x+0.15,3.25)--(\x+0.15,-2.25)--cycle;
                }
            \end{tikzpicture}
            \end{minipage}
            &
            \begin{minipage}{0.1\linewidth}
                \begin{center}
                    \begin{tikzpicture}[scale=0.8]
                        \node[font=\huge] at (0,0) {\hspace{-0.5em}$=$};
                    \end{tikzpicture}
                \end{center}
            \end{minipage}
            &
            \begin{minipage}{0.4\linewidth}
            \centering
            \begin{tikzpicture}[scale=0.8]
                \draw[->,very thick,red](-2.2,3.8)--(2.2,3.8) node [above=0.2cm] {$\omega$};
                \foreach \x in {-2,-1,0,1,2}
                {
                    \draw (\x,-2.3)--(\x, 3.3);
                }
                \foreach \y in {-1,0,1,2}
                {
                    \draw (-2.5,\y-0.5)--(2.5,\y+0.5);
                }
                \draw (-2.5,2.5)--(0.5,2.5+3/5);
                \draw (-0.5,-1.5-3/5)--(2.5,-1.5);
                \foreach \x in {-2,-1,0}
                {
                    \filldraw[fill=white](\x,3+\x*0.2)circle(0.1);
                    \fill[OliveGreen!60](\x,3+\x*0.2)circle(0.1)node[below right, OliveGreen] {$Z$};
                    
                }
                \foreach \x in {0,1,2}
                {
                    \filldraw[fill=white](\x,-2+\x*0.2)circle(0.1);
                    \fill[OliveGreen!60](\x,-2+\x*0.2)circle(0.1)node[below right, OliveGreen]{$Z$};
                }
                \foreach \y in {-1,0,1,2}
                {
                    \foreach \x in {-2,-1,0,1,2}
                    {
                        \filldraw[fill=white](\x,\y+\x*0.2)circle(0.1);
                        \fill[OliveGreen!60](\x,\y+\x*0.2)circle(0.1)node[below right, OliveGreen] {$Z$};
                    }
                    \filldraw[fill=white](0, \y)circle(0.1);
                    \fill[OliveGreen!60](0, \y)circle(0.1) node [below right, OliveGreen] {$Z$};
                }
                \foreach \y in {-1.5,-0.5,1.5,2.5}
                {
                    \draw[very thick,dashed] (-3,\y)--(-2.7,\y);
                    \draw[very thick,dashed](3,\y)--(2.7,\y);
                }
                \draw[very thick,red,dashed] (-3,0.5)--(-2.7,0.5);
                \draw[very thick,red,dashed] (3,0.5)--(2.7,0.5);
                \foreach \y in {-1,0,1,2}
                    {
                        \fill [opacity=0.15,green,rounded corners] (-2-0.25,\y-0.2*2.25-0.25)--(-2-0.25,\y-0.2*2.25+0.25)--(2+0.25,\y+0.2*2.25+0.25)--(2+0.25,\y+0.2*2.25-0.25)--cycle;
                    }
                \fill [opacity=0.15,green,rounded corners] (0-0.25,-2-0.2*0.25-0.25)--(0-0.25,-2-0.2*0.25+0.25)--(2+0.25,-2+0.2*2.25+0.25)--(2+0.25,-2+0.2*2.25-0.25)--cycle;
                \fill [opacity=0.15,green,rounded corners] (-2-0.25,3-0.2*2.25-0.25)--(-2-0.25,3-0.2*2.25+0.25)--(0+0.25,3+0.2*0.25+0.25)--(0+0.25,3+0.2*0.25-0.25)--cycle;
            \end{tikzpicture}
        \end{minipage}
        \end{tabular}
    \end{center}
    \caption{The left panel denotes piling up the vertical logical loop operators in the horizontal direction, and the right one does the loop operator stretched in the horizontal direction. These two configurations of the logical operators are equivalent to each other.}
    \label{fig:finite_size_effect_twisted}
\end{figure}

Next, we consider a plaquette Ising model on twisted lattice spins as depicted in Fig.\,~\ref{fig:twisted_foliated_lattice_spins}.
While we employ the periodic boundary condition along the vertical direction, we take the \textit{twisted} boundary condition along the horizontal direction.
This lattice has a foliation structure along $x^1$-direction but does not have along $x^2$-direction.
The Hamiltonian is given by
\begin{align}
\label{eq:twisted_Ising_plaquette}
    H = -\sum_{p:~\mathrm{plaquettes}} X_p\ ,
\end{align}
where the local term $X_{p}$ is defined in Fig.\,\ref{fig:twisted_plaquette_ising_local_operator}.
As depicted in Fig.\,\ref{fig:logical_operators_in_twisted_plaquette_ising_model}, we have two types of logical operators in this model. 
The first one illustrated in Fig.\,\ref{fig:twisted_plaquette_ising_logical_loop} is a loop operator embedded in a leaf of the foliation. 
The second one becomes an operator which fills all the lattice points like Fig.\,\ref{fig:twisted_plaquette_ising_logical_filling} because of the twisted boundary condition.
There is an identification of these logical operators as shown in Fig.\,\ref{fig:finite_size_effect_twisted} coming from the finiteness of the spatial lattice size. 
The GSD is given by
\begin{align}
    \mathrm{GSD} = 2^{\beta_{1}^{\omega} L_\omega}\ .
\end{align}
This result is consistent with the one obtained by the field theory \eqref{eq:GV_action_step4}.

\clearpage

\subsection{Two Independent Codimension One Foliations}
In the last subsection, we have presented the two lattice models with a single codimension one foliation.
In this subsection, we generalize the previous discussion to include the lattice models with two independent codimension one foliations. 
To this end, we consider the field theory given by the sum of the actions \eqref{eq:GV_action_step4} with $\omega$ and $\tilde{\omega}$, 
where these two normal 1-form fields $(\omega ,\tilde{\omega} )$ are linearly independent of each other at every point:
\begin{align}
    S = \frac{k}{2\pi} \int 
    \Bigl[ b\wedge \d c - \lambda \wedge (\d\phi \wedge \omega - \omega \wedge b) \Bigr] 
    + \frac{\tilde{k}}{2\pi} \int 
    \Bigl[ \tilde{b} \wedge \d\tilde{c} - \tilde{\lambda} \wedge (\d\tilde{\phi} \wedge \tilde{\omega} - \tilde{\omega} \wedge \tilde{b} ) \Bigr] \ .
    \label{eq:two_GV_BF}
\end{align}
The Hilbert space of this theory is the tensor product of
ones of the single foliation case, and the GSD is given by
\begin{align}
\label{eq:GSD_two_BF}
    \mathrm{GSD} = k^{\beta_1^\omega L_{\omega}}\times \tilde{k}^{\beta_1^{\tilde{\omega}} L_{\tilde{\omega}}}\ (=\infty)\ .
\end{align}
In the following, we discuss some lattice models which realize this GSD.

\subsubsection{Two ($1+1$)-dimensional Ising Stacks}
\begin{figure}[t]
    \begin{center}
    \begin{minipage}[b]{0.5\linewidth}
    \begin{center}
        \begin{tikzpicture}[scale=0.8]
            \draw[->,very thick,red](-2.2,2.8)--(2.2,2.8) node [above=0.2cm] {$\omega$};
            \draw[->,very thick,blue](-2.8,-2.2)--(-2.8,2.2) node [left=0.2cm] {$\tilde{\omega}$};
            \draw[step=1cm](-2.5,-2.5) grid (2.5,2.5);
            \foreach \x in {-2,-1,0,1,2}
            {
                \foreach \y in {-2,-1,0,1,2}
                {
                    \filldraw[fill=white](\x-0.15,\y)circle(0.1);
                    \filldraw[fill=black](\x+0.15,\y)circle(0.1);
                }
                \fill[opacity=0.15, red, rounded corners] (\x-0.25,-2.25)--(\x-0.25,2.25)--(\x+0.25,2.25)--(\x+0.25,-2.25)--cycle;
            }
            \foreach \y in {-2,-1,0,1,2}
            {
                \fill[opacity=0.15, blue, rounded corners] (-2.25,\y-0.25)--(-2.25,\y+0.25)--(2.25,\y+0.25)--(2.25,\y-0.25)--cycle;
            }
        \end{tikzpicture}
        \subcaption{}
        \label{fig:2_way_foliated_lattice_spins}
        \end{center}
        \end{minipage}%
        \begin{minipage}[b]{0.5\linewidth}
        \begin{center}
            \begin{minipage}{0.5\linewidth}
            \begin{center}
                \begin{tikzpicture}[scale=0.9]
                    \node [] at (-1.5,0.5) {$X_e=$};
                    \draw[step=1cm] (-0.5,-0.5) grid (0.5, 1.5);
                    \filldraw[fill=blue!80](-0.15,0)circle(0.1);
                    \filldraw[fill=blue!80](-0.15,1)circle(0.1);
                    \filldraw[fill=black](0.15,0)circle(0.1);
                    \filldraw[fill=black](0.15,1)circle(0.1);
                    \node [below right, blue!80] at (0,0) {$X_{2}$};
                    \node [above right, blue!80] at (0,1) {$X_{1}$};
                    \fill[opacity=0.15,red,rounded corners](-0.25,-0.25)--(-0.25,1.25)--(0.25,1.25)--(0.25,-0.25)--cycle;
                    \node [right, red!80] at (0.2,0.5) {$e$};
                \end{tikzpicture}    
                \end{center}
            \end{minipage}%
            \begin{minipage}{0.5\linewidth}
            \begin{center}
                \begin{tikzpicture}[scale=0.9]
                    \node [] at (-1.5,0) {$X_{\tilde{e}} =$};
                    \draw[step=1cm] (-0.5,-0.5) grid (1.5, 0.5);
                    \filldraw[fill=white](-0.15,0)circle(0.1);
                    \filldraw[fill=white](1-0.15,0)circle(0.1);
                    \filldraw[fill=blue!80](0.15,0)circle(0.1);
                    \filldraw[fill=blue!80](1+0.15,0)circle(0.1);
                    \node [above left, blue!80] at (0,0) {$X_{\tilde{1}}$};
                    \node [above right, blue!80] at (1,0) {$X_{\tilde{2}}$};
                    \fill[opacity=0.15,red,rounded corners](-0.25,-0.25)--(-0.25,0.25)--(1.25,0.25)--(1.25,-0.25)--cycle;
                    \node [right, red!80] at (0.2,0.5) {$\tilde{e}$};
                \end{tikzpicture}    
                \end{center}
            \end{minipage}
            \vspace{1.5\baselineskip}
        \subcaption{}
        \label{fig:two_ising_stacks_local_operator}
        \end{center}
        \end{minipage}
        \caption{\textbf{(a)} Lattice spins hosting two independent foliation structures in the model \eqref{eq:two_Ising_Hamiltonian}. The physical degrees of freedom are qubits denoted by circles ($\circ$ and $\bullet$). We have the two independent normal 1-forms $\omega$ and $\tilde{\omega}$, which point in the horizontal and vertical directions, respectively. Also, the red and blue rectangles denote the leaves associated with the normal 1-form $\omega$ and $\tilde{\omega}$, respectively. \textbf{(b)} Local terms $X_{e}$ and $X_{\tilde{e}}$ in the two Ising stacked Hamiltonian \eqref{eq:two_Ising_Hamiltonian}.
        The local terms are expressed as the products $X_{e} = X_1 X_2$ and $X_{\tilde{e}} = X_{\tilde{1}} X_{\tilde{2}}$, where $X_i$'s and $X_{\tilde{i}}$'s are $X$ operators acting on the two endpoints of the edges $e$ and $\tilde{e}$, respectively.}
    \end{center}
\end{figure}

To construct the lattice model whose low energy field theory is described by \eqref{eq:two_GV_BF}, we have to consider a lattice theory with two independent codimension one foliations.
We can construct such theory simply by placing two copies of the Ising stacking model discussed in the last subsection, along two different directions as depicted in Fig.\,\ref{fig:2_way_foliated_lattice_spins}.
Putting two qubits at each vertex, the Hamiltonian is given by
\begin{align}
\label{eq:two_Ising_Hamiltonian}
    H = -\sum_{e\,\in\, E_{\omega}} X_e- \sum_{\tilde{e}\,\in\, E_{\tilde{\omega}}} X_{\tilde{e}}\ ,
\end{align}
where $E_{\omega}$ and $E_{\tilde{\omega}}$ are the sets of all edges transverse to $\omega$ and $\tilde{\omega}$, respectively. Also, the local operators $X_e^\omega$ and $X_{\tilde{e}}^{\tilde{\omega}}$ are defined in Fig.\,\ref{fig:two_ising_stacks_local_operator}.

\begin{figure}[t]
    \begin{center}
        \begin{tabular}{cc}
            \begin{minipage}{0.4\linewidth}
            \centering
                \begin{tikzpicture}[scale=0.8]
                \fill[opacity=0.2, red, rounded corners] (0-0.45,-2.3)--(0-0.45,2.3)--(0+0.45,2.3)--(0+0.45,-2.3)--cycle;
                    \draw[->,very thick,red](-2.2,2.8)--(2.2,2.8) node [above=0.1cm] {$\omega$};
                    \draw[->,very thick,blue](-2.8,-2.2)--(-2.8,2.2) node [left=0.2cm] {$\tilde{\omega}$};
                    \draw[step=1cm](-2.5,-2.5) grid (2.5,2.5);
                    \foreach \y in {-2,-1,0,1,2}
                    {
                        \foreach \x in {-2,-1,0,1,2}
                        {
                            \filldraw[fill=white](\x-0.15,\y)circle(0.1);
                            \filldraw[fill=black](\x+0.15,\y)circle(0.1);
                        }
                        \fill[fill=OliveGreen!60](0-0.15,\y)circle(0.1);
                        \node[below left,OliveGreen] at (0-0.15,\y) {$Z$};
                    }
                    \fill[opacity=0.2, green, rounded corners] (0-0.25,-2.25)--(0-0.25,2.25)--(0+0.25,2.25)--(0+0.25,-2.25)--cycle;
                \end{tikzpicture}
            \end{minipage}
            &
            \begin{minipage}{0.4\linewidth}
                \centering
                    \begin{tikzpicture}[scale=0.8]
                     \fill[opacity=0.15, blue, rounded corners] (-2.3,-0.45)--(-2.3,+0.45)--(2.3,+0.45)--(2.3,-0.45)--cycle;
                        \draw[->,very thick,red](-2.2,2.8)--(2.2,2.8) node [above=0.1cm] {$\omega$};
                        \draw[->,very thick,blue](-2.8,-2.2)--(-2.8,2.2) node [left=0.2cm] {$\tilde{\omega}$};
                        \draw[step=1cm](-2.5,-2.5) grid (2.5,2.5);
                        \foreach \x in {-2,-1,0,1,2}
                        {
                            \foreach \y in {-2,-1,0,1,2}
                            {
                                \filldraw[fill=white](\x-0.15,\y)circle(0.1);
                                \filldraw[fill=black](\x+0.15,\y)circle(0.1);
                            }
                            \fill[fill=OliveGreen!60](\x+0.15,0)circle(0.1);
                            \node [below right,OliveGreen] at (\x+0.15,0) {$Z$};
                        }
                        \fill[opacity=0.2, green, rounded corners] (-2.25,-0.25)--(-2.25,0.25)--(2.25,0.25)--(2.25,-0.25)--cycle;
                    \end{tikzpicture}
            \end{minipage}
        \end{tabular}
        \caption{Two types of logical operators in the two Ising stacked theory~\eqref{eq:two_Ising_Hamiltonian}. 
        The configuration of string of $Z$ operators in the left and right panels represent the logical operators which are embedded in leaves associated to the normal 1-forms $\omega$ and $\tilde{\omega}$, respectively. }
        \label{fig:two_ising_stacks_logical_operator}
    \end{center}
\end{figure}
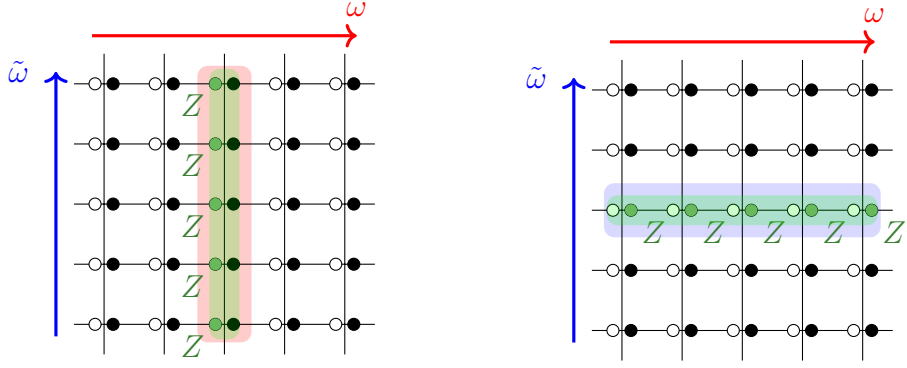

Similar to the previous argument, we discuss the ground state property of the model.
Since all the local terms in the Hamiltonian commute with each other, this model can be regarded as a stabilizer code.
Logical operators of this model are generated from those of two Ising stacks.
While the local $X$ operators become logical $X$ operators, only rigid $Z$-loop operators depicted in Fig.\,\ref{fig:two_ising_stacks_logical_operator} become logical $Z$ operators. The GSD of this model is 
\begin{align}
    \mathrm{GSD} = 2^{\beta_1^{\omega} L_{\omega} + \beta_1^{\tilde{\omega}} L_{\tilde{\omega}}}\ , 
\end{align}
where $\beta_1^{\omega}$ and $\beta_1^{\tilde{\omega}}$ are first Betti numbers of leaves transverse to $\omega$ and $\tilde{\omega}$, respectively. Also, $L_\omega$ and $L_{\tilde{\omega}}$ are lattice sizes along $\omega$ and $\tilde{\omega}$-directions, respectively.
This GSD is just a multiple of GSDs of the two Ising stacks with foliation structures given by $\omega$ and $\tilde{\omega}$.
The ground state property of this model shows agreement with that of the field theory \eqref{eq:GSD_two_BF}.

Similar to the single codimension one foliation case, fractionalized quasiparticles in this model are endpoints of rigid $Z$-string operators.
Since we now consider two independent foliations, we have two types of quasiparticles;
One is mobile only transverse to the $\omega$-direction, and the other one is mobile only transverse to the $\tilde{\omega}$-direction.
Thus, we have two types of lineon excitations coming from two independent codimension one foliation structures.

\subsubsection{Plaquette Ising Model}
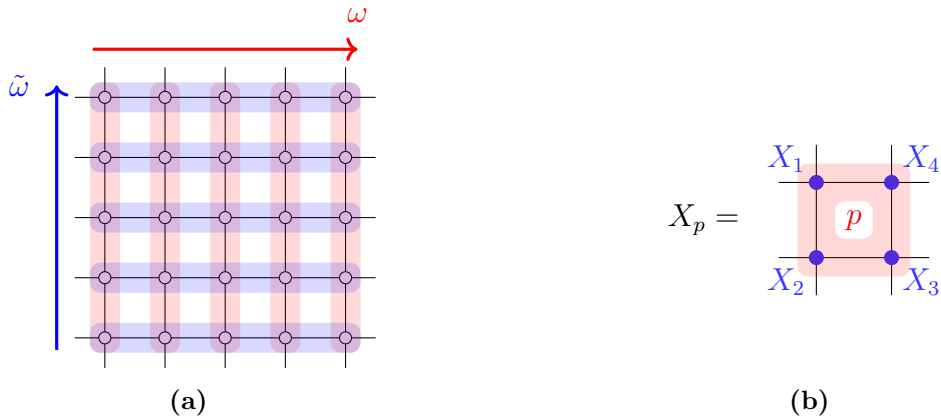
\begin{figure}[t]
    \begin{center}
    \begin{minipage}[b]{0.5\linewidth}
    \begin{center}
        \begin{tikzpicture}[scale=0.8]
            \draw[->,very thick,red](-2.2,2.8)--(2.2,2.8) node [above=0.2cm] {$\omega$};
            \draw[->,very thick,blue](-2.8,-2.2)--(-2.8,2.2) node [left=0.2cm] {$\tilde{\omega}$};
            \draw[step=1cm](-2.5,-2.5) grid (2.5,2.5);
            \foreach \x in {-2,-1,0,1,2}
            {
                \foreach \y in {-2,-1,0,1,2}
                {
                    \filldraw[fill=white](\x,\y)circle(0.1);
                }
                \fill[opacity=0.15,red, rounded corners] (\x-0.25,-2.25)--(\x-0.25,2.25)--(\x+0.25,2.25)--(\x+0.25,-2.25)--cycle;
            }
            \foreach \y in {-2,-1,0,1,2}
            {
                \fill[opacity=0.15,blue,rounded corners] (-2.25,\y-0.25)--(-2.25,\y+0.25)--(2.25,\y+0.25)--(2.25,\y-0.25)--cycle;
            }
        \end{tikzpicture}
        \subcaption{}
        \label{fig:two_foliated_lattice_spins}
        \end{center}
        \end{minipage}%
        \begin{minipage}[b]{0.5\linewidth}
        \begin{center}
        \begin{tikzpicture}
            \node [] at (-1.5,0.5) {$X_p=$};
            \draw[step=1cm] (-0.5,-0.5) grid (1.5, 1.5);
            \fill[fill=blue!80](0,1)circle(0.1);
            \node [above left, blue!80] at (0,1) {$X_1$};
            \fill[fill=blue!80](0,0)circle(0.1);
            \node [below left, blue!80] at (0,0) {$X_2$};
            \fill[fill=blue!80](1,1)circle(0.1);
            \node [above right, blue!80] at (1,1) {$X_4$};
            \fill[fill=blue!80](1,0)circle(0.1);
            \node [below right, blue!80] at (1,0) {$X_3$};
            \fill [opacity=0.155, red, rounded corners] (-0.25,-0.25)--(-0.25,1.25)--(1.25,1.25)--(1.25,-0.25)--cycle;
            \fill [white, rounded corners] (0.25,0.25)--(0.25,0.75)--(0.75,0.75)--(0.75,0.25)--cycle;
            \node [red] at (0.5, 0.5) {$p$};
        \end{tikzpicture}
        \vspace{1.5\baselineskip}
        \subcaption{}
        \label{fig:plaquette_ising_local_operator}
    \end{center}
    \end{minipage}
    \caption{\textbf{(a)} Lattice spins hosting two independent foliation structures in the model \eqref{eq:plaquette_Ising_model}. The physical degrees of freedom are qubits denoted by circles ($\circ$).
    We have the two independent normal 1-forms $\omega$ and $\tilde{\omega}$, which point in the horizontal and vertical directions, respectively. The red and blue rectangles means the leaves associated to the normal 1-form $\omega$ and $\tilde{\omega}$, respectively. \textbf{(b)} The term $X_{p}$ given in the $(2+1)$-dimensional Plaquette Ising Model \eqref{eq:plaquette_Ising_model}, which is expressed as the
        product $X_p = X_1 X_2 X_3 X_4$, where $X_i$'s are $X$ operators acting on the corner of the plaquette $p$.}
    \end{center}
\end{figure}

We can also consider the $(2+1)$-dimensional plaquette Ising model on $T^2$ with two trivial foliations as the UV lattice theory of the low energy field theory \eqref{eq:two_GV_BF}.%
\footnote{See e.g.,~\cite{plqt_ising_2004,Johnston:2016mbz} for earlier expositions on the related model. }

Compared with the previous cases, this model exhibits more intriguing UV completion of the BF-like Godbillon-Vey field theory since we cannot write this theory as a mere stacking of lower dimensional models.
The physical degrees of freedom are qubits, one placed on each vertex of a lattice which hosts two linearly independent foliation structures (Fig.\,\ref{fig:two_foliated_lattice_spins}).
The Hamiltonian of the plaquette Ising model is defined by
\begin{align}
\label{eq:plaquette_Ising_model}
    H = -\sum_{p\ :\ \mathrm{plaquettes}} X_p\ , 
\end{align}
where the local term $X_p$ is defined on each plaquette of the lattice as Fig.\,\ref{fig:plaquette_ising_local_operator}.

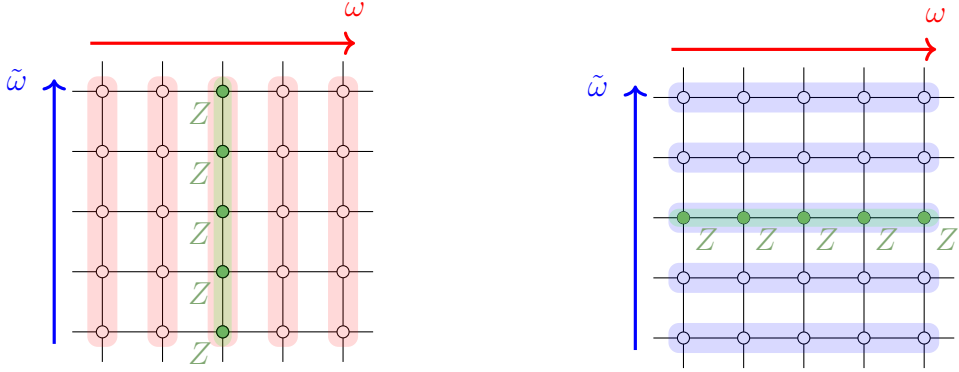
\begin{figure}[t]
    \begin{center}
        \begin{tabular}{cc}
            \begin{minipage}{0.4\linewidth}
                \begin{tikzpicture}[scale=0.8]
                    \draw[->,very thick,red](-2.2,2.8)--(2.2,2.8) node [above=0.2cm] {$\omega$};
                    \draw[->,very thick,blue](-2.8,-2.2)--(-2.8,2.2) node [left=0.2cm] {$\tilde{\omega}$};
                    \draw[step=1cm](-2.5,-2.5) grid (2.5,2.5);
                    \foreach \y in {-2,-1,0,1,2}
                    {
                        \foreach \x in {-2,-1,0,1,2}
                        {
                            \filldraw[fill=white](\x,\y)circle(0.1);
                        }
                    }
                    \foreach \x in {-2,-1,0,1,2}{
                    \fill[opacity=0.15,red, rounded corners] (\x-0.25,-2.25)--(\x-0.25,2.25)--(\x+0.25,2.25)--(\x+0.25,-2.25)--cycle;
                    }
                    \foreach \y in {-2,-1,0,1,2}{
                        \filldraw[fill=OliveGreen!60](0,\y)circle(0.1);
                        \node[below left,OliveGreen!60] at (0,\y) {$Z$};
                        }
                    \fill[opacity=0.15,green, rounded corners] (0-0.15,-2.25)--(0-0.15,2.25)--(0+0.15,2.25)--(0+0.15,-2.25)--cycle;
                \end{tikzpicture}
            \end{minipage}
            &
            \begin{minipage}{0.4\linewidth}
                \begin{center}
                    \begin{tikzpicture}[scale=0.8]
                        \draw[->,very thick,red](-2.2,2.8)--(2.2,2.8) node [above=0.2cm] {$\omega$};
                        \draw[->,very thick,blue](-2.8,-2.2)--(-2.8,2.2) node [left=0.2cm] {$\tilde{\omega}$};
                        \draw[step=1cm](-2.5,-2.5) grid (2.5,2.5);
                        \foreach \x in {-2,-1,0,1,2}
                        {
                            \foreach \y in {-2,-1,0,1,2}
                            {
                                \filldraw[fill=white](\x,\y)circle(0.1);
                            }
                        }
                        \foreach \y in {-2,-1,0,1,2}{
                        \fill[opacity=0.15,blue, rounded corners] (-2.25,\y-0.25)--(-2.25,\y+0.25)--(2.25,\y+0.25)--(2.25,\y-0.25)--cycle;
                        }
                        \foreach \x in {-2,-1,0,1,2}{
                        \fill[OliveGreen!60](\x,0)circle(0.1) node [below right, OliveGreen!60] {$Z$};
                        }
                        \fill[opacity=0.15,green, rounded corners] (-2.25,-0.15)--(-2.25,0.15)--(2.25,0.15)--(2.25,-0.15)--cycle;
                    \end{tikzpicture}
                \end{center}
            \end{minipage}
        \end{tabular}
        \caption{Two types of logical operators in the Ising plaquette theory \eqref{eq:plaquette_Ising_model}. 
        The logical operators are depicted as green dots in 
        the left and right panels,
        which are embedded in leaves associated to the normal 1-forms $\omega$ and $\tilde{\omega}$, respectively. }
        \label{fig:plaquette_ising_logical_operator}
    \end{center}
\end{figure}

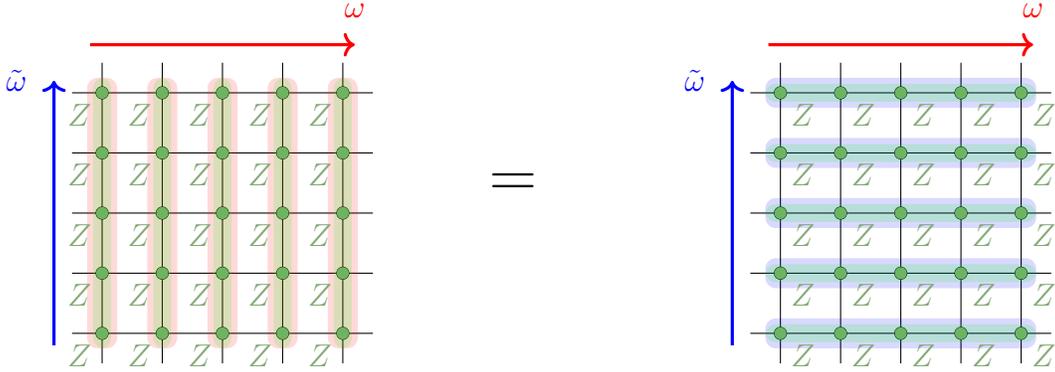
\begin{figure}[t]
    \begin{center}
        \begin{tabular}{ccc}
            \begin{minipage}{0.4\linewidth}
            \centering
                \begin{tikzpicture}[scale=0.8]
                    \draw[->,very thick,red](-2.2,2.8)--(2.2,2.8) node [above=0.2cm] {$\omega$};
                    \draw[->,very thick,blue](-2.8,-2.2)--(-2.8,2.2) node [left=0.2cm] {$\tilde{\omega}$};
                    \draw[step=1cm](-2.5,-2.5) grid (2.5,2.5);
                    \foreach \x in {-2,-1,0,1,2}
                    {
                        \foreach \y in {-2,-1,0,1,2}
                        {
                            \filldraw[fill=white](\x,\y)circle(0.1);
                        }
                        \fill[opacity=0.15,red, rounded corners] (\x-0.25,-2.25)--(\x-0.25,2.25)--(\x+0.25,2.25)--(\x+0.25,-2.25)--cycle;
                    }
                    \foreach \x in {-2,-1,0,1,2}{
                        \foreach \y in {-2,-1,0,1,2}{
                            \fill[OliveGreen!60](\x,\y)circle(0.1) node[below left,OliveGreen!60] at (\x,\y) {$Z$};
                        }
                        \fill[opacity=0.15,green, rounded corners] (\x-0.15,-2.25)--(\x-0.15,2.25)--(\x+0.15,2.25)--(\x+0.15,-2.25)--cycle;
                    }
                \end{tikzpicture}
            \end{minipage}
            &
            \begin{minipage}{0.1\linewidth}
                \begin{center}
                    \begin{tikzpicture}[scale=0.8]
                        \node[font=\huge](0,0) {\hspace{-0.5em}$=$};
                    \end{tikzpicture}
                \end{center}
            \end{minipage}
            &
            \begin{minipage}{0.4\linewidth}
                \begin{center}
                    \begin{tikzpicture}[scale=0.8]
                        \draw[->,very thick,red](-2.2,2.8)--(2.2,2.8) node [above=0.2cm] {$\omega$};
                        \draw[->,very thick,blue](-2.8,-2.2)--(-2.8,2.2) node [left=0.2cm] {$\tilde{\omega}$};
                        \draw[step=1cm](-2.5,-2.5) grid (2.5,2.5);
                        \foreach \y in {-2,-1,0,1,2}
                        {
                            \foreach \x in {-2,-1,0,1,2}
                            {
                                \filldraw[fill=white](\x,\y)circle(0.1);
                            }
                            \fill[opacity=0.15,blue, rounded corners] (-2.25,\y-0.25)--(-2.25,\y+0.25)--(2.25,\y+0.25)--(2.25,\y-0.25)--cycle;    
                        }
                        \foreach \y in {-2,-1,0,1,2}{
                            \foreach \x in {-2,-1,0,1,2}{
                                \fill[OliveGreen!60](\x,\y)circle(0.1) node [below right,OliveGreen!60] at (\x,\y) {$Z$};   
                            }
                            \fill[opacity=0.15,green, rounded corners] (-2.25,\y-0.15)--(-2.25,\y+0.15)--(2.25,\y+0.15)--(2.25,\y-0.15)--cycle;    
                        }
                    \end{tikzpicture}
                \end{center}
            \end{minipage}
        \end{tabular}
        \caption{The left panel dictates stacking the vertical logical loop operators stacked in the horizontal direction whereas the right one does the loop operator stretched in the horizontal direction piled up in the vertical direction. These two distributions of the logical operators 
        are identical with one another in the ground state.}
        \label{fig:finite_size_effect}
    \end{center}
\end{figure}

We cannot write this theory as a stack of lower dimensional theories.
However, we can easily compute the structure of the ground state of this model.
All of the local terms in the Hamiltonian commute with each other, thus this model is again recognized as a stabilizer code model.
Moreover, all of the local terms are self-inverse, hence we can repeat the discussion of the Ising stacking case in the previous section. 
The resulting GSD of this model is given by
\begin{align}
    \mathrm{GSD} = 2^{\beta_1^\omega L_\omega + \beta_1^{\tilde{\omega}} L_{\tilde{\omega}} - 1}\ .
\end{align}
Here, $\beta_1^\omega$ and $\beta_1^{\tilde{\omega}}$ are first Betti numbers of leaves transverse to $\omega$ and $\tilde{\omega}$, respectively. Also, $L_\omega$ and $L_{\tilde{\omega}}$ are the lattice sizes along $\omega$ and $\tilde{\omega}$-direction, respectively.

The form of logical operators resembles those of two stacks of Ising model except for the fact that the finite size effect stems from the finiteness of a lattice.
All of the $X$ operators become logical operators in this case.
However, only rigid $Z$-loop operators extending along $\omega$-transverse direction and $\tilde{\omega}$-transverse direction like Fig.\,\ref{fig:plaquette_ising_logical_operator} become logical $Z$ operators.
There is one difference between logical operators of two Ising stacks model and plaquette Ising model.
As shown in Fig.\,\ref{fig:finite_size_effect}, the filling up of $\omega$-transverse logical $Z$ operators and that of $\tilde{\omega}$-transverse logical $Z$ operators are the same, since the size of the lattice which we are considering is finite.
This finite size effect causes a constant shift in the GSD, which cannot be seen in the continuum theory where $L_{\omega}$ and $L_{\tilde{\omega}}$ are taken as infinities. 

The fractionalized quasiparticles in this theory
are also given by the endpoints of rigid $Z$-string
operators transverse either to $\omega$ or
$\tilde{\omega}$-direction.
Similar to the two Ising stacks case, they are lineon excitations
which are only mobile along $\omega$-transverse 
or $\tilde{\omega}$-transverse direction.

\section{Coupling U(1) BF-like Godbillon-Vey Field Theory to Matter Theories}
\label{sec:coupling_matter_theory}
In this section, we investigate matter field theories coupled to the Godbillon-Vey field theories, particularly paying our attention to the U(1) BF-like theory \eqref{eq:GV_action_step4}.%
\footnote{
In principle, we can also extend the discussions in this section to any field theories where gauge fields have gauge symmetries of \eqref{eq:GVBF_gauge_1} and \eqref{eq:GVBF_gauge_3}. Thus, we can consider the coupled matter theories with the invertible $\mathbf{R}$ gauge theory \eqref{eq:GV_action} or the invertible U(1) gauge theory \eqref{eq:GV_action_step3}.
}
As such examples, we present two matter theories where dynamical matter fields are scalar or fermion ones and are coupled to the U(1) gauge field $c$. Although we work with the Euclidean spacetime throughout this section, subsequent discussions are valid even for Lorentzian spacetime. 

\subsection{Scalar QED-like Theory}
Let us consider a matter theory with a classical current $J$ satisfying \eqref{eq:continuous_equation} and \eqref{eq:current_omega_ortho}:
\[
    \d\star J = 0\  ,\quad   \omega \wedge \star J = 0\ .
\]
We can construct such a theory by considering the action in $(2+1)$ dimensions:
\begin{align}
S = \int \Bigl[ \left((\d+\i\Lambda\omega) \Phi^* \right) \wedge \star\left((\d-\i\Lambda\omega) \Phi \right)  + V\left(|\Phi|^2\right) + F(\Lambda \omega) \Bigr] \ ,
\end{align}
where $\Phi$ is a complex scalar field with the potential $V(|\Phi|^2)$, 
$\Lambda$ is a real scalar field and $\omega$ is a background foliation 1-form. 
$F(\Lambda \omega)$ is some gauge invariant function depending only on $\Lambda \omega$ which satisfies
\begin{align}
    \label{eq:F_condition}
    \left.\partial_\mu \left(\frac{\delta \int F(X)}{\delta X_\mu} \right|_{X=\Lambda\omega} \right)= 0\  .  
\end{align}
Its typical example is the kinetic term of $\Lambda \omega$:
\begin{align}
    F_{\text{kin}}(\Lambda \omega)= \d(\Lambda\omega) \wedge \star \d (\Lambda \omega)\ , 
\end{align}
which indeed satisfies the above conditions: 
\begin{align}
    \left.\partial_\mu \left(\frac{\delta \int F_{\text{kin}}(X)}{\delta X_\mu} \right|_{X^{\sigma}=\Lambda\omega^{\sigma}} \right) = -4 \partial_\mu \partial_\nu \left.(\partial^\nu X^\mu - \partial^\mu X^\nu)\right|_{X^\sigma = \Lambda\omega^\sigma} = 0\ .
\end{align}
In the following, we work with the the expression of the Lagrangian defined on a local patch: 
\begin{align}
\label{eq:scalar_qed_theory_lag}
    \mathcal{L} = (\partial_\mu + \i \Lambda\omega_\mu)\Phi^{*} (\partial^\mu - \i\Lambda\omega^\mu)\Phi + V\left(|\Phi|^2\right) + F(\Lambda \omega)\ .
\end{align}
The equations of motions for $\Phi$ and $\Lambda$ are given by
\begin{align}
\label{eq:EOM_scalarQED}
\begin{split}
&(\partial_\mu - \i\Lambda \omega_\mu)(\partial^\mu - \i\Lambda \omega^\mu) \Phi - \left(\left.\frac{\partial V(X)}{\partial X}\right|_{X=|\Phi|^2}\right) \Phi = 0\ ,\\
&\i\left(\Phi^* \del_\mu \Phi - \del_\mu \Phi^* \Phi \right) \omega^\mu + 2\Lambda \omega_\mu \omega^\mu \Phi^* \Phi + \left(\left.\frac{\delta F(X)}{\delta X^\mu}\right|_{X^\nu=\Lambda\omega^\nu}\right) \omega^\mu = 0\ ,
\end{split}
\end{align}
respectively.
In the scalar QED-like theory \eqref{eq:scalar_qed_theory_lag}, we have the following U(1) global symmetry:
\begin{align}
    \Phi \rightarrow e^{\i\alpha} \Phi\ , \quad \Phi^* \rightarrow e^{-\i\alpha} \Phi^* \quad (\alpha: \mathrm{constant})\ .
\end{align}
The first equation of motion implies that there is a conserved current corresponding to this U(1) global symmetry:
\begin{align}
    J^\mu = \i\left(\Phi^* \partial^\mu \Phi - \partial^\mu \Phi^* \Phi\right) + 2\Lambda \omega^\mu \Phi^* \Phi + \left.\frac{\delta F(X)}{\delta X_\mu} \right|_{X_\nu=\Lambda\omega_\nu}\ ,
\end{align}
which satisfies the standard conservation law:
\begin{align}
\label{eq:conserved_law_scalar_classical}
    \partial_\mu J^\mu = 0\ . 
\end{align}
By employing the second equation of motion, we can show that this conserved current is orthogonal to the normal 1-form field $\omega$:
\begin{align}
\label{eq:mobility_scalar_classical}
    \omega_\mu J^\mu = 0\ ,
\end{align}
which indicates that this current describes flow of codimension one subdimensional particles. 
Note that generalization the discussion to any spacetime dimension is straightforward. Indeed, 
if we set the spacetime dimensions to be $d+1$, the field theory
describes the particles which flow in the $(d-1)$-dimensional 
spatial submanifold.

So far, we have only treated the scalar QED-like theory at the classical level. 
We can also find the quantum analogs of these classical equations \eqref{eq:conserved_law_scalar_classical}-\eqref{eq:mobility_scalar_classical}. 
By using the Ward-Takahashi identity, the counterpart of the classical conservation law \eqref{eq:conserved_law_scalar_classical} is found to be
\begin{align}
\langle \del_\mu J^\mu \rangle = 0\ .
\end{align}
On the other hand, we can also find the quantum level equality corresponding to the classical mobility constraint \eqref{eq:current_omega_ortho} by varying $\Lambda$:
\begin{align}
\langle \omega_\mu J^\mu \rangle \propto \int \mathcal{D}\Phi^* \mathcal{D}\Phi \mathcal{D}\Lambda \ \frac{\delta}{\delta \Lambda} e^{-S} = 0\ .
\label{eq:quantum_current_omega_ortho}
\end{align}
Finally, we present the full Lagrangian where the scalar matter field is coupled to the U(1) BF-like theory \eqref{eq:GV_action_step4}:
\begin{align}
\begin{aligned}
    \mathcal{L} &=  (\partial_\mu + \i\Lambda\omega_\mu + \i c_\mu )\Phi^* (\partial^\mu - \i\Lambda \omega^\mu - \i c^\mu) \Phi + V\left(|\Phi|^2\right) + F(\Lambda \omega)\\
    &\quad\quad\quad +\frac{\i k}{2\pi} \epsilon^{\mu\nu\rho} \left(b_\mu \partial_\nu c_\rho - \lambda_\mu (\partial_\nu\phi \omega_\rho - \omega_\nu b_\rho)\right) \ ,
    \end{aligned}
\end{align}
whose gauge symmetries are
\begin{align}
\begin{split}
    &c_\mu \rightarrow c_\mu + u\,\omega_\mu\ , \quad \lambda_\mu \rightarrow \lambda_\mu + u\xi_\mu - \partial_\mu u \ ,\quad \Lambda \rightarrow \Lambda - u\ , \\
    &b_\mu \rightarrow b_\mu-\partial_\mu\theta\ , \quad \phi \rightarrow \phi + \theta\ , \\
    &c_\mu \rightarrow c_\mu + \partial_\mu\tilde{\theta}\ , \quad \Phi \rightarrow e^{\i\tilde{\theta}}\Phi\ , \quad \Phi^* \rightarrow e^{-\i\tilde{\theta}}\Phi^*\ ,  \\
    &\lambda_\mu \rightarrow \rho^{-1}\lambda_\mu\ , \quad \omega_\mu \rightarrow \rho\, \omega_\mu\ , \\
    &\lambda_\mu \rightarrow \lambda_\mu + v\,\omega_\mu\ .
\end{split}
\end{align}
We also comment that the combination of the field $c + \Lambda\omega$ behaves as the ordinary U(1) gauge field. Therefore, we can also include gauge invariant terms as ordinary U(1) gauge theories.
For instance, we can construct the following Maxwell term: 
\begin{align}
    \frac{1}{4} \d (c + \Lambda \omega ) \wedge \star \d (c + \Lambda \omega)\ . 
\end{align}

\subsection{Fermion QED-like Theory}
In this subsection, we discuss a fermionic analog of the scalar QED-like theory in the last subsection
which is coupled to the BF-like theory \eqref{eq:GV_action_step4}. 
The Lagrangian is given by
\begin{align}
    \label{eq:lagrangian_foliation_fermion}
    \mathcal{L} =  \bar{\psi} \gamma^\mu (\del_\mu - \i \Lambda \omega_\mu) \psi + V(\psi, \bar{\psi}) + F(\Lambda \omega)\  , 
\end{align}
where $V(\psi, \bar{\psi})$ consists of quadratic forms of $\psi$ and $\bar{\psi}$, e.g., the mass term or the Thirring coupling like term. Also, $F(\Lambda \omega)$ is some gauge invariant function which satisfies \eqref{eq:F_condition}. 
The equations of motion are given by 
\begin{align}
    \gamma^\mu (\partial_\mu - \i\Lambda \omega_\mu) \psi + \frac{\delta V(\psi,\bar{\psi})}{\delta \bar{\psi}} \psi = 0 ,\qquad
     \omega_\mu\,\bar{\psi} \gamma^\mu \psi   = 0 \ .
\end{align}
Now we have the vector U(1) global symmetry:%
\footnote{If we drop off the potential term $V(\psi, \bar{\psi})$ from the Lagrangian, then we also have the axial U(1) symmetry.}
\begin{align}
        \psi \rightarrow e^{\i\alpha} \psi\ , \quad \bar{\psi} \rightarrow \bar{\psi} e^{-\i\alpha} \quad (\alpha: \mathrm{constant})\ , 
\end{align}
whose classical current is given by
\begin{align}
    J^\mu = \bar{\psi}\gamma^\mu \psi + \left.\frac{\delta F(X)}{\delta X_\mu} \right|_{X_\nu=\Lambda\omega_\nu}\  . 
\end{align}
In a similar manner to the scalar QED-like theory, the above equation of motions leads to  conservation law and mobility constraints on the U(1) current:
\begin{align}\label{eq:constraints_fermion_classical}
    \partial_{\mu}J^\mu = 0 \ , \quad \omega_{\mu} J^{\mu}=0\ . 
\end{align}
These classical behaviors are exactly what we want the current
to satisfy when coupling with the gauge fields in Godbillon-Vey
field theory.%
\footnote{Similar to the argument in the previous subsection, generalization of the matter theory to any spacetime dimension is straightforward.}

Finally, we also present the full Lagrangian where the fermionic matter field is coupled to the U(1) BF-like theory \eqref{eq:GV_action_step4}:
\begin{align}
\label{eq:full_QED_like_Lagrangian_coupled_fermion}
\begin{aligned}
    \mathcal{L} &=  \bar{\psi}\gamma^{\mu}(\partial_\mu - \i\Lambda \omega_\mu - \i c_\mu) \psi + V\left(|\Phi|^2\right) + F(\Lambda \omega)\\
    &\quad\quad +\frac{\i k}{2\pi} \epsilon^{\mu\nu\rho} \left( b_\mu \partial_\nu c_\rho - \lambda_\mu (\partial_\nu\phi \omega_\rho - \omega_\nu b_\rho)\right) \ ,
    \end{aligned}
\end{align}
whose gauge redundancies are listed as follows:
\begin{align}
\begin{split}
    &c_\mu \rightarrow c_\mu + u\,\omega_\mu\ , \quad \lambda_\mu \rightarrow \lambda_\mu + u\xi_\mu - \partial_\mu u \ ,\quad \Lambda \rightarrow \Lambda - u\ , \\
    &b_\mu \rightarrow b_\mu-\partial_\mu\theta\ , \quad \phi \rightarrow \phi + \theta\ , \\
    &c_\mu \rightarrow c_\mu + \partial_\mu\tilde{\theta}\ , \quad \psi \rightarrow e^{\i\tilde{\theta}}\psi\ , \quad \bar{\psi} \rightarrow \bar{\psi}e^{-\i\tilde{\theta}}\ ,  \\
    &\lambda_\mu \rightarrow \rho^{-1}\lambda_\mu\ , \quad \omega_\mu \rightarrow \rho\, \omega_\mu\ , \\
    &\lambda_\mu \rightarrow \lambda_\mu + v\,\omega_\mu\ .
\end{split}
\end{align}

\subsubsection*{Relation to Ordinary QED}
\begin{figure}[t]
    \begin{center}
    \begin{tabular}{ccc}
    \begin{minipage}{0.4\linewidth}
    \begin{center}
    \begin{tikzpicture}
        \draw[->] (-2.5,0)--(2.5,0);
        \draw[->] (0,-2.5)--(0,2.5);
        \draw[->,red, ultra thick] (0,0)--({1.5*cos(60)},{1.5*sin(60)}) node[right] {\small$a$};
        \draw[red!50,ultra thick,->] ({1.3*cos(80)},{1.3*sin(80)}) arc(80:110:1.3);
        \draw[red!50,ultra thick,->] ({1.3*cos(40)},{1.3*sin(40)}) arc(40:10:1.3);
        \draw[red!50,ultra thick,<->] ({2*cos(60)},{2*sin(60)})--({3*cos(60)},{3*sin(60)}); 
    \end{tikzpicture}
    \end{center}
    \end{minipage}
    &
    \begin{minipage}{0.1\linewidth}
    \begin{center}
    \begin{tikzpicture}
        \draw[->,very thick] (-0.5,0)--(0.5,0);
    \end{tikzpicture}
    \end{center}
    \end{minipage}
    &
    \begin{minipage}{0.4\linewidth}
    \begin{center}
    \begin{tikzpicture}
        \draw[->] (-2.5,0)--(2.5,0);
        \draw[->] (0,-2.5)--(0,2.5);
        \draw[->,red, ultra thick] (0,0)--({1.5*cos(60)},{1.5*sin(60)}) node[right] {\small$a$};
        \foreach \r in {0.5,1.5,2}{
        \draw[OliveGreen!30] (0,0) circle(\r);
        }
        \draw[blue, thick] (0,0) circle(1);
        \draw[->,blue, thick] (0,0)--({1*cos(60)},{1*sin(60)}) node[above left] {\small$\omega$};
        \draw[blue,ultra thick,->] ({0.8*cos(80)},{0.8*sin(80)}) arc(80:110:0.8);
        \draw[blue,ultra thick,->] ({0.8*cos(40)},{0.8*sin(40)}) arc(40:10:0.8);
        \draw[OliveGreen] (0,0) to [bend right] (0.2,-0.1);
        \draw[OliveGreen] (1.3,-0.1) to [bend right] (1.5,0);
        \node [OliveGreen] at (0.75, -0.2) {\small$\Lambda$};
        \draw[<->, OliveGreen,ultra thick] (0.1, -0.5) -- (1.9, -0.5);
    \end{tikzpicture}
    \end{center}
    \end{minipage}
    \end{tabular}
    \end{center}
    \caption{Coordinate transformation of the path integral. Instead of integrating the gauge field $a$ itself~(left), we decompose it into the magnitude part $\Lambda$ and the polarization part $\omega$, and then integrate separately $\Lambda$ and $\omega$~(right).}
    \label{fig:path_int_coordinate_transf}
\end{figure}
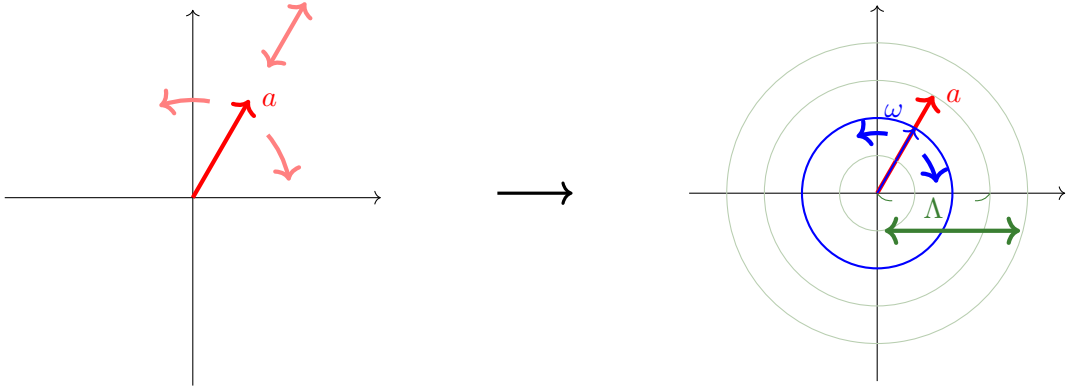

Here, we argue some connection between our QED-like theory \eqref{eq:full_QED_like_Lagrangian_coupled_fermion} and the ordinary QED.
Recall that the ordinary QED Lagrangian with a massive fermion is given by
\begin{align}
    \mathcal{L}[a,\bar{\psi}, \psi] = \bar{\psi} \gamma^\mu (\partial_\mu - \i a_\mu) \psi - m\bar{\psi}\psi + \frac{1}{4} (\partial_\mu a_\nu - \partial_\nu a_\mu) (\partial^\mu a^\nu - \partial^\nu a^\mu)\  , 
\end{align}
and its partition function is by
\begin{align}
    Z = \int \mathcal{D}a \mathcal{D}\bar{\psi} \mathcal{D} \psi e^{-S[a, \bar{\psi}, \psi]} \ , \quad S[a,\bar{\psi},\psi]:=\int \d^{d+1}x\,  \mathcal{L}[a,\bar{\psi}, \psi]\ , 
\end{align}
By decomposing the gauge field $a$ into the magnitude part $\Lambda$ and the angular part $\omega$, we change the path integral variables as (see Fig.\,\ref{fig:path_int_coordinate_transf})
\begin{align}
    \label{eq:path_int_coordinate_transf}
    a_\mu = \Lambda\omega_\mu\ , \quad
    \mathcal{D}a \rightarrow \mathcal{J} \, \mathcal{D}\Lambda \mathcal{D}\omega\ .
\end{align}
where $\mathcal{J}$ is the Jacobian coming from the path integral measure.
Under this transformation, the partition function can be rewritten as
\begin{align}
\label{eq:pathintegral_qed}
    Z 
    = \int \mathcal{D}\omega \mathcal{D}\Lambda \mathcal{D} \bar{\psi} \mathcal{D}\psi e^{-S[\Lambda, \bar{\psi}, \psi, \omega] + \log\mathcal{J}[\Lambda, \omega]} \ .
\end{align}
If the Jacobian is written as
\begin{align}
    \log\mathcal{J} = \int \d^{d+1} x\ F_{\text{Jac}}(\Lambda\omega)\ ,
\end{align}
where $F(\Lambda\omega)$ satisfies the condition \eqref{eq:F_condition}, then the effective action becomes
\begin{align}
    -S[\Lambda, \bar{\psi}, \psi,\omega] + \log\mathcal{J} = - \bar{\psi} \gamma^\mu (\del_\mu - \i \Lambda \omega_\mu) \psi + m\bar{\psi}\psi + F_{\text{Jac}}(\Lambda \omega)\  , 
\end{align}
which take the same form as the subdimensional matter theory \eqref{eq:lagrangian_foliation_fermion}.
Indeed, if we rewrite the path integral \eqref{eq:pathintegral_qed} in the following way
\begin{align}
    Z = \int \mathcal{D}\omega\, \widetilde{Z}[\omega]\ , \quad  \widetilde{Z}[\omega]:= \int \mathcal{D}\Lambda \mathcal{D} \bar{\psi} \mathcal{D}\psi e^{-S[\Lambda, \bar{\psi}, \psi, \omega] + \log\mathcal{J}[\Lambda, \omega]}\ .
\end{align}
We can interpret the tilded partition function $\widetilde{Z}[\omega]$ as the subdimensional matter theory~\eqref{eq:lagrangian_foliation_fermion}, where $\omega$ is treated as a background field. 
This may provide some hints on how to realize subdimensional particles by experiments.
It would be interesting to develop this idea in more detail.

\section{Conclusion and Future Directions}
\label{sec:conclusion}
In this paper, we explored the field theories which capture some notable features of subdimensional particles such as mobility constraints and the subextensive GSD in a manner distinct from the conventional foliated quantum field theories. 
Our starting point was the Godbillon-Vey invariant, a mathematical invariant of the manifold equipped with the foliation structure. 
In section~\ref{sec:GV_field_theory_for_single_foliation}, we successfully constructed the U(1) BF-like field theory \eqref{eq:GV_action_step4} by modifying the Godbillon-Vey field theory \eqref{eq:GV_action}, following the strategy \eqref{eq:strategy}. 
We found that this BF-like theory possesses the subsystem 1-form symmetry, where the symmetry operators are topological only along a leaf of the foliation. 
Additionally, we also showed that the GSD of our theory exhibits the desired subextensive behaviour. 
In section~\ref{sec:lattice_spin_models}, we proposed the several UV lattice models of the (continuum) U(1) BF-like theory~\eqref{eq:GV_action_step4} and its variants. 
We demonstrated that these lattice models behave similarly to the continuum theories in the sense that their GSDs become identical in the low energy limit. 
In section~\ref{sec:coupling_matter_theory}, we proposed the new dynamical matter theories which are coupled with the Godbillon-Vey theory.

There are several research directions regarding this work.
Firstly, while our continuum theory can describe some lattice spin models which have subdimensional particle excitations, it cannot completely realize the behavior of the X-cube model.
Although we can obtain the field theoretical framework like~\eqref{eq:GV_action_step4} to exhibit the same GSD as the one of the X-cube model, it is difficult to reproduce the full properties of logical operators such as the statistics of fractional excitations.
Hence, it is intriguing to construct the field theory that captures properties of the X-cube model starting form the Godbillon-Vey invariant.
Secondly, in 
mathematical literature, higher codimension foliations are also discussed
while we only considered the Godbillon-Vey invariant for the codimension one foliation. 

As discussed in appendix \ref{sec:higher_codimension}, we expect that the field theory with the higher codimension foliation structure has a non-Abelian gauge redundancy rather than the Abelian one. 
It would be interesting to study new matter phases with the higher codimension foliations.     
Thirdly, there may be ways to realize subdimentional particles by experiments.
Our discussion in section~\ref{sec:coupling_matter_theory} may shed light on this aspect.
For example, it could be realized by charged particles under an external gauge field along a fixed direction while changing its magnitude, although we may need to make an effort at the estimation of quantum fluctuations.

Our approach may have applications beyond describing foliation matter phases, potentially extending to cosmological models and string theories.
In cosmological models, a timelike foliation structure is often required to allow the $3+1$ decomposition of the spacetime manifold.
Consequently, there may be a scope for foliation field theory incorporating this timelike foliation structure.
Although timelike and spacelike foliation differ in their canonical structures, this distinction could yield significant effect on cosmological models. Also, the foliation structures could potentially emerge in string theory setups. For instance, when an infinite number of branes are stacked within the compactified space, a foliation structure may naturally arise and give rise to some non-trivial effects on our real world. This is because parameters such as Yukawa couplings in effective four-dimensional spacetime can be determined by the geometrical data of the compactified manifold. It would be intriguing to build up string phenomenological models which depend on the foliation structures of the compactified manifold.

\section*{Acknowledgement}
We thank Kantaro Ohmori, Tetsuya Onogi, Shutaro Shimamura, and Qiang Jia for helpful discussions.
H. E. is supported by KAKENHI-PROJECT-23H01097.
M. H. is supported by MEXT Q-LEAP, JSPS Grant-in-Aid for Transformative Research Areas (A) ``Extreme Universe" JP21H05190 [D01], JSPS KAKENHI Grant Number 22H01222, JST PRESTO Grant Number JPMJPR2117.
T. N. is supported by JST SPRING, Grant Number JPMJSP2110, and RIKEN Junior Research Associate Program.
S. S. is supported by Grant-in-Aid for JSPS Fellows No. 23KJ1533.

\appendix 
\section{Canonical quantization of the U(1) BF-like Theory}
\label{sec:canonical_quantization}
In section \ref{sec:GV_field_theory_for_single_foliation}, we successfully construct the U(1) BF-like field theory \eqref{eq:GV_action_step4} motivated by the Godbillon-Vey invariant. In this appendix, we present the complete discussion on canonical quantization of this theory \eqref{eq:GV_action_step4} which was skipped in the main text due to the technical reasons. The below discussions closely follow~\cite{Lin:1990kx} where the canonical quantization of $(2+1)$-dimensional Chern-Simons theory is discussed. To perform the canonical quantization, it is convenient to rewrite the Lagrangian \eqref{eq:GV_action_step4} with the normal 1-form $\omega=\omega_{1}\d x^{1}$ in the following symmetric manner:
\begin{align}
    \mathcal{L}=\frac{k}{4\pi}\,  \left(\epsilon^{\mu\nu\rho}b_{\mu}\partial_{\nu}c_{\rho}+\epsilon^{\mu\nu\rho}c_{\mu}\partial_{\nu}b_{\rho}-\epsilon^{\mu\nu 1}\omega_{1}\lambda_{\mu}\partial_{\nu}\phi+\epsilon^{\mu\nu 1}\partial_{\nu}(\omega_{1}\lambda_{\mu})\phi-2\,\epsilon^{\mu\nu 1}\omega_{1}\lambda_{\mu}b_{\nu}\right)\ . 
\end{align}
By using the above expression, we can define the canonical momentum for each dynamical field\footnote{We set $\epsilon^{012}=+1$.}:
\begin{align}\label{eq:canonical_momenta}
    \begin{aligned}
    &\pi_b^\mu := \frac{\partial \mathcal{L}}{\partial \dot{b}_\mu} = \frac{k}{4\pi} \epsilon^{0\mu \nu} c_\nu \ , \qquad \pi_c^\mu := \frac{\partial \mathcal{L}}{\partial \dot{c}_\mu} = \frac{k}{4 \pi} \epsilon^{0\mu\nu} b_\nu \ , \\ 
    &\pi_\lambda^0 := \frac{\partial \mathcal{L}}{\partial \dot{\lambda}_0} = 0\ , \qquad \pi_\lambda^2 := \frac{\partial \mathcal{L}}{\partial \dot{\lambda}_2} = \frac{k}{4\pi}\omega_{1}\phi\ , \qquad     \pi_\phi := \frac{\partial \mathcal{L}}{\partial \dot{\phi}} = -\frac{k}{4\pi}\omega_1 \lambda_2  \ . 
    \end{aligned}
\end{align}
The Poisson brackets $\{\, \cdot\ ,\ \cdot\, \}_{\text{PB}}$ among these canonical variables are given by
\begin{align}
\label{eq:canonical relations}
\begin{aligned}
    &\{X_{\mu}(\bm{x}),\pi_{X}^{\nu}(\bm{y})\}_{\text{PB}}=\delta_{\mu}^{\nu}\, \delta^{2}(\bm{x}-\bm{y})\ , \qquad X=b\,,\, c \, , \\
    &\{\lambda_{0}(\bm{x}),\pi_{\lambda}^{0}(\bm{y})\}_{\text{PB}}=\{\lambda_{2}(\bm{x}),\pi_{\lambda}^{2}(\bm{y})\}_{\text{PB}}= \delta^{2}(\bm{x}-\bm{y})\ , \qquad  \\
    &\{\phi(\bm{x}),\pi_{\phi}(\bm{y})\}_{\text{PB}}= \delta^{2}(\bm{x}-\bm{y})\ ,
\end{aligned}
\end{align}
where $\bm{x}$ and $\bm{y}$ represent the two-dimensional space coordinates.
\subsection{Primary and Secondary Constraints}
We first need to find out all constraints in the U(1) BF-like theory, which can be mainly classified into primary and secondary constraints. From the expressions of canonical momenta \eqref{eq:canonical_momenta}, we can obtain the following primary constraints ($i=1, 2$):
\begin{align}\label{eq: primary constraint1}
\begin{aligned}
    &\pi_b^0 \approx 0\ ,\qquad \pi_c^0 \approx 0\ , \qquad \pi_\lambda^0 \approx 0 \ ,  
    \end{aligned}
    \end{align}
\begin{align}\label{eq: primary constraint2}
\begin{aligned}
    &\chi_b^i := \pi_b^i - \frac{k}{4\pi} \epsilon^{0ij} c_j \approx 0 \ ,\qquad \chi_c^i := \pi_c^i - \frac{k}{4\pi} \epsilon^{0ij} b_j \approx 0\ ,  \\
    &\chi_\lambda^2 := \pi_\lambda^2 - \frac{k}{4\pi} \omega_1\phi \approx 0\ , \qquad 
    \chi_\phi := \pi_\phi + \frac{k}{4\pi} \omega_1 \lambda_2 \approx 0\ .
\end{aligned}
\end{align}
Here, the symbol ``$\approx$'' denotes the \emph{weak} equality. In other words, $f\approx g$ for two functions $f$ and $g$ defined on the phase space means that $f$ and $g$ are equal to each other only if the constraints are satisfied. On the one hand, we refer to the ordinary equality ``='' as the \emph{strong} equality, meaning that it holds at every point in the phase space. In the rest of this section, we distinguish these two different equalities. We can then define the Hamiltonian $H$ as follows:
\begin{align}
    \begin{aligned}
    H &=  \int \D^2x \left[-\frac{k}{4\pi}\left(\epsilon^{0ij} \left(b_0 \partial_i c_j + b_i \del_j c_0+c_0 \partial_i b_j + c_i \del_j b_0\right) \right.\right.\\
    &\left.\left. \qquad \qquad\qquad\quad  + \omega_1 \lambda_0 \partial_2 \phi-\partial_2(\omega_1 \lambda_0)  \phi + 2 \omega_1 \lambda_0  b_2 - 2\omega_1 \lambda_2 b_0\right)\right. \\
    &\left.\qquad\qquad\quad+\sum_{X=b,c} \left( \theta_0^X \pi_X^0 + \theta_i^X \chi_X^i \right)  + \theta^\lambda_0 \pi_\lambda^0 + \theta^\lambda_2 \chi_\lambda^2 + \theta^\phi \chi_\phi \right] ,
    \end{aligned} 
\end{align}
where the $\theta$'s in the final line denote the Lagrange multipliers associated to the primary constraints \eqref{eq: primary constraint1}-\eqref{eq: primary constraint2}. 

We also must require that all constraints satisfy, what is called, the \emph{consistency conditions}, i.e., constraints must be invariant under the time evolution. These consistency conditions may give rise to new types of constraints, which are often referred to as the secondary constraints. Indeed, by requiring that the primary constraints $\pi_{b}^{0}$, $\pi_{c}^{0}$ and $\pi_{\lambda}^{0}$ are invariant under the time evolution (e.g., $\dot{\pi}_{b}^{0}:=\left\{\pi_{b}^{0}, H \right\}_{\text{PB}}\approx 0$), we can obtain the following secondary constraints:
\begin{align}
    \label{eq:secondary_constraint1}
    \begin{aligned}
    &\xi_{b}^{0}:= \dot{\pi}_{b}^{0}=\frac{k}{2\pi}\epsilon^{0ij}\partial_{i}c_{j}-\frac{k}{2\pi}\omega_{1}\lambda_{2}\approx 0 \ , \\  
    &\xi_{c}^{0}:= \dot{\pi}_{c}^{0}=\frac{k}{2\pi}\epsilon^{0ij}\partial_{i}b_{j}\approx 0\ , \qquad  \xi_{\lambda}^{0}:= \omega_{1}^{-1}\dot{\pi}_{\lambda}^{2}=\frac{k}{2\pi}\partial_{2}\phi +\frac{k}{2\pi}b_{2}\approx 0 \ . 
    \end{aligned}
\end{align}
We should notice that the consistency conditions on the primary constraint functions $\chi$'s in \eqref{eq: primary constraint2} do not produce new constraints if we choose nice Lagrange multipliers from the beginning. 
For instance, the consistency condition on $b_{1}$ asserts
\begin{align}
    \dot{\chi}_{b}^{1}=\left\{\chi_{b}, H\right\}_{\text{PB}}=\frac{k}{2\pi}(\partial_{2}c_{0}-\theta_{2}^{c})\approx 0\ . 
\end{align}
This implies that if we take a Lagrange multiplier $\theta_{2}^{c}$ such that $\theta_{2}^{c}\approx \partial_{2}c_{0}$, the consistency condition is automatically satisfied and no secondary constraints appear. As the same reasoning, we can conclude that is we choose some of the Lagrange multipliers such that
\begin{align}
\label{eq:condition_multiplier}
    \begin{aligned}
    \theta_{i}^{c}\approx \partial_{i}c_{0}-\delta_{i}^{1}\omega_{1}\lambda_{0}\ \ , \ \ \theta_{i}^{b}\approx \partial_{i}b_{0}\ \ , \ \  \theta_{2}^{\lambda}\approx \partial_{2}\lambda_{0}+\lambda_{0}\,\partial_{2}\log \omega_{1} \ \ , \ \  \theta^{\phi}\approx -b_{0} \ , 
    \end{aligned}
\end{align}
there are no secondary constraints stemming from the consistency conditions of $\chi$'s.  

Although we seem to have non-trivial constraints from the consistency conditions on secondary constraints \eqref{eq:secondary_constraint1}, we have no further meaningful constraints. This is because the consistency conditions on secondary constraints are automatically satisfied thanks to \eqref{eq:condition_multiplier}. Indeed, the time derivatives of second constraint functions can be computed as
\begin{align}
    \begin{aligned}
        \dot{\xi}_{b}^{0}\approx \frac{k}{2\pi}\left(\epsilon^{0ij} \partial_{i}\theta_{j}^{c}-\omega_{1}\theta_{2}^{\lambda}\right)\ , \qquad \dot{\xi}_{c}^{0}\approx \frac{k}{2\pi}\epsilon^{0 i \ell}\partial_{i}\theta_{\ell}^b\ , \qquad \dot{\xi}_{\lambda}^{0}\approx \frac{k}{2\pi}\omega_{1}(\partial_{2}\theta^{\phi}+\theta_{2}^{b})\ , 
    \end{aligned}
\end{align}
and we can readily check that all these vanish on the constrained surface by using \eqref{eq:condition_multiplier}. 
From the discussions so far, we can conclude that all the constraints appearing in the U(1) BF-like theory \eqref{eq:GV_action_step4} are the primary constraints \eqref{eq: primary constraint1}-\eqref{eq: primary constraint2}, and secondary constraints \eqref{eq:secondary_constraint1}.

\subsection{First, Second Class Constraints and Dirac Bracket}
In the previous subsection, we succeeded in finding out the all constraints of the U(1) BF-like theory. We next classify the obtained constraints into the first class and second one to introduce the Dirac bracket \cite{Dirac_1950}. Firstly, the primary constraints \eqref{eq: primary constraint1} clearly belong to the first class constraints since they weakly commute with the all other constraints. Also, at first glance, the secondary constraints \eqref{eq:secondary_constraint1} seem to be second class. However, they can be linearly mapped to the first class constraints after the following linear transformations:
\begin{align}
    \begin{aligned}
        \tilde{\xi}_{b}^{0}&:=\xi_{b}^{0}+\partial_{1}\chi_{b}^{1}+\partial_{2}\chi_{b}^{2} + \chi_\phi  
        =\frac{k}{4\pi}\epsilon^{0ij}\partial_{i}c_{j}+\partial_{i}\pi^{i}_{b}-\frac{k}{4\pi}\omega_{1}\lambda_{2}+\pi_{\phi}\ ,  \\ \tilde{\xi}_{c}^{0}&:=\xi_{c}^{0}+\partial_{1}\chi_{c}^{1}+\partial_{2}\chi_{c}^{2} 
        =\frac{k}{4\pi}\epsilon^{0ij}\partial_{i}b_{j}+\partial_{i}\pi^{i}_{c}\ ,  \\
        \tilde{\xi}_{\lambda}^{0}&:=\xi_{\lambda}^{0}+\partial_{2}\left(\omega_{1}^{-1}\chi_{c}^{1}\right)+\chi_c^1 
        =\frac{k}{4\pi}\partial_{2}\phi+\partial_{2}(\omega_{1}^{-1}\pi_{\lambda}^{2})+\pi_{c}^{1}+\frac{k}{4\pi}b_{2}\ . 
    \end{aligned}
\end{align}
Indeed, we can check that the Poisson brackets between $\tilde{\xi}_{b}^{0}$, $\tilde{\xi}_{c}^{0}$ and $\tilde{\xi}_{\lambda}^{0}$, and all constraint functions vanishes. This phenomenon can be also seen in the contest of the canonical quantization of the Chern-Simons theory \cite{Lin:1990kx}. In summary, the all second class constraints of the U(1) BF-like theory are $\chi$'s listed as \eqref{eq: primary constraint2}. 

Under the preparation in the above, we can compute the Dirac bracket $\{\, \cdot , \cdot \,\}_{\text{DB}}$ defined by 
\begin{align}
    \{ A(\bm{x}) , B(\bm{y}) \}_{\text{DB}}:= \{ A(\bm{x}) , B(\bm{y}) \}_{\text{PB}}-\int \d^{2}\bm{z} \d^{2}\bm{w} \{ A(\bm{x}) , \chi_{I}(\bm{z}) \}_{\text{PB}} C_{IJ}^{-1}(\bm{z}, \bm{w}) \{\chi_{J}(\bm{w}), B(\bm{y}) \}_{\text{PB}}\  , 
\end{align}
where $I$ and $J$ run over the all second class constraint functions, and $C_{IJ}(\bm{x}, \bm{y})$ is given by
\begin{align}
    C_{IJ}(\bm{x}, \bm{y}):= \{\chi_{I}(\bm{x}), \chi_{J}(\bm{y}) \}_{\text{PB}}\ . 
\end{align}
Exploiting the canonical relations \eqref{eq:canonical relations}, some of the important Dirac brackets can be computed as
\begin{align}
    &\{b_{i} (\bm{x}),c_{j}(\bm{y})\}_{\text{DB}}=\frac{2\pi}{k}\epsilon_{0ij}\, \delta^{2}(\bm{x}-\bm{y})\ , 
    &\{\phi (\bm{x}), \lambda_{2}(\bm{y})\}_{\text{DB}}=\frac{2\pi}{k}\omega_{1}^{-1}\, \delta^{2}(\bm{x}-\bm{y})\ ,
\end{align}
\begin{align}
\label{eq:b0p0_Dorac_bracket}
    &\{b_{0} (\bm{x}),\pi_{b}^{0}(\bm{y})\}_{\text{DB}}=\{c_{0} (\bm{x}),\pi_{c}^{0}(\bm{y})\}_{\text{DB}}=\{\lambda_{0} (\bm{x}),\pi_{\lambda}^{0}(\bm{y})\}_{\text{DB}}=\delta^{2}(\bm{x}-\bm{y})\ . 
\end{align}
We next deal with the first class constraints \eqref{eq: primary constraint1}. These constraints are closely related to the gauge redundancies, hence we impose the following gauge fixing conditions:
\begin{align}
    b_{0}\approx 0\ , \quad c_{0}\approx 0 \ ,\quad \lambda_{0}\approx 0\ .
\end{align}
Remark that once we fix the gauge, the first class constraints turn into the second ones in general. As a result, we need to introduce a new Dirac bracket where the constraint functions in \eqref{eq: primary constraint1} also contribute to the matrix $C_{IJ}$. 
After some calculations in this new brackets, we find that all Dirac brackets in \eqref{eq:b0p0_Dorac_bracket} vanish. To pass from the classical theory to the quantum one, we must replace the Dirac bracket to the commutator, and arrive at the following canonical commutation relations:
\begin{align}
    [\,b_{i} (\bm{x}),c_{j}(\bm{y})\,]=\frac{2\pi\i}{k}\epsilon_{0ij}\, \delta^{2}(\bm{x}-\bm{y})\ , \quad [\,\phi (\bm{x}), \lambda_{2}(\bm{y})\,]=\frac{2\pi\i}{k}\omega_{1}^{-1}\, \delta^{2}(\bm{x}-\bm{y})\ .
\end{align}
This completes the rigorous proof of the formula \eqref{eq:com_relation} used in the main text.

\section{Precise Formulation of Godbillon-Vey Field Theories}
\label{sec:precise_formulation}
In this section, we carefully investigate the gauge invariance of the Godbillon-Vey field theories. First, we consider the original Godbillon-Vey action \eqref{eq:GV_action}
\begin{align}
    S = C \int_{M^3} 
    \Bigl[ \eta \wedge \d \eta - \lambda \wedge(\d \omega - \omega \wedge \eta) \Bigr] \ .
\end{align}
This action has the three types of the gauge symmetries given by \eqref{eq:GV_gauge_shift}-\eqref{eq:GV_gauge_lambda}.
The invariance under \eqref{eq:GV_gauge_lambda} can be obviously seen by the integrability condition \eqref{eq:omega_integrability}.
On the other hand, there are some subtlities in the proof of the invariance under \eqref{eq:GV_gauge_shift}-\eqref{eq:GV_gauge_exact}. To see this, under the gauge transformation \eqref{eq:GV_gauge_shift}, the action transforms as
\begin{align}
    S \rightarrow S + C \int_{M^3} \d (\eta \wedge u\omega + 2u\d\omega) \ .
\end{align}
If we could naively drop the last \textit{apparent} surface term, we would obtain the desired gauge invariance. We, however, should be aware whether these apparent ``exact'' forms are really exact.
We can say that the second term in the right hand side is truly exact only if the form $\eta \wedge u\omega + 2u\d\omega$ is a well-defined (namely, gauge invariant) differential form on the spacetime manifold $M^3$.
However, we cannot say so because under the gauge transformation \eqref{eq:GV_gauge_exact}, the form $\eta \wedge u\omega + 2u\d\omega$ transforms as
\begin{align}\label{eq:exact_term_gauge_transformation1}
    \eta \wedge u\omega + 2u \d \omega \rightarrow \eta \wedge u\omega + 2u\d \omega + u \d (\log c) \wedge \omega.
\end{align}
Hence, the interpretation of the second term as an integration of an exact form breaks down. 
A similar problem also happens for the gauge redundancy \eqref{eq:GV_gauge_exact}. Indeed, the action transforms under \eqref{eq:GV_gauge_exact} as
\begin{align}
    S \ \rightarrow\ S - C \int_{M^3} \d ( (\log c)\,  \d \eta )\ .
\end{align}
However, the last apparent surface term is not well-defined as an integration since the form $(\log c)\,  \d \eta $ is not invariant under the gauge redundancy \eqref{eq:GV_gauge_shift}:
\begin{align}\label{eq:exact_term_gauge_transformation2}
    (\log c)\, \d \eta \ \rightarrow\ (\log c) \, \d \eta + (\log c)\, \d (u\omega)\ .
\end{align}

In the following, we elaborate on why these break downs occur. Although we have treated the field $\eta$ (or $\omega$, other fields as well) as if it was a 1-form field defined on $M^3$, it should be, strictly speaking, expressed as
\begin{align}
    \eta = s^* \eta_{P}\ ,
\end{align}
where $s^*$ is a pullback by a section of a principal bundle (i.e., $s:M^{3}\rightarrow P$, where $P$ is the principal bundle) and $\eta_P$ is a 1-form field defined on the principal bundle. Suppose that the transformation of the action takes the following form:
\begin{align}\label{eq:transformation_action_section_rep}
    S \ \rightarrow\  S + \int_{M^3} s^*(\d \Xi_{P})\ ,
\end{align}
where $\Xi_P$ is some differential form defined on the principal bundle $P$. 
If we can take $s$ to be a global section, then we can drop the last term in the right hand side since the pullback by a global section and the external derivative commute to each other: 
\begin{align}\label{eq:commutative_property}
    s^*(\d \Xi_{P}) = \d (s^* \Xi_{P})\ ,
\end{align}
and we can exploit the Stokes theorem without any issue.

However, we should remark that such a global section does not always exist, and instead we usually work with a patchwork of local sections $\{s_{i}\}$, which are defined only on local patches (i.e., $s_{i}: U_i \rightarrow P\vert_{U_i}$, where $U_i\subset M^{3}$ is a local patch labelled by the index $i$, and $P\vert_{U_i}$ is the local trivialization of $P$ on $U_i$). In such case, we should define $\eta$ for each local patch:
\begin{align}
    \eta(x) = (s_i^*\, \eta_{P}) (x) \ , \quad x \in U_i \ .
\end{align}
We emphasize that this $\eta$ is not a well-defined form on $M^3$, since $(s_i^* \eta_{P})(x)$ and $(s_j^* \eta_{P})(x)$ are different by a gauge transformation function for $x \in U_i \cap U_j$. 
This fact makes $\Xi$ in \eqref{eq:transformation_action_section_rep} ill-defined as a differential form on $M^3$, and we cannot neglect the apparent surface term in general. Nevertheless, we fortunately have a chance to remedy this difficulty. If $\Xi_P$ is a gauge invariant form, then we can uniquely define its pullback $\tilde{s}^*\, \Xi_P$ as
\begin{align}
    (\tilde{s}^*\, \Xi_{P})(x) := (s_i^* \, \Xi_{P}) (x)\ , \quad x \in U_i\ .
\end{align}
This definition manifestly ensures that $\tilde{s}^* \, \Xi_{P}$ is independent of the choice of the local sections:
\begin{align}
    (s_i^* \, \Xi_{P})(x) = (s_j^* \, \Xi_{P})(x)\ ,\quad x \in U_i \cap U_j\ .
\end{align}
Thanks to the commutable property: 
\begin{align}
    \tilde{s}^*(\d \Xi_P) = \d (\tilde{s}^* \Xi_P)\ , 
\end{align}
and the Stokes theorem, we can drop the surface term and prove the gauge invariance in a strict manner.

The above discussion implies that the Godbillon-Vey action is not well-defined as a gauge invariant action since the surface term is not gauge invariant as seen in \eqref{eq:exact_term_gauge_transformation1} and \eqref{eq:exact_term_gauge_transformation2}. Nevertheless, we can construct a well-defined Godbillon-Vey field theory in the same spirit as \cite{Dijkgraaf:1989pz}.
Instead of working in a three-dimensional manifold $M^{3}$, we prepare a four-dimensional manifold $M^4$ whose boundary is $M^{3}$ and consider the action\footnote{\label{fn:cobordism}
We can always take such a manifold because all closed orientable three-dimensional manifolds are null-cobordant.
} \footnote{Note that the normal 1-form $\omega$ is not extended to $M^4$. If we do so, then we have to take a four-dimensional manifold with a foliation structure. Thus, we cannot use the fact from the cobordism theory described in footnote \ref{fn:cobordism}.}
\begin{align}
    S' = C \left(\int_{M^4} \d \eta \wedge \d \eta - \int_{M^3} \lambda \wedge (\d \omega - \omega \wedge \eta)\right)\ .
\end{align}
This action precisely possesses the three types of gauge symmetries, and particularly the symmetries \eqref{eq:GV_gauge_exact}-\eqref{eq:GV_gauge_lambda} hold without any surface terms. Also, under \eqref{eq:GV_gauge_shift}, the action transforms as
\begin{align}
    S' \ \rightarrow\  S' + C \int_{M^3} \d (2\eta \wedge u\omega + 2u\d \omega)\ ,
\end{align}
where the form inside the external derivative is gauge invariant unlike the original action. Therefore, the pullback by a local section $\tilde{s}^*(2\eta_{P} \wedge u_{P}\omega_{P} + 2u_{P}\d\omega_{P})$ is well-defined. As explained above, we can then use the Stokes theorem to drop the second term in the right hand side. Also this theory $S'$ should be independent of four-dimensional extensions. this conditions do not give rise to non-trivial conditions on the coefficient $C$ since the Chern number on a closed manifold $N^{4}$ is always zero for $\mathbf{R}_{+}$ gauge theories:
\begin{align}
    \int_{N^4} \d \eta \wedge \d \eta = 0\ .
\end{align}

In the same manner, we propose a four-dimensional extension of the U(1) BF like theory \eqref{eq:action_U(1)_GVBF} in the gauge invariant way:\footnote{The same discussion is applicable for other Godbillon-Vey field theories like \eqref{eq:GV_action_step3}.}
\begin{align}
    \frac{k}{2\pi} \left( \int_{M^4} \d b \wedge \d c - \int_{M^3} \lambda \wedge (\d\phi \wedge \omega - \omega \wedge b)\right).
\end{align}
Note that in the case of U(1) gauge group, the Chern number becomes non-trivial, which leads to the level quantization $k\in\mathbf{Z}$.

\section{Field Theory of Higher Codimension Foliation}
\label{sec:higher_codimension}
In the main text, we constructed U(1) BF-like theory based on the Godbillon-Vey invariant with a codimension one foliation.
In this appendix, we generalize such discussions to include higher codimension foliations. 
In section \ref{sec:GV_field_theory_for_single_foliation}, the integrability condition \eqref{eq:omega_integrability} plays the crucial role in the construction of the Godbillon-Vey characteristic class \eqref{eq:Godbillon-Vey_definition}. We fortunately have the integrablity condition for codimension $n$ foliations which is given by
\begin{align}\label{eq:_integrable_condition_for_multi_fol}
    \d  \omega_{i}=-\sum_{j=1}^{n}\theta_{ij}\wedge \omega_{j}\ , \quad i=1\, , 2\, , \cdots n \ , 
\end{align}
where $\omega_{i}$ is a normal 1-form field and $\Theta_{ij}$ is some 1-form field. For simplicity of our notations, we rewrite the above condition \eqref{eq:_integrable_condition_for_multi_fol} in the following way:
\begin{align}\label{eq:Frobenius_cond_multi_fol}
    \d \bm{\omega}=-\bm{\Theta}\wedge \bm{\omega} \ , 
\end{align}
where $\bm{\omega}$ and $\bm{\Theta}$ are defined by
\begin{align}
\begin{aligned}
    \bm{\omega}:=
    \begin{pmatrix}
        \omega_{1} \\
        \omega_{2} \\ 
        \vdots \\
        \omega_{n}
    \end{pmatrix}\ , \quad 
    \bm{\Theta}:=
    \begin{pmatrix} 
  \theta_{11} & \theta_{12} & \dots  & \theta_{1n} \\
  \theta_{21} & \theta_{22} & \dots  & \theta_{2n} \\
  \vdots & \vdots & \ddots & \vdots \\
  \theta_{n1} & \theta_{n2} & \dots  & \theta_{nn}
\end{pmatrix}\ . 
\end{aligned}
\end{align}
Most importantly, the integrability condition \eqref{eq:Frobenius_cond_multi_fol} is invariant under the following non-Abelian gauge transformation:
\begin{align}\label{eq: gauge trsf nonAbelian}
    \bm{\omega}\mapsto \bm{\omega}':= U \bm{\omega}\ , \quad \bm{\Theta}\mapsto \bm{\Theta}':= U\,\bm{\Theta}\,U^{-1}+U\,\d\, U^{-1}\ ,  
\end{align}
where $U\in\text{GL}(n,\mathbf{R})$. The gauge invariance can be easily checked as follows:
\begin{align}
    \begin{aligned}
         \d \bm{\omega}'+\bm{\Theta}'\wedge \bm{\omega}'&=\d U \wedge \bm{\omega} +U\D \bm{\omega}+U\,\bm{\Theta}\,U^{-1}\wedge U\bm{\omega}+U\,\d\, U^{-1}\wedge U\bm{\omega} \\
         &=\d U \wedge \bm{\omega} +U(\d \bm{\omega}+\bm{\Theta}\wedge\bm{\omega})+U\,\d\, U^{-1}\wedge U\bm{\omega} \\
          &=\d U \wedge \bm{\omega} -\d U \wedge \bm{\omega}\\
          &=0\ . 
    \end{aligned}
\end{align}
In particular, if we restrict to $n=1$, this gauge transformation is reduced to \eqref{eq:GV_gauge_exact}. Also, the physical meaning of this gauge redundancy is that the foliation structure characterized by $\bm{\omega}$ is unchanged under the linear transformation of $\bm{\omega}$.
From the expression of gauge transformations \eqref{eq: gauge trsf nonAbelian}, we can naively construct the ``Godbillon-Vey number'' for codimension $n$ foliations in terms of this gauge connection $\bm{\Theta}$ as follows:
\begin{align}\label{eq:GV_higher_codim}
    \text{GV}[\bm{\Theta}] := \frac{k}{4\pi}\int \text{Tr}\left[\bm{\Theta}\wedge \D \bm{\Theta}+\frac{2}{3}\bm{\Theta}\wedge\bm{\Theta}\wedge\bm{\Theta}\right]\ , \quad k \in \mathbf{Z}\ ,
\end{align}
which is similar to the $\text{GL}(n,\mathbf{R})$ Chern-Simons theory.
However, this number for $n>1$ is not truly an invariant of the foliation structure. In other words, this number depends on the specific choice of the gauge connection $\bm{\Theta}$.
Indeed, if we pick another $\tilde{\bm{\Theta}}$, which satisfies the condition \eqref{eq:Frobenius_cond_multi_fol}, the above ``Godbillon-Vey number'' is changed.

This situation is largely different from the codimension one foliation where the Godbillon-Vey number is invariant under the $\omega$ gauge transformation. For the codimension one foliation case, the Godbillon-Vey number is invariant under the $\omega$-dimensional shift gauge symmetry \eqref{eq:GV_gauge_shift}, which forces the field theory to be invertible phase. On the other hand, the ``Godbillon-Vey number'' for a higher codimension foliation is not invariant under such a shift symmetry, which implies the existence of non-trivial phase. Namely, if we promote this number to a physical action in a similar manner to \eqref{eq:GV_invariant}:
\begin{align}
\begin{aligned}
    S &= \frac{k}{4\pi} \int \text{Tr} \left[ \bm{\Theta} \wedge \d \bm{\Theta} + \frac{2}{3} \bm{\Theta} \wedge \bm{\Theta} \wedge \bm{\Theta} \right] - \bm{\lambda}^{\text{T}} \wedge (\d \bm{\omega} + \bm{\Theta} \wedge \bm{\omega} ) \\
    & = \frac{k}{4\pi}\int \theta_{ij} \wedge \d \theta_{ji} + \frac{2}{3} \theta_{ij} \wedge \theta_{jk} \wedge \theta_{ki} - \lambda_i \wedge (\d \omega_i + \theta_{ij} \wedge \omega_j )\ ,
    \end{aligned}
\end{align}
it could have room for a nontrivial ground state Hilbert space structure.
The equations of motion are
\begin{align}
    \begin{aligned}
    &\d \theta_{ij} + 2\theta_{ik} \wedge \theta_{kj} - \lambda_i \wedge \omega_j = 0\ ,  \\
    &\d\omega_{i} + \theta_{ij} \wedge \omega_j = 0\ .
    \end{aligned}
\end{align}
Particularly, the first equation can be rewritten as follows:
\begin{align}
    \mathcal{F}^\theta_{ij}\vert_{\omega_j^T} := (\d\theta_{ij} + 2 \theta_{ik} \wedge \theta_{kj} )\vert_{\omega_j^\text{T}} = 0\  , 
\end{align}
where, $|_{\omega_{j}^\text{T}}$ stands for the restriction of the form to the transverse directions to the foliation normal field $\omega_{j}$. This implies the the existence of subsystem 1-form symmetries. 
Note that the gauge group $\text{GL}(n,\mathbf{R})$ has maximal compact subgroup $\text{O}(n)$. Therefore, the level $k$ is quantized even though we do not follow the strategy \eqref{eq:strategy}.
This can be seen by%
\footnote{Of course, we can complexify the gauge group just like performed in section \ref{sec:GV_field_theory_for_single_foliation}. For this case, the polar decomposition for GL$(n, \mathbf{C})$ becomes $ \text{GL}(n,\mathbf{C}) \overset{\text{top.}}{\cong} \text{U}(n) \times \mathbf{R}^{n^2}$.
}
\begin{align}
    \text{GL}(n,\mathbf{R}) \overset{\text{top.}}{\cong} \text{O}(n) \times \mathbf{R}^{\frac{n(n+1)}{2}}\ ,
\end{align}
where ``top.'' means the this isomorphism is not a group isomorphism but a topological isomorphism (namely, homeomorphism). However, due to the difficulty of splitting compact gauge fields and non-compact ones, it seems hard to obtain the subextensive GSD like section \ref{subsec:subGSD}. We leave further investigation of this Godbillon-Vey field theories for higher codimension foliations which reproduce the property of the subextensive GSD to the future work.

\section{Global Spacetime Symmetries and G-structures}
\label{sec:g-structure}
In this appendix, we elaborate on a relationship between the spacetime symmetries in field theories and background mathematical structures in terms of \textit{G-structure}\footnote{
The interested reader is referred to 
\cite{kobayashi1972transformation} or \url{https://en.wikipedia.org/wiki/G-structure_on_a_manifold} for more details.} \cite{bams/1183527777} and its generalization.
The concept of a G-structure encompasses a wide range of geometrical structures, including foliation and Riemannian structures. We provide its precise definition later in this section.

We start with recalling the spacetime symmetries in field theories. The spacetime symmetry on a $(d+1)$-dimensional manifold $M^{d+1}$ is a diffeomorphism 
\begin{align}
    f : M^{d+1}\rightarrow M^{d+1}\ ,
\end{align}
under which the background fields associated with mathematical structures are invariant.
In the following, we describe what we mean by this statement in details.  
Under a general orientation preserving diffeomorphism $f$, the action of a field theory should be invariant:
\begin{align}\label{eq:invariant_condition}
    S[a,b,\cdots; X,Y,\cdots]\ = \ S[f^* a, f^*b, \cdots; f^* X, f^* Y, \cdots]\ ,
\end{align}
where $a,b,\cdots$ are physical degrees of freedom such as a scalar field $\phi$ or a gauge field $A$, and $X,Y,\cdots$ are background fields corresponding to the given mathematical structures such as a foliation structure $\langle \omega \rangle$ or a metric $g$. Also, $f^*$ is an induced map acting on each field associated to the diffeomorphism $f$.
In particular, when the diffeomorphism $f$ keeps the background fields $X, Y, \cdots$: 
\begin{align}
     f^* X=X\ , \quad f^* Y=Y\ , \cdots\ ,
\end{align}
the invariant condition \eqref{eq:invariant_condition} becomes
\begin{align}
    S[a,b,\cdots; X,Y,\cdots] = S[f^* a, f^* b,  \cdots; X, Y,  \cdots ]\ .
\end{align}
This is nothing but the spacetime symmetries often referred in physics literature. Remark that the background fields $X, Y, \cdots$ serve as just parameters in the sense that they belong to the trivial representations under the diffeomorphism $f$.
These diffeomorphisms forms an automorphism group preserving mathematical structures. 

\subsection{Examples of G-structures in Field Theories}
For the concreteness, we provide some examples of field theories and spacetime symmetries below.

\subsubsection*{Example 1. Topological field theories}
In ordinary topological field theories such as the 
Chern-Simons theory \cite{cmp/1104161738} or topological BF theory \cite{Horowitz:1989ng}, the non-trivial mathematical structure is an orientation $\mu$ of the spacetime manifold.
Thus, the automorphism group is an orientation preserving diffeomorphism group on $M^{d+1}$:
\begin{align}
    \text{Auto} (M^{d+1}; \mu) = \{\,f \in \text{Diff}(M^{d+1})\, \vert\, f^* \mu = \mu\,\}\ .
\end{align}
Of course, the global structure of the automorphism group depends on the details of a spacetime manifold $M^{d+1}$ and mathematical structures on it. 
However, the Lie algebra associated with the spacetime symmetry only depends on local information, and it can be written by
\begin{align}\label{eq:Lie_algebra_sym1}
    \mathfrak{gl}(d+1,\mathbf{R}) \oplus \mathfrak{t}^{d+1}\ ,
\end{align}
where $\mathfrak{gl}$ and $\mathfrak{t}^{d+1}$ mean the Lie algebras corresponding to general linear group and translations, respectively. 
As usual, by exponentiating this algebra, we obtain the Lie group expression which is valid in a local coordinate patch:
\begin{align}\label{eq:exponentiated_form}
    \text{GL}(d+1, \mathbf{R})^+ \ltimes \mathbf{R}^{d+1}_{\,\text{translation}} \ ,
\end{align}
where $\text{GL}(d+1, \mathbf{R})^+$ is a positive determinant general linear group, and $\mathbf{R}^{d+1}_{\,\text{translation}}$ is the group of translations.

\subsubsection*{Example 2. Field theories with broken time reversal symmetry} Another case is a field theory possessing the Lorentz symmetry while missing the time-reversal symmetry. The most familiar example to high energy physicists is the standard model where the CP symmetry is broken (or equivalently, T symmetry is broken due to the CPT theorem).
Here, the mathematical structures are a pseudo Riemannian metric $g$, the orientation of the spacetime manifold $\mu$, and the orientation of the time direction $\xi$. The automorphism group is described by
\begin{align}
    \text{Auto} (M^{d+1}; g, \mu, \xi) = \{\,f \in \text{Diff}(M^{d+1})\, \vert\, f^*g = g\, , \, f^* \mu = \mu\,, \, f^* \xi = \xi\, \}\ .
\end{align}
The Lie algebra corresponding to the above automorphism group is
\begin{align}
    \mathfrak{so}(3,1) \oplus \mathfrak{t}^{d+1}\ ,
\end{align}
and the spacetime symmetry group in local coordinate patch is\footnote{The superscript $+$ here should not be confused with other $+$'s such as $\text{GL}(d+1, \mathbf{R})^+$. The $+$ here means that the $(0,0)$-component of the matrix should be positive, and other $+$'s mean that the determinant of the matrix should be positive.}
\begin{align}
    \text{SO}(3,1)^+ \ltimes \mathbf{R}^{d+1}_{\,\text{translation}}\ .
\end{align}

\subsubsection*{Example 3. Conformal field theories} 
In (Euclidean) conformal field theories (CFTs), in addition to the orientation $\mu$, we have a mathematical structure $[g]$ which is the equivalence class of the metric $g$ defined by the following identifications under Weyl transformations: 
\begin{align}
    g \sim g' \overset{\text{def.}}{\Longleftrightarrow} g' = e^\rho g\ , \quad \rho\in \mathbf{R}\ . 
\end{align}
For these CFTs, we have the following automorphism group:
\begin{align}
    \text{Auto} (M^{d+1}; [g], \mu) = \{\,f \in \text{Diff}(M^{d+1})\, \vert\, f^*[g] = [g]\, , \, f^* \mu = \mu\, \}\ .
\end{align}
The corresponding Lie algebra is
\begin{align}\label{eq:cft_absense_defect}
    \mathfrak{so}(d+2,1)\cong \left(\mathbf{M} \oplus  \mathbf{D} \oplus \mathbf{K}\right) \oplus\mathbf{P}\ , 
\end{align}
where $\mathbf{M}$, $\mathbf{D}$, $\mathbf{K}$ and $\mathbf{P}(=\mathfrak{t}^{d+1})$ are the conformal generators for rotation, dilatation, special conformal transformation and translation symmetries, respectively. 
The spacetime symmetry can be locally described by 
\begin{align}
    \text{SO}(d+2,1)^+\ .
\end{align}

\subsubsection*{Example 4. Foliation field theories}
In the foliation field theories like Godbillon-Vey field theories, the given mathematical structures are the orientation $\mu$ and the codimension $n$ foliation structure $\langle \omega_i \rangle$ where $i$ runs over $1, 2, \cdots, n$. The automorphism group for a foliation field theory is 
\begin{align}
    \text{Auto} (M^{d+1}; \langle \omega_{i}\rangle, \mu) = \{\,f \in \text{Diff}(M^{d+1})\, \vert\, f^*\langle \omega_{i}\rangle = \langle \omega_{i}\rangle\, , \, f^* \mu = \mu\, \}\ .
\end{align}
The corresponding Lie algebra is
\begin{align}
    \mathfrak{gl}(d+1-n,n) \oplus \mathfrak{t}^{d+1} \ , 
\end{align}
and the spacetime symmetry becomes 
\begin{align}
    \text{GL}(d+1-n,n)^+ \ltimes \mathbf{R}^{d+1}_{\text{translation}}\ .
\end{align}
Here, we define the linear group $\text{GL}(d+1-n,n)^+$ by
\begin{align}
\begin{aligned}
    &\text{GL}(d+1-n,n)^+ \\
    &= \left\{ 
    \begin{pmatrix}
    A & 0 \\
    B & C \\
    \end{pmatrix}
    \in \text{GL}(d+1,\mathbf{R})^+ \middle\vert 
    \begin{matrix}
    A \in \text{GL}(d+1-n,\mathbf{R})\ , \ C \in \text{GL}(n,\mathbf{R}) \\
    B: n \times (d+1-n)\ \text{matrix}
    \end{matrix}
     \right\}\ ,
\end{aligned}
\end{align}
and $\mathfrak{gl}(d+1-n,n)$ is the corresponding Lie algebra.

As seen from these examples, we can convince ourselves that for each mathematical structure, the spacetime symmetry is decomposed into a semi-direct product of a matrix group and a translation group. 
Such matrix group parts can be recast in terms of the G-structure, which characterizes mathematical structures by the reduction of the structure group $\text{GL}(d+1,\mathbf{R})$ of the frame bundle. 
This correspondence is ensured by the fact that a part of the symmetry transformation induces the action of the structure group in the frame bundle.
In summary, we can understand the spacetime symmetries with given mathematical structures in terms of G-structures.

\subsection{Reduction of Translational Symmetries in Defect Field Theories}
Until here, we argued the connections between matrix parts of spacetime symmetries and G-structures. We can naturally extend the above discussions to the case where not only the matrix groups but also translation groups are partly broken to their subgroups. These situations can naturally occurs in the context of physics. For instance, the spacetime can host lower-dimensional defects in CFTs, where the full conformal symmetry is broken to the subgroup which does not change the configurations of defects \cite{Billo:2016cpy, Gadde:2016fbj}. In the rest of this appendix, we will see these examples in details.
\subsubsection*{Example 1. Field theories on spacetime manifold with a puncture} 
The simplest example is the case where a spacetime manifold $M^{d+1}$ admits a puncture $x_{0} \in M^{d+1}$ (namely, pointed space).
    In the absence of any mathematical structures other than this puncture,\footnote{The orientation was needed for physical theories to construct the action integral. We do not need it only for considering the automorphism of the manifold.} the general spacetime symmetry which does not change the location of $x_0$ is written as
    \begin{align}
        \text{Auto}(M^{d+1}; x_{0})=\{\,f \in \text{Diff}(M^{d+1})\, \vert\,  f(x_{0}) = x_{0} \,\}\ .
    \end{align}
    The Lie algebra of this symmetry transformation behaves differently depending on the position.
    On the fixed point $x_{0}$, the symmetry structure can be described by the following Lie algebra:
    \begin{align}
        \mathfrak{gl}(d+1,\mathbf{R})\ . 
    \end{align}
    Note that the translation symmetry in the right hand side of \eqref{eq:Lie_algebra_sym1} is fully broken because the translations change the location of the point $x_{0}$. 
    Also, the form of the symmetry group structure around $x_0$ is given by
    \begin{align}
        \text{GL}(d+1,\mathbf{R})\ .
    \end{align}
    On the other hand, around any point other than $x_0$, the Lie algebra and its exponentiated group of the spacetime symmetry are just the same as \eqref{eq:Lie_algebra_sym1} and \eqref{eq:exponentiated_form}. 

    \subsubsection*{Example 2. Field theories in the presence of  defects} 
    The second example is a spacetime manifold with a $q$-dimensional defect $N^q \subset M^{d+1}$. If we do not have any other mathematical structures, the spacetime symmetry is the form of
    \begin{align}
        \text{Auto}(M^{d+1}; N^q) = \{\, f\in \text{Diff}(M^{d+1})\, \vert\,  f(N^q) = N^q\,\}\ .
    \end{align}
    The structure of this symmetry transformation around the point on $N^q$ is described by
    \begin{align}
        \mathfrak{gl}(q,\mathbf{R}) \oplus \mathfrak{gl}(d-q+1,\mathbf{R}) \oplus \mathfrak{t}^{q}\ .
    \end{align}
    Also, the symmetry group structure around a point on $N^q$ is given by exponentiating the Lie algebra:
    \begin{align}
        \left(\text{GL}(q,\mathbf{R}) \times \text{GL}(d-q+1,\mathbf{R})\right) \ltimes \mathbf{R}^{q}\ , 
    \end{align}
    where the matrix group $(\text{GL}(q,\mathbf{R}) \times \text{GL}(d-q+1,\mathbf{R}))$ is given by
    \begin{align}
    \begin{aligned}
        &(\text{GL}(q,\mathbf{R}) \times \text{GL}(d-q+1,\mathbf{R})) \\
        &\quad\quad  =\left\{ 
        \begin{pmatrix}
            A & 0 \\
            0 & B
        \end{pmatrix}
        \in \text{GL}(d+1,\mathbf{R})\,
        \,\middle\vert\, A \in \text{GL}(q, \mathbf{R}) \ , \text{GL}(d-q+1, \mathbf{R}) \right\}\ .
    \end{aligned}
    \end{align}
    On the other hand, around any point apart from $N^q$, the Lie algebra and its exponentiated group of the spacetime symmetry are just the same as \eqref{eq:Lie_algebra_sym1} and \eqref{eq:exponentiated_form}.
    \subsubsection*{Example 3. Defect conformal field theories}
    The final example is the defect CFT. When we have a $(p+1)$-dimensional (typically) planer or spherical defect $\mathcal{D}^{p+1}$ into a $(d+1)$-dimensional Euclidean CFT on a spacetime $M^{d+1}$, the spacetime symmetry can be written as
    \begin{align}
        \text{Auto}(M^{d+1}; \mathcal{D}^{p+1}, [g])=\{\, f\in\text{Diff}(M^{d+1})\, \vert\, f(\mathcal{D}^{p+1}) = \mathcal{D}^{p+1}, \ f([g]) = [g]\, \}\ .
    \end{align}
    The Lie algebra of this spacetime symmetry around $\mathcal{D}^{p+1}$ becomes
    \begin{align}
        \mathfrak{so}(p+2, 1)\oplus\mathfrak{so}(d-p)\ , 
    \end{align}
    where the first part corresponds to the conformal algebra along with the conformal defect $\mathcal{D}^{p+1}$, and the second one is the Lie algebra of the rotational symmetry around the defect. 
    On the other hand, for other points, the Lie algebra for a spacetime symmetry is just the same as \eqref{eq:cft_absense_defect}. These symmetry structures correspond to the fact that the bulk and defect local primary operators can be characterized by the irreducible representations of $\mathfrak{so}(d+2, 1)$ and $\mathfrak{so}(p+2, 1)\oplus\mathfrak{so}(d-p)$, respectively. As usual, by exponentiating these Lie algebras, we can obtain the local expression of these spacetime symmetries.

\bibliographystyle{utphys}
\bibliography{fracton_realization}

\end{document}